# A comprehensive study on the processing of Co:ZnO ceramics: defect chemistry engineering and grain growth kinetics

R. T. da Silva[a], J. M. Morbec[b‡], G. Rahman[c†], and H. B. de Carvalho[d]*

[a] *Universidade Federal de Ouro Preto – UFOP, 35400-000 Ouro Preto, MG, Brazil.*
[b] *School of Chemical and Physical Sciences, Keele University, Keele ST5 5BG, United Kingdom*
[c] *Department of Physics, Quaid-i-Azam University, 45320 Islamabad, Pakistan.*
[d] *Universidade Federal de Alfenas - UNIFAL, 37130-000 Alfenas, Brazil.*

*Corresponding Authors*:
‡ j.morbec@keele.ac.uk
† gulrahman@qau.edu.pk.
* hugo.carvalho@unifal-mg.edu.br.

**ABSTRACT:** In this report we present a systematic study on the preparation of Co:ZnO ceramics via standard solid-state route from different Co precursors ($Co_3O_4$, CoO and metallic Co) and atmospheres ($O_2$ and Ar). Particular emphasis was done on the defect chemistry engineering and on the sintering growth kinetics. First-principles calculations based on density functional theory were employed to determine the formation energy of the main point defects in ZnO and Co:ZnO systems. Based on the theoretical results a set of chemical reactions was proposed. A detailed microstructural characterization was performed in order to determine the degree of Co incorporation into the ZnO lattice. The samples prepared in Ar atmosphere and from metallic Co presents the highest Co solubility limit (lower apparent Co incorporation activation energy) due to the incongruent ZnO decomposition. The determination of the parameters of the sintering growth kinetics reveals that $Co_3O_4$ is the best sintering additive in order to achieve higher densities in both sintering atmospheres. The results give evidences that the sintering in $O_2$ is effective in promoting zinc vacancies in the ZnO structure, while the sintering in Ar promotes zinc interstitial defects. Our findings give valuable contribution to the understanding of the preparation of Co-doped ZnO ceramics and the sintering growth kinetics, what would allow to improve the state of the art on the processing of the material at both bulk and nanometric scales.

## 1 INTRODUCTION

ZnO is one of the truly multifunctional material due to its extensive and interesting physical and electrochemical properties. The optimization of the ZnO properties for a particular application, the ZnO functionalization, takes place mainly via defect engineering, by doping or even by the introduction of defects into the structural lattice in a well-controlled manner [1]. Specifically, transition metal (TM) doped ZnO has been explored as a dilute magnetic semiconductor (DMS) to be used as spin injection layer into spintronic semiconductor devices. Spintronics is attracting considerable attention in the last decades by potentially implement new data storage and quantum computing devices [2-5]. However, the obtained results, regarding the nature of the usual observed room temperature ferromagnetism (RTFM), are very controversial. In addition to the trivial extrinsic origins of the RTFM that result from the ferromagnetic secondary phases, the main theoretical models that are currently available for describing the origins and properties of the observed RTFM are all linked to structural defects [6-11]. This linkage explains why the ferromagnetic properties are rarely reproducible. Before elucidating the mechanisms of intrinsic FM, the measured magnetic hysteresis loops must be classified as intrinsic or extrinsic. Contamination by magnetic elements [12], measurement errors [13] and segregated secondary phases [14, 15] can all cause ferromagnetic signals.

From the theoretical point of view, different models have been proposed to explain the usual observed RTFM in DMS's. Here, the main accepted model for insulating systems is the bound magnetic polaron (BMP) theory [16] were the ferromagnetic exchange coupling among the TM doping elements is mediated by shallow donor electrons. These defects form bound magnetic polarons, which overlap to create a spin-split impurity band. For the particular case of the Co-doped ZnO, the $Co^{2+}$ ions can hybridize effectively with shallow-donor impurity bands in ZnO because the states related to the complex $Co^+$ ($Co^{2+}$ + $e^-_{donor}$) are also of shallow-donor character [17]. Considering the question about the nature of the necessary shallow-donor defect, several experimental reports argue that the RTFM in the TM-doped ZnO system, explained in terms of the BMP model, is associated to oxygen vacancies ($V_O$) [18-21]. In fact, the defect state related to $V_O$ in the wurtzite ZnO structure has a donor character, however it is a deep-donor state [22, 23]. Although states of deeper-donors can hybridize to magnetic dopants, their smaller Bohr radius would require relatively higher dopant and defect concentrations to achieve a necessary spatial overlapping, however at such short-range antiferromagnetic superexchange interactions would take place. Instead, the most promising shallow-donor defect in the ZnO structure is the zinc interstitials ($Zn_i$) [22, 24]. From the experimental point of view, there are a growing number of reports in the literature giving clear evidences of the relation between the observed RTFM and defects at the zinc sites [11, 25-27].

Nanostructured TM-doped ZnO has also been considered for biomedical applications due to its low-toxicity in bioimaging and drug delivery systems [28], and as antibacterial agent [29] . However, it is a well-known fact that the incorporation of dopant at nanoscale is a very difficult task [30], even for highly soluble dopants, the incorporation of a significant amount of dopant atoms during synthesis is not straightforward. Even when dopants are incorporated, their concentration is typically an order of magnitude less than in the growth solution [31]. These results have led to theoretical efforts to understand the mechanisms that control the doping process [32]. An interesting strategy to overcome this problem is the top-down approach, where the preparation of nanostructured materials is performed via, for example, mechanical milling of bulk materials [33, 34]. At the bulk scale a homogeneous distribution of the dopants can be achieved before the subsequent reduction in the dimensionality. The gridding would lead also to surfaces of the nanoparticles with a high degree of defects, that can enhance particular desired properties, like the necessary visible fluorescence in bioimaging [35, 36].

Therefore, the study and determination of the phase diagram of ZnO and TM oxides is important to assure the synthesis of TM-doped ZnO ($Zn_{1-x}TM_xO$) magnetic semiconductor. Besides, the controlling of the presence and the density of specific point defects is necessary in order to tune the desired properties and functionalities in the development of the aforementioned technologies. Also, the knowledge of the sintering kinetics parameters and its correlations are fundamental in the processing of bulk material, and in its eventual reduction of dimensionality down to the nanoscale. In this context, the aim of the present report is to give further contribution in the understanding of the Co incorporation process into the wurtzite ZnO (*w*-ZnO) lattice, and how it can affect some important properties of the material, such as their sintering kinetics. Here Co:ZnO samples were prepared from different Co precursor ($Co_3O_4$, CoO, and metallic Co) in different sintering atmosphere (oxygen and argon). First-principles calculations were performed to give support to the defect chemical analysis and insight into the mechanisms of the Co incorporation in the ZnO. A detailed structural characterization was employed by conjugating several different techniques to determine the structures of the samples, the Co apparent incorporation activation energy, and the grain growth kinetic parameters.

## 2 EXPERIMENTAL AND THEORETICAL METHODS

Polycrystalline Co:ZnO samples were prepared via solid-state reaction method. Stoichiometric amounts of high purity powders of ZnO (Alfa Aesar 99.99% purity), and different Co precursor sources: $Co_3O_4$ (Alfa Aeser, 99.7% purity), CoO (Sigma Aldrich 99.99% purity), metallic Co (Sigma Aldrich, 99.9% purity). The samples prepared with $Co_3O_4$, CoO, and metallic Co was labeled as $Co_3O_4$:ZnO, CoO:ZnO, and $m$Co:ZnO, respectively. All the samples were prepared with Co atomic concentration of 8 at.% in respect of the cationic sites of a potential single phase $Zn_{1-x}Co_xO$ (Co-doped $w$-ZnO), it means $x = N_{Co}/(N_{Co} + N_{Zn}) = 0.08$ ($N$ is the number of atoms). The precursors were manually mixed and ground in a planetary ball mill (Retsch PM 100) using tungsten carbide jar and spheres at rotation speed of 500 RPM for 4 h, the ratio between the masses of the spheres and the powder was kept at 13:1. Isopropyl alcohol was also added to the samples in order to optimize a grinding process. The resulting powder was dried and sintered in oxygen ($O_2$) or argon (Ar) atmospheres in the temperature range of 600 to 1200 °C for 4 h with heat/cooling rate of 10 °C/min. Flowchart in Figure 1 illustrates the procedure employed in the preparation of the samples.

The crystal structures of the Co:ZnO powders were investigated by X-ray diffraction (XRD) performed in the range of $2\theta = 15°-120°$ in steps of 0.02° at 7 s/step using Cu-K$\alpha$ radiation ($\lambda = 1.542$ Å) in a Rigaku Ultima IV diffractometer. The determination of the lattice parameters and the occupation factor over the structure were evaluated using the Rietveld method as implemented by the General Structure Analysis System (GSAS) software package with the graphical user interface EXPGUI [37, 38]. Raman spectroscopy was also used to add information about the Co doping and the resulting lattice disorder, as well as to analyze the formation of segregated secondary phases. Photoluminescence (PL) measurements were performed in order to directly check the Co incorporation into the *w*-ZnO lattice. Both Raman and PL measurements were carried out at room temperature on a modular spectrometer consisting of an Olympus B-X41 microscope and a Horiba iHR550 monochromator at the backscattered photon detection geometry. A 532 nm B&W Tek solid-state laser was used as source of excitation. The morphology and the grain size distribution of the sintered samples were evaluated via scanning electron microscope (SEM) in a Hitachi S-4800 FEG-SEM.

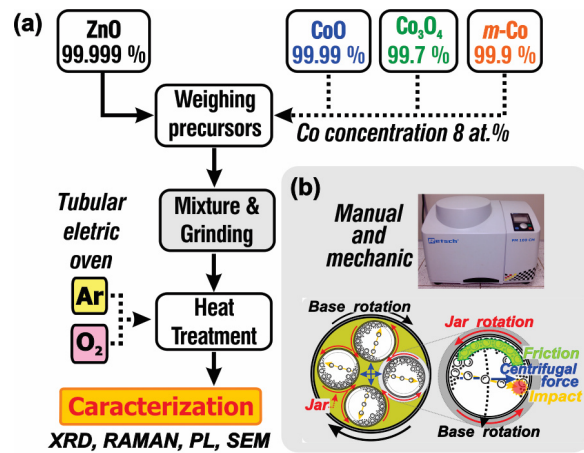

**Figure 1.** (a) Flowchart of the preparation of the Co:ZnO samples. (b) Schematic representation of the rotating ball milling showing impact milling actions.

The defects formation energies in the Co:ZnO system were also investigated by means of first-principles calculations based on density functional theory (DFT) [39]. We used the Siesta code [40], which employs norm-conserving Troullier–Martins pseudopotentials and linear combinations of atomic orbitals [41]. We used norm-conserving Troullier-Marting pseudopotentials and a double-zet set with polarization functions (DZP) for all atoms with a real-space energy cutoff of 200 Ry. We considered *w*-ZnO cluster with 80 atoms (40 Zn and 40 O), simulated within the supercell approach with a vacuum of ~10 Å between the cluster and its image. The dangling bonds at the surface were kept unsaturated and the atoms were allowed to relax to their minimum energy configurations. All atomic positions were fully relaxed until the forces on each atom were smaller than 0.02 eV Å$^{-1}$.

## 3 RESULTS AND SISCUSSION

### 3.1 *First principles calculations and defect chemistry*

The defect chemistry is a power toll in the understanding of the properties of materials and in their development, especially for ceramics. Nevertheless, it is a difficult task to determine which exact reaction occurs during a specific processing. The case of ZnO is not different. Considering the cobalt doping of the *w*-ZnO lattice using different precursors and different temperatures of sintering and atmospheres, several different possible defect chemical reactions are possible. To get the right direction and reach the right chemical reaction set of equations, some theoretical and experimental considerations have to be taken into account.

Figure 2(a) presents the calculated formation energies ($E_f$) for the main point defects in *w*-ZnO (most common reported): oxygen vacancy ($V_O$), zinc interstitial ($Zn_i$), zinc vacancy ($V_{Zn}$) [42, 43], for the cobalt incorporation into the *w*-ZnO lattice ($Co_{Zn}$), and the same point defects in the presence of cobalt already incorporated in the *w*-ZnO ($V_O$+Co, $Zn_i$+Co and $V_{Zn}$+Co). Zinc oxide is an unintentional *n*-type semiconductor, and its conductivity was considered for a long time to be due to zinc excess in the non-stoichiometric compound $Zn_{1+\delta}O$, with zinc interstitials ($Zn_i$) been the dominating lattice defects [44]. However, $Zn_i$, in spite of been a sallow-donor defect [11, 23], has a relatively high $E_f$ in the pristine *w*-ZnO lattice in both zinc- and oxygen-rich conditions. Further theoretical results also demonstrate $Zn_i$ has a low migration barrier of 0.57 eV [23]. Besides, $V_O$ have a relatively low $E_f$, but it is still high in Zn-rich condition (Figure 2). Different from $Zn_i$, $V_O$ is a deep-donor defect [11, 23], and as a consequence it cannot provide electrons to the conduction-band by thermal excitation. Consequently, the observed *n*-type conductivity cannot be attributed to the $V_O$. In fact, the nature of the *n*-type conductivity in ZnO is still a matter of debate. However, unintentional hydrogen doping in ZnO has been considered a more likely electron source [45]. The $E_f$ of $V_{Zn}$ is relatively higher as compared to that of the $V_O$. $V_{Zn}$ are deep-acceptor defects [11, 23], and thus it is unlikely that $V_{Zn}$ can play any role in *p*-type conductivity. Our results also demonstrate that the Co incorporation into the *w*-ZnO are energetically favorable in both zinc- and oxygen-rich conditions, what explains the achieved relatively high Co solubility limit in the $w$-ZnO lattice [46]. Considering the Co-doped $w$-ZnO, the obtained $E_f$ for the analyzed point defects ($Zn_i$, $V_O$, and $V_{Zn}$) changes drastically, all of them are now energetically favorable. To address more properly the changes Figure 2(b) presents the difference in the $E_f$ for each point defect between the pristine and the Co-doped $w$-ZnO systems. We observe that the main differences are related to the O-rich condition, and that $V_{Zn}$ is the most affected defect by the Co-doping in both conditions. These results indicate that the processing of the Co:ZnO system would be quite different from that of pure $w$-ZnO, and it would lead to a quite different defect structure and physical-chemical properties that could be engineered in order to achieve functionalities that cannot be possible, or at least difficult, in the undoped $w$-ZnO.

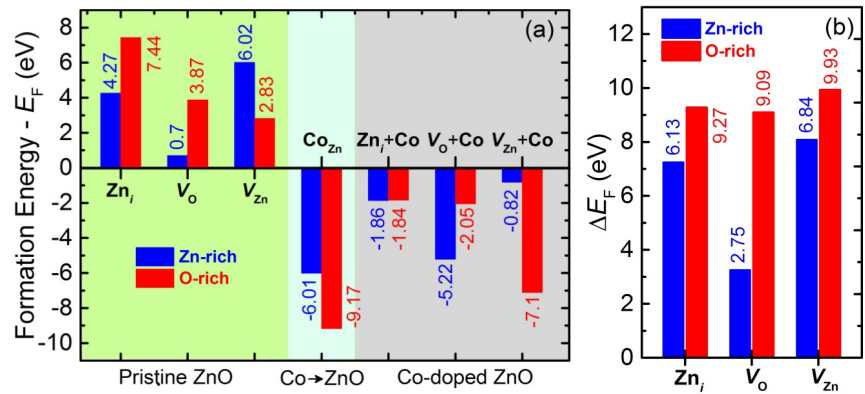

**Figure 2.** (a) Calculated formation energies ($E_f$) in Zn-rich and O-rich conditions for native point defects in pristine ZnO ($Zn_i$, $V_O$, $V_{Zn}$ and $Co_{Zn}$) and Co-doped $w$-ZnO ($Zn_i$+Co, $V_O$+Co, and $V_{Zn}$+Co). (b) Difference between the defect formation energy for pristine ZnO and the Co-doped ZnO ($\Delta E_f = E_{f,ZnO} - E_{f,Co:ZnO}$).

Facing these results, we proceed into defining the set of more plausible chemical reactions for our set of Co-doped $w$-ZnO samples. In spite of the calculated high $E_f$ for $Zn_i$ in both conditions (4.27 and 7.44 eV at Zn-rich and O-rich conditions, respectively), experimentally it was observed that under heat treatment between 300–800 °C and at lower oxygen partial pressures, $w$-ZnO decomposes incongruently by first formatting $V_O$, that has a lower $E_f$ of only 0.7 eV at Zn-rich condition, on the surfaces of the material. Once the zinc atoms become mobile ($Zn_i$ are much more mobile than $V_O$ [23]), the zinc can be dissolved into the ZnO lattice as $Zn_i$ according to Equation (1) [47, 48]. With the presence of the Co, both defects, $Zn_i$ and $V_O$, are energetically feasible, and the ZnO incongruent decomposition (Equation (1)) become more intense. Our previous report on the Co-doped ZnO samples confirms, via photoluminescence and *dc*-electrical measurements, an increase of the $Zn_i$ concentration under heat treatment performed in reducing atmosphere [11]. The *w*-ZnO incongruent decomposition ends up leading to Zn-rich environment, allowing us to consider the sintering performed in Ar in this work comparable to the Zn-rich condition in our theoretical DFT calculations presented above. On the other hand, the sintering in $O_2$ may lead to the oxygen adsorption and the formation of mainly $V_{Zn}$ as described in Equation (2). Our calculated $E_f$ of the $V_{Zn}$ is relatively high, 2.83 eV, which makes the process described in Equation (2) unprobeable. In fact, this process would lead to the creation of holes ($h^{\bullet}$) in the valence band decreasing the *w*-ZnO *n*-type conductivity, which has not yet reported. However, with the Co the $E_f$ of the $V_{Zn}$ changes drastically, assuming the value of –7.1 eV. Thus, we consider the reaction described in Equation (2) in support to the following analysis.

$$ZnO \leftrightarrow Zn_i^{\bullet\bullet} + \frac{1}{2}O_2 + 2e' \quad (1)$$

$$\frac{1}{2}O_2 \leftrightarrow O_O^x + V_{Zn}'' + 2h^{\bullet} \quad (2)$$

The CoO-ZnO solid solution is described in Equation (3). Once the CoO has the same *w*-ZnO stoichiometry, the reaction is independent on the processing atmosphere and on the *w*-ZnO previous defect structure. The Co incorporation into the *w*-ZnO via CoO as a precursor also does not promote any additional specific point defects. In turn, for the $Co_3O_4$-ZnO solid solution one has to take into account the sintering atmosphere. The $Co_3O_4$ crystallize in a normal spinel structure with $Co^{2+}$ and $Co^{3+}$ located at tetrahedral and octahedral sites, respectively [49]; and for a defect chemistry analysis it can be thought as a composition of CoO and $Co_2O_3$, $Co_3O_4 \leftrightarrow CoO + Co_2O_3$ [50]. In Ar (low oxygen pressure) with relatively high $Zn_i$ concentration (Equation (1)), the Co incorporation (from $Co_3O_4$) is properly described via Equations (3) and (4). Here the Co incorporation into the *w*-ZnO lattice leads to the increase of the electron concentration (*n*). Besides, in $O_2$ atmosphere the main process is related to the oxygen incorporation with the Co taking the Zn sites leading to formation of $V_{Zn}$. However, we have also to consider that experimentally it is observed in Co-doped $w$-ZnO samples prepared form $Co_3O_4$ in O-rich condition that the $Co^{3+}$, taking the place of the $Zn^{2+}$ in the $w$-ZnO lattice, undergoes a reduction process assuming the +2 oxidation state by taking a free electron from conduction-band, reducing the $w$-ZnO $n$-type conductivity [6, 8, 11, 46]. In such case, the entire process is described by Equations (2), (3) and (5).

$$CoO \xrightarrow{ZnO} Co_{Zn}^{x} + O_{O}^{x} \qquad (3)$$

$$Co_2O_3 \xrightarrow{ZnO} 2Co_{Zn}^{\cdot} + 2e' + 2O_{O}^{x} + \frac{1}{2}O_2 \qquad (4)$$

$$Co_2O_3 + 2e' \xrightarrow{ZnO} 2Co_{Zn}^{x} + 3O_{O}^{x} + V_{Zn}'' \qquad (5)$$

Considering the metallic Co insertion into the $w$-ZnO lattice the scenario is quite different, the process involves the oxidation of the Co atom. In Ar atmosphere, the Co incorporation takes place by sitting in the Zn site and forming $V_O$. Here, we have to consider a probable oxidation of the metallic Co once placed in an oxide matrix, and that, as pointed before, $V_O$ is a deep-donor defect, its ionization is quite unlikely. In such situation, the Co-doping of the $w$-ZnO lattice can be described by Equation (6) with no changes in the $w$-ZnO $n$-type conductivity. In the $O_2$ atmosphere we infer that the main process corresponds to the oxygen adsorption described in Equation (2), followed for the Co filling the zinc vacant site, with the holes compensated by the electrons related to the Co oxidation process. These processes are simply described by Equation (7).

$$Co \xrightarrow{ZnO} Co_{Zn}'' + V_{O}^{\cdot\cdot} \rightarrow Co_{Zn}^{x} + V_{O}^{x} \qquad (6)$$

$$3Co + 2O_2 \rightarrow Co_3O_4 \xrightarrow{ZnO} Co_{Zn}^{x} + 2Co_{Zn}^{\cdot} + 4O_{O}^{x} + V_{Zn}'' \qquad (7)$$

In summary, the defect chemistry of the Co-doped $w$-ZnO is very rich and diverse, depending on the precursor, CoO, $Co_3O_4$ or metallic Co, and on the atmosphere. Besides the total Co incorporation into the $w$-ZnO lattice, different point defects are introduced, leading to changes of the electron band structure, corresponding to changes in the $w$-ZnO electrical transport, optical and magnetic properties.

3.2 *Studies on the Co incorporation into the w-ZnO lattice:*

Figure 3 presents the XRD patterns for the *m*Co:ZnO set of samples prepared in $O_2$ (Figure 3(a)) and in Ar (Figure 3(b)) up to 1000 °C. The XRD data for all the Co:ZnO set of samples are presented in the supplementary file (Figures S1, S2, and S3). The observed diffraction pattern for all the samples sintered in $O_2$ and Ar at 1000 ºC and at higher temperatures reveals, within the detection limit of the technique, only the characteristic diffraction peaks corresponding to those for *w*-ZnO lattice (ICDD crystal chart no. 36-1451). No segregated phases were detected, indicating that at around 1000 °C and higher temperatures the Co atoms are taking the place of the Zn in the *w*-ZnO lattice forming the ternary compound $Zn_{1-x}Co_xO$ (substitutive doping). For samples sintered at temperatures below 1000 ºC, in both atmospheres, it is observed, besides the *w*-ZnO diffraction pattern, the presence of diffraction peaks associated with metallic Co and Co oxides (CoO and $Co_3O_4$). Qualitatively, it is observed that the diffraction peaks related to the secondary phases decreases in intensity with increasing temperature of sintering. However, the dynamics of the Co incorporation into the *w*-ZnO lattice is distinct as considering the atmosphere of sintering and the different precursors, thus indicating that the incorporation is quite sensitive to these parameters.

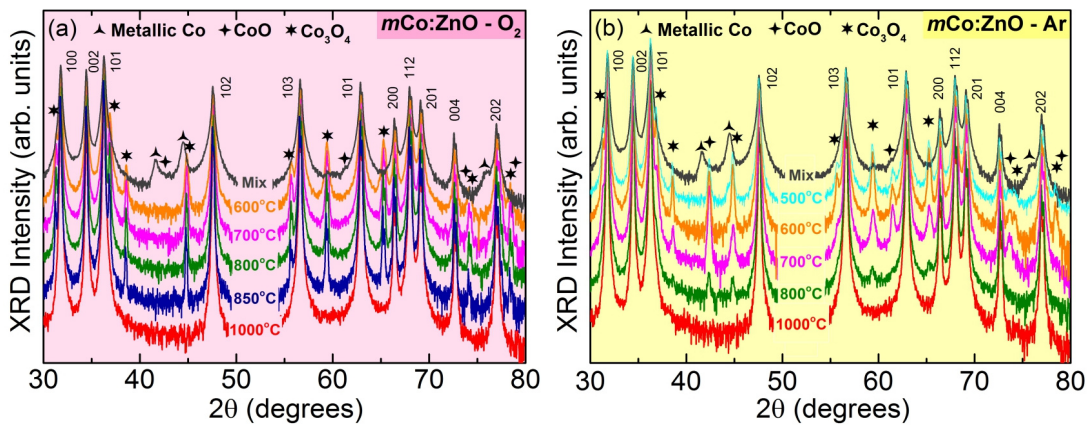

**Figure 3.** X-Ray diffraction pattern for the *m*Co-ZnO set of samples prepared in (a) oxygen ($O_2$) and (b) argon (Ar) atmospheres up to 1000 °C.

Rietveld refinement of the XRD data was performed to quantify the lattice parameters, the phase percentages (*w*-ZnO, metallic Co, CoO and $Co_3O_4$) and the elemental occupation factors into the $w$-ZnO structure for the elements Zn, Co and O for each set of samples. The obtained results are presented in Table S1–S6 in the supplementary file. We observe that the occupation factor for the Co into the $w$-ZnO lattice, which is assumed as the effective concentration of cobalt incorporated into the ZnO matrix ($x_E$), increases as the temperature of sintering increases. It reaches the maximum nominal concentration of 8 at.% for temperatures around 900–1000 °C. Figure 4 shows the calculated unit cell volume ($V$ in Table S2, S4 and S6) as a function of the obtained effective Co concentration incorporated in the $w$-ZnO lattice in our samples ($x_E$ in Table S1, S3 and S5). We observe a slight linear increase of $V$ as a function of $x_E$, as expected by the Vegard's law for solid state solutions, indicating the Co incorporation into the $w$-ZnO lattice. Nevertheless, the increase in the cell volume is not expected, since the ionic radii of tetrahedrally coordinated $Co^{2+}$ ($r_{Co}$ = 0.58 Å) is slightly smaller than that for $Zn^{2+}$ in the $w$-ZnO structure ($r_{Zn}$ = 0.60 Å) [51]. Such kind of behavior was already reported by Kolesnik *et al.* [52]. Here we attribute this unexpected result to the induced structural defects promoted by the Co incorporation. However, further studies are necessary to address this point.

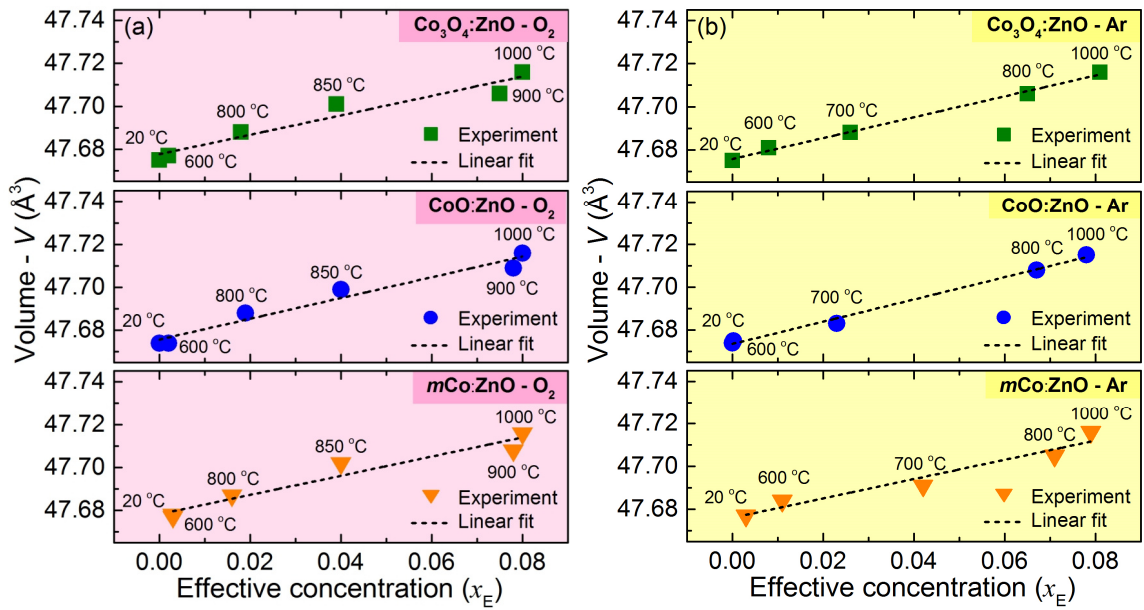

**Figure 4.** Variation of the $w$-ZnO lattice volume ($V$) as a function of the Co effective concentration ($x_E$) incorporated into the $w$-ZnO lattice for the set of samples prepared in (a) oxygen ($O_2$) and (b) argon (Ar) atmospheres. The linear behavior of $V \times x_E$ is consistent with the Vegard's law.

In order to complement the structural characterization of the samples we performed Raman scattering spectroscopy measurements. Figure 5 presents the obtained spectra for the *m*Co:ZnO set of samples prepared in $O_2$ (Figure 5(a)) and in Ar (Figure 5(b)) up to 1000 °C. Figure S4 in the supplementary file present the measured spectra for the CoO:ZnO and $Co_3O_4$:ZnO set of samples. The modes at 98, 330, 384 and 435 $cm^{-1}$ are associated with first and second order modes characteristic of the *w*-ZnO and correspond to the modes $E_{2L}$, $2E_{2L}$ at the M-point of the Brillouin zone (BZ), $A_1$(TO) e $E_{2H}$, respectively [46, 53]. The modes related to cobalt oxides are also shown in Figure 5. The indexation of the cobalt oxide modes was performed after measure the spectrum for the raw precursors (Figure S5). An important observed feature is the presence of modes above 700 $cm^{-1}$ and the relative high intensity of the mode at ~522 $cm^{-1}$ (open stars in Figure 5). As can be seen in Figure S5, these features are not present in the raw cobalt oxide precursors, instead, they are clearly observed in the spectra for Zn-doped cobalt oxides, indicating an interdiffusion of the cations between the *w*-ZnO and cobalt oxides (CoO and $Co_3O_4$) during sample processing.

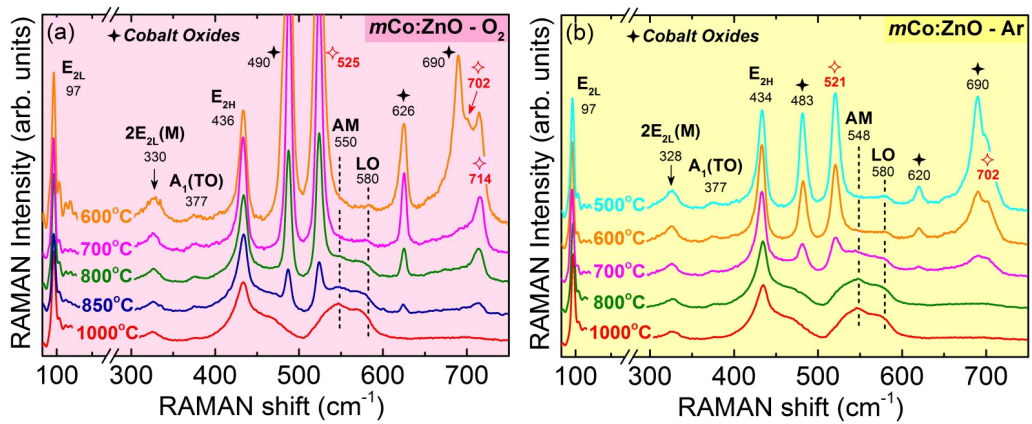

**Figure 5.** Raman scattering spectra for the *m*Co:ZnO set of samples prepared in (a) oxygen ($O_2$) and (b) argon (Ar) atmospheres up to 1000 °C. The spectra were acquired at room temperature and are normalized by integrated area of the $E_{2H}$ vibrational mode obtained by gaussian fit of the spectra. The stars (symbols) indicate the cobalt oxide related modes. The modes marked with the open stars indicate the Zn-doping of the CoO and $Co_3O_4$.

As the temperature of sintering increases the modes related to the cobalt oxides decreases indicating a progressive Co incorporation into the *w*-ZnO lattice. It is also important to note the emergence of a broadband between 500 and 600 $cm^{-1}$ as the temperature of sintering increases. This broadband consists of the convolution of several modes, but the main ones are centered ~550 and ~580 $cm^{-1}$. The peak in 580 $cm^{-1}$ is assigned to the overlapping between the $A_1$(LO) and $E_1$(LO) modes, however the peak in 550 cm$^{-1}$ cannot be attributed in any of the $w$-ZnO vibrational modes, it is indexed as AM (additional mode) in the spectra. The observation of this broad band and its dependence on the dopant concentration is well report on the literature for doped $w$-ZnO with different elements, including Mn [7, 9, 54] and Co [6, 32, 46], and its nature is correlate to structural disorder/distortions induced by the incorporation of dopants into the $w$-ZnO lattice [55]. Consequently, the observation of the broadband for our samples and its increase as the temperature of sintering is raised is an indication that a growing fraction of the Co atoms is successively incorporated into the $w$-ZnO lattice as the temperature of sintering increases.

Figure 6 presents the PL spectra of the $m$Co:ZnO set of samples prepared in O$_2$ (Figure 6(a)) and in Ar (Figure 6(b)) up to 1000 °C obtained at room temperature and excited with a 532 nm (2.33 eV) laser light. The spectra for the other set of samples, Co$_3$O$_4$:ZnO and CoO:ZnO, are presented if Figures S6 and S7, respectively. The emission band at 1.8 eV (690 nm) is associated with relaxations from excited states to $^2$E($^2$G) state, followed by a transition from $^2$E($^2$G) to the $^4$A$_2$($^4$F) ground state of Co$^{2+}$ in the 3$d^7$ high-spin configuration in a tetrahedral crystalline field formed by the O$^{2-}$ ions in its neighborhood [56, 57]. The identification of the PL spectra related to the Co$^{2+}$ inner transitions is an evidence that the Co ions in our samples are undoubtedly incorporated into the $w$-ZnO lattice. As the temperature of sintering increases the amount of the Co ions incorporated the Zn tetrahedral site of the $w$-ZnO lattice increases, leading to the increase of the PL intensity as function of the temperature of sintering observed in Figure 6(a) and 6(b). The calculated integrated area under the PL spectrum for each temperature plotted as a function of $x_E$ obtained from the Rietveld structural refinement is presented in Figure 6(c) and 6(d). It is observed a linear behavior, this result support and validates the calculated values for $x_E$ for the different temperatures of sintering. We can also observe a redshift of the maximum of the PL spectrum as function of the temperature of sintering. Since this emission is related to the structural crystalline field, the redshift can be understood in terms of the structural disorder/distortions introduced by the increasing of the Co concentration incorporated into the $w$-ZnO lattice, in concordance with the Raman results, and the calculated changes in the cell parameters obtained via the Rietveld structural refinement.

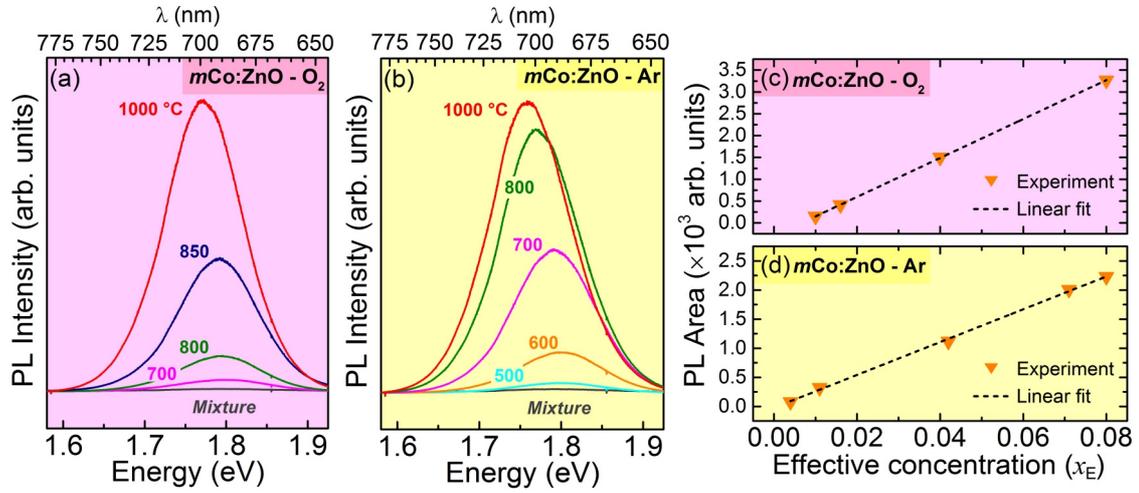

**Figure 6.** Room temperature PL spectra for part of the set of samples $m$Co:ZnO prepared in (a) oxygen ($O_2$) and (b) argon (Ar) atmospheres. It was used a 532 nm laser for excitation with optical power density of 50 kW/cm$^2$. Integrated area underneath the spectra plotted as a function of the calculated effective cobalt concentration ($x_E$) for the samples prepared in (c) $O_2$ and in (d) Ar.

As previously highlighted, the fraction of incorporated Co into the $w$-ZnO lattice depends on the temperature of sintering, on the atmosphere, and on the Co precursor source, which indicates different chemical dynamics for each set of parameters as discussed in the previous section. Here, a very important factor that can be extracted from our results is the apparent activation energy ($Q_i$) for the Co incorporation into the $w$-ZnO lattice. Figure 7 present the Arrhenius plot of the $x_E$ versus the absolute temperature of sintering for each used atmosphere. The obtained $Q_i$ for each sample is presented also in Figure 7. These results show a substantial difference between the samples sintered in Ar and $O_2$. The $Q_i$ for samples sintered in Ar are relatively low as compared to the values obtained for the set of samples processed in $O_2$. Considering the precursors, metallic Co stands out among the three used ones presenting the lowest $Q_i$ in both atmospheres.

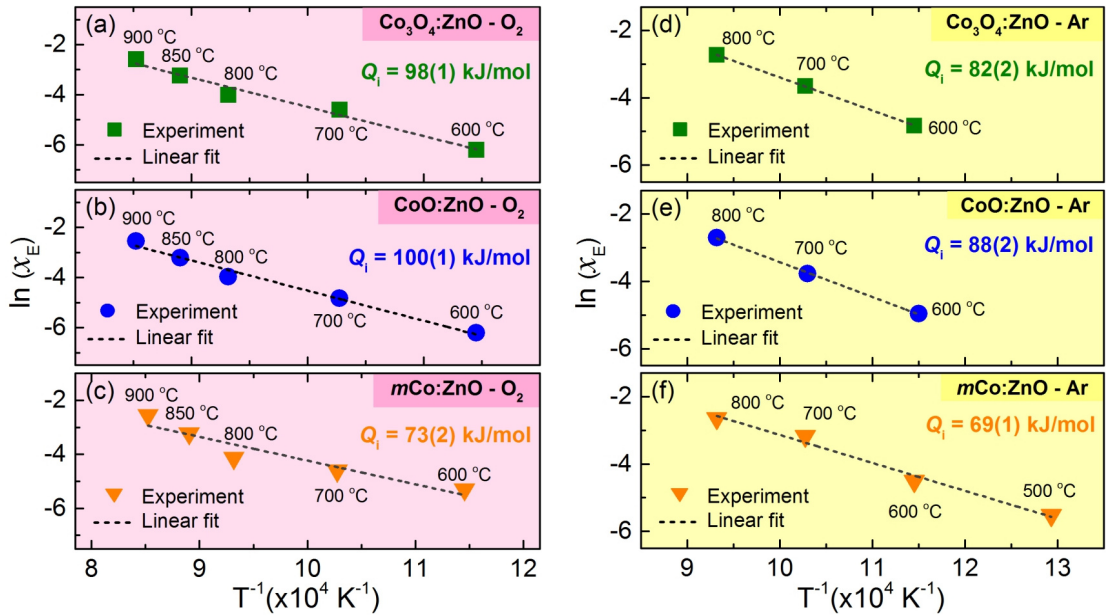

**Figure 7.** Arrhenius plot. Logarithm of the effective Co concentration ($x_E$) *versus* the inverse of the temperature of sintering for the set of samples prepared in (a) oxygen ($O_2$) and (b) argon (Ar) atmospheres. In the inset it is presented the apparent activation energy ($Q_i$) for the Co incorporation into the $w$-ZnO lattice for each case.

From the Rietveld refinement we can extract information to understand the obtained $Q_i$ values. First, at low temperature (600 °C) and under $O_2$ atmosphere, the CoO oxidize almost completely to $Co_3O_4$. These explain why the $Q_i$ for the CoO:ZnO and $Co_3O_4$:ZnO set of samples sintered in $O_2$ are quite similar. Besides, at Ar atmosphere, in a lesser extent as compared to the $O_2$ atmosphere, the CoO (CoO:ZnO set of samples) and the metallic Co ($m$Co:ZnO) also oxidizes, CoO to $Co_3O_4$ and metallic Co to CoO and to $Co_3O_4$ (Table S3 and S5). The oxidation of the Co precursors in Ar indicates that the ZnO decomposes, most probably via the chemical defect reaction described by Equation (1). Therefore, we can infer that the decomposition of the ZnO in Ar atmosphere favors the Co incorporation into the $w$-ZnO lattice, leading to the relatively lower $Q_i$ for these set of samples. Among the samples sintered at the same atmosphere the CoO:ZnO set of samples presents the higher $Q_i$ as compared with the others precursor sources. This behavior can be addressed to the CoO higher melting point (1930 °C) as compared to that of the $Co_3O_4$ (895 °C) and the metallic Co (1495 °C). It is noteworthy that the fraction of CoO remains in the powder mixture even at higher temperature (800 °C) due to the reduction of the formed $Co_3O_4$ again into CoO in Ar atmosphere at this temperature range (Table S3) [58]. Finally, as mentioned before, the $m$Co:ZnO set of samples presents the lowest $Q_i$ for both $O_2$ and Ar atmospheres. For instance, at 700 °C in Ar atmosphere the Co solubility limit achieved with metallic Co is 4.2 at.%, while for CoO and $Co_3O_4$ it is 2.3 and 2.6 at.%, respectively (Tables S1, S3 and S5). Taking again into account the difference between the melting temperatures for the $Co_3O_4$ (895 °C) and for the metallic Co (1495 °C), and considering the entire process of the material dissociation, changing of oxidation sate, and lattice incorporation, we can conclude that the oxidation of the metallic Co is considerably less energetically expensive than the reduction of the cobalt in the $Co_3O_4$ during the Co incorporation into the $w$-ZnO lattice.

3.3 *Studies on the Grain Growth Kinetics of the Co:ZnO Ceramics*

In this section we present the studies concerning the grain growth kinetics for the Co:ZnO set of samples prepared with the different precursors: $Co_3O_4$ ($Co_3O_4$:ZnO), CoO (CoO:ZnO) and metallic Co ($m$Co:ZnO). As a reference, a pure ZnO (ZnO) set of samples were processed at the same condition as the Co:ZnO samples. Figure S8 presents representative SEM images acquired over the surface of the compacted resulted powder mixtures (ZnO, $Co_3O_4$:ZnO, CoO:ZnO and $m$Co:ZnO) after the high-energy ball milling process, and before the sintering for comparison. Representative SEM imagens of the set of sintered samples at different temperatures (600–1200 °C) in $O_2$ and Ar for 4 hours are presented in Figures S9, S11, S13 and S15 for the ZnO, $Co_3O_4$:ZnO, CoO:ZnO and $m$Co:ZnO, respectively. The histograms concerning the distribution of the main particle diameters are presented in Figures S10, S12, S14 and S16, and the parameters obtained after a statistical analysis performed via log-normal fit of the histograms are presented in Tables S7, S8, S9 and S10 also for the ZnO, $Co_3O_4$:ZnO, CoO:ZnO and $m$Co:ZnO, respectively. The statistical analysis gives us the main diameter ($L$) of the particles and the geometric standard deviation ($\sigma_g$) of the distributions. Grain sizes ($G$) is obtained directly from the $L$ as described by Mendelson [59]:

$$G = 1.56\,L. \tag{8}$$

The grain size statistical analysis for each mixture presented in Figure S8 reveals a sharp distribution with a mean diameter (*L*) around 100 nm, confirming the nanostructured nature of the starting powders and the efficiency of the high-energy ball milling process in reduce the dimensionality of the grain powder mixtures down to the nanoscale. The obtained statistical numbers are quite similar, revealing a homogeneous starting point for the entire set of studied samples. This condition is very important in order to compare the grain growth parameters for each studied cobalt precursor ($Co_3O_4$, $CoO$ and *m*Co). Figure 8 present the grain size (*G*) for the set of samples sintered in (a) $O_2$ and (b) Ar for 4 hours. It is observed that the increase of the grain size with temperature has an exponential character, and that the $Co_3O_4$:ZnO/ZnO set of samples present the highest/lowest growth rate in both $O_2$ and Ar atmospheres. This result indicates that the $Co_3O_4$ is a good additive for the ZnO sintering in order to achieve larger grains and densities.

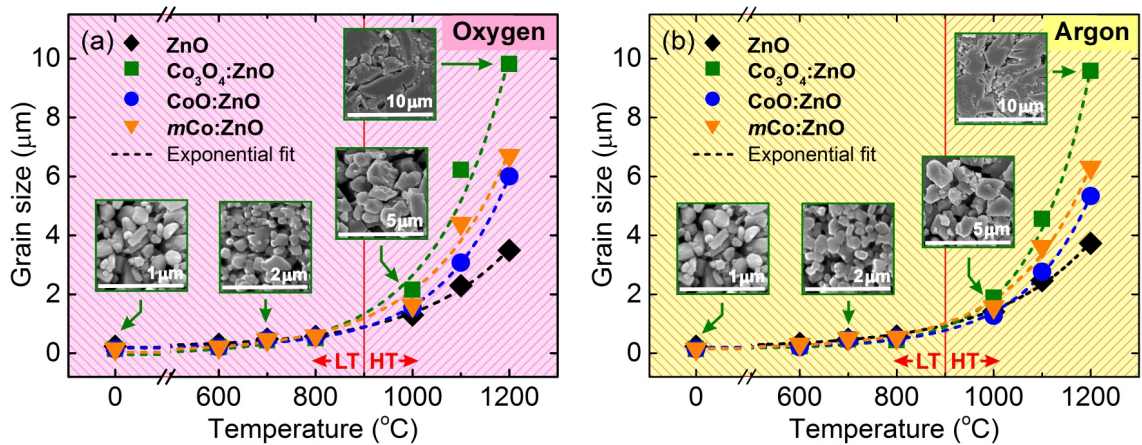

**Figure 8.** Grain size (*G*) of the samples sintered in (a) oxygen ($O_2$) and (b) argon (Ar). The grain size corresponds to the main diameter obtained after a statistical analysis performed via log-normal fit of the size distribution histograms. The data was fit via an exponential function to highlight the exponential character of the grain growth. In the inset is presented representative electron micrographs of the $Co_3O_4$:ZnO set of samples showing the evolution of the grains size as a function of the sintering temperature. The vertical red line at 900 °C separates the low temperature (LT) form the high temperature (HT) grain growth regimes.

Assuming a negligible initial grain size, the grain growth kinetics can be determined using the simplified phenomenological kinetics proposed by T. Senda and R. C. Bradt through the Equation (9) [60]. It can be rewriting as in Equation (10) and (11).

$$G^n = K_0\, t\, \exp(-Q_g / RT). \tag{9}$$

$$\log G = \left(\frac{1}{n}\right)\log t + \left(\frac{1}{n}\right)\left[\log K_0 - 0.434\left(\frac{Q_g}{RT}\right)\right]. \quad (10)$$

$$\log\left(\frac{G^n}{t}\right) = \log K_0 - 0.434\left(\frac{Q_g}{RT}\right). \quad (11)$$

Here the main parameters are the grain growth kinetic exponent ($n$), and the grain growth apparent activation energy ($Q_g$). $K_0$ is a pre-exponential constant, $t$ is the time in hours, $R$ is the gas constant (8.31 J/mol·K), and $T$ is the absolute temperature. $n$ can be determined by the inverse of slope of the curve as defined in Equation (10) for a set of samples processed at the same $T$ and different $t$. In turn, with the value of $n$ in hands, $Q_g$ can be calculated from the slope of the Arrhenius plot of the $\log(G^n/t)$ versus $1/T$ for a set of samples processed at the same $t$ and different $T$ (Equation (10)). Higher $n$-values and lower $Q_g$ leads to higher grain growth rates, $n$ is related to the mass transport mechanism, while $Q_g$ holds its standard meaning [61].

The sintering, and the consequent grain growth, of a starting nanostructured powder has to be analysed carefully [48]. Figure S17 and S18 present representative micrographs for the pure ZnO and for the $m$Co:ZnO samples, respectively. The micrographs show the samples before and after sintering in the $O_2$ and Ar atmospheres in the temperatures of 700 °C and 1200 °C. We observe for the samples sintered at 700 °C a relatively small shrinkage of the powder after the sintering, besides, for the samples sintered at 1200 °C the shrinkage is quite considerably. From this result we can infer that for the starting nanostructured powders at lower temperatures the coarsening dominates the grain growth process (rearrangement of the particles and formation of necks), while at high temperatures, with grains already of the size of the order of micrometres, the densification process takes place. Consequently, we can state, as it is expected, that the growth mechanism at low temperatures is distinct from that at higher temperatures. For small particles and material of high vapor pressure the evaporation-condensation sintering process is favourable, and at low temperatures the relatively low activation energy process, like the surface diffusion, dominates, what leads to coarsening [62]. At high temperatures, process of high activation energy, like grain boundary and bulk diffusion, dominates [60]. Nevertheless, sintering involves several mechanisms operating simultaneously over overlapping stages. In fact, there are numerous studies on the grain growth kinetics of ZnO, these studies have revealed that the rate controlling mechanism is the $Zn^{2+}$ diffusion, which proceeds via surface at the LT regime [62] and via bulk at the HT regime [60, 63]. Therefore, here the grain growth process is analysed in two different regimes, one below around 900 °C, referred as low temperature regime (LT), and other above 900 °C, referred as high temperature regime (HT).

Figure 9 shows the plot of the log($G$) *versus* log($t$) (Equation (10)) for the entire set of samples obtained at the 700 °C (LT) and 1200 °C (HT). Representative SEM imagens of the entire set of samples sintered at different times (1, 1.5, 2, 2.5 and 3h), and the histograms concerning the distribution of the main particle diameters are presented in Figures S19 to S34. The correspondent obtained statistical parameters are presented in Tables S11 to S18 of the supplementary file. Figure 10 shows the Arrhenius plot of log ($G^n/t$) *versus* (1/$T$) (Equation (11)) for pure ZnO and Co:ZnO set of samples sintered for 4h in $O_2$ and Ar.

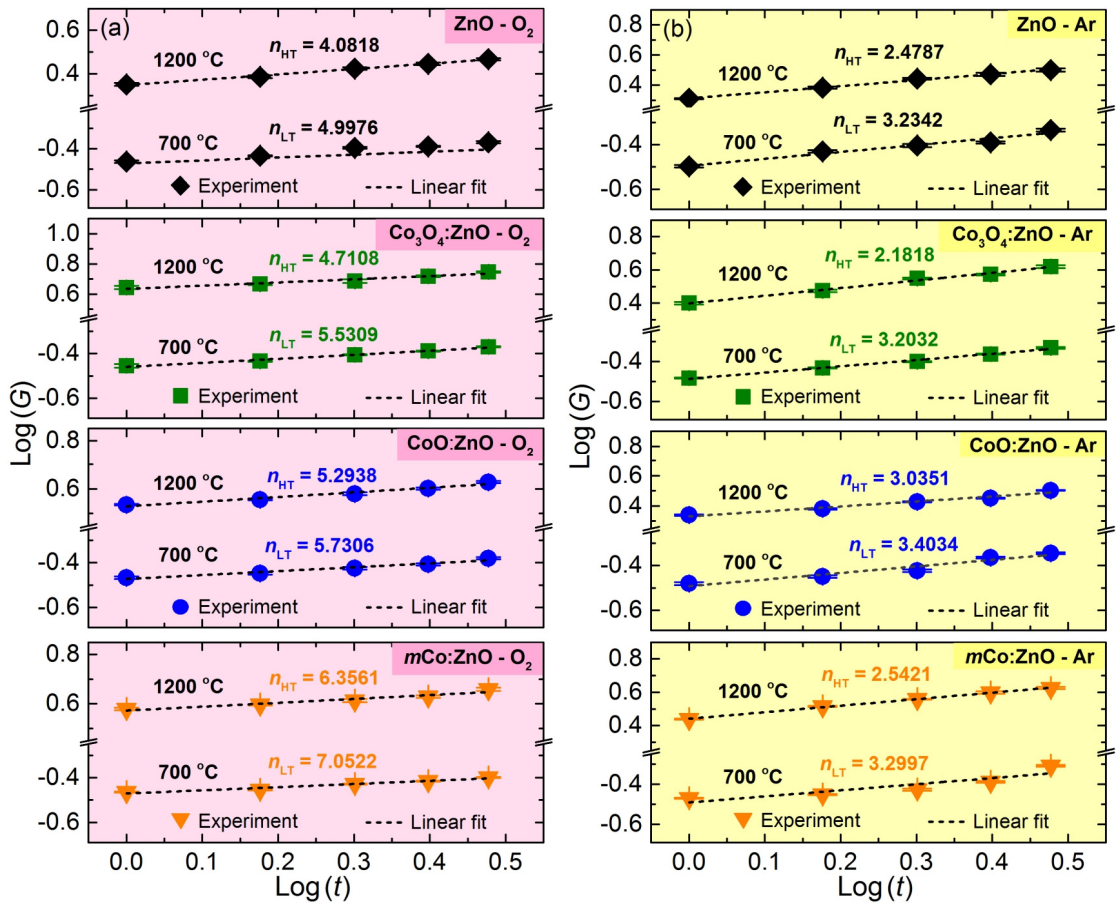

**Figure 9.** Isothermal grain growth of ZnO and Co:ZnO set of samples sintered in (a) oxygen ($O_2$) and (b) argon (Ar).

The obtained *n*-values and the grain growth apparent activation energy ($Q_g$) after a linear fit of the experimental points are presented in Table 1 and Table 2. First, we call attention to that the *n*-values for samples processed at 700 °C is quite different from those for samples processed at 1200 °C in all cases, and it is evident from Figure 10 the difference between the slope of the curves in the different temperature ranges, confirming the statement of the LT and HT grain growth regimes associated to different grain growth mechanisms. Comparing the *n* and $Q_g$ values at the LT and HT regime for the same atmosphere we note that $Q_g$ is larger at HT regime, as we would expect, while *n* is larger at the LT regime. Besides, both parameters *n* and $Q_g$ are larger for the Co:ZnO set of samples are higher than those for the pure ZnO sintered in $O_2$ and Ar atmospheres. Both parameters are also larger for samples sintered in $O_2$ as compared to those for samples sintered in Ar.

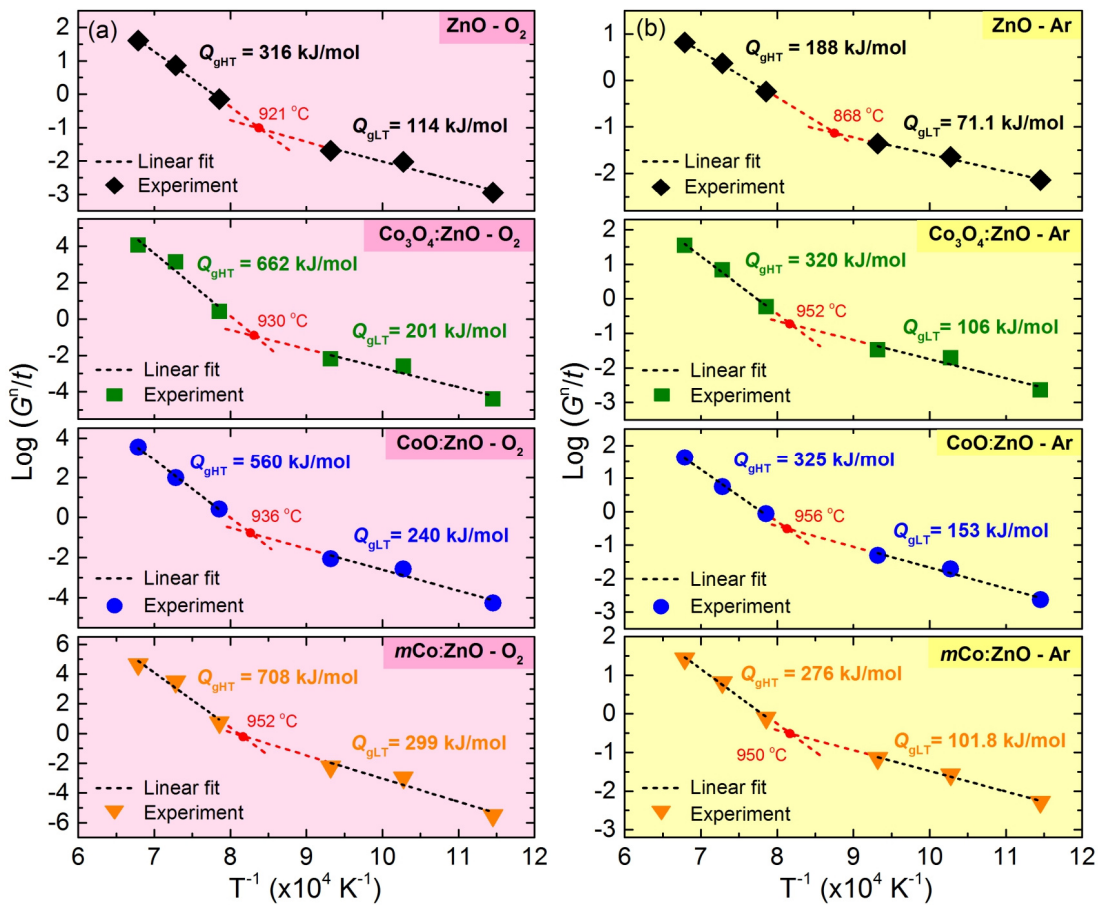

**Figure 10.** Arrhenius plots for the grain growth of ZnO and Co:ZnO set of samples sintered in (a) oxygen ($O_2$) and (b) argon (Ar).

First, it is important to point that the obtained $n$ and $Q_g$ for the pure ZnO system is relatively close to those reported in the literature [60, 64-66]. The differences in $n$ and $Q_g$ at LT and HT ranges are quite understood in terms of the change of the main growth process in the different temperature ranges. As stated before, $n$ is related to mass transport process, and at high temperature processes corresponding to higher $Q_g$ are activated. For the Co:ZnO it is plausible to infer that the Co incorporation into the $w$-ZnO lattice takes place mainly at the surface of the $w$-ZnO grains, changing drastically the diffusion and subsequent grain boundary mobility, what leads to the increasing of the $Q_g$ for those samples. Besides, the samples sintered in $O_2$ present higher $n$-values due probably to the promoted higher density of $V_{Zn}$, in the same way, the addition of Co to the system favor the formation of both $V_{Zn}$ and $Zn_i$ defects (full analysis presented in section 3.1) increasing $Zn^{2+}$ diffusion, and leading to a faster growth rate. Finally, both $n$ and $Q_g$ is lower for the samples sintered in Ar. The relative decrease in $n$ can be interpreted by the suppression of the $V_{Zn}$, while the decrease in $Q_g$ can be attributed to a higher Co solubility into the $w$-ZnO lattice (full analysis presented in section 3.2), what would lead to more homogeneous distribution into the $w$-ZnO grain and not to a higher concentration at grain surface. The lower $Q_g$ for the $m$Co:ZnO set of samples, which presents the lower $Q_i$ (Figure 7), corroborates this statement.

**Table 1.** Grain growth kinetic parameters obtained in the range of temperatures below 900 °C (LT regime): grain growth kinetic exponent ($n$), grain growth apparent activation energy ($Q_{gLT}$), and log $K_0$ (for completeness).

| | Sample | log $K_0$ | $n$ | $Q_{gLT}$ (kJ/mol) |
|---|---|---|---|---|
| *Oxygen* | ZnO | 4(1) | 4.9976 | 114(3) |
| | CoO:ZnO | 8(3) | 5.7306 | 240(2) |
| | $Co_3O_4$:ZnO | 7(3) | 5.5309 | 201(1) |
| | $m$Co:ZnO | 13(4) | 7.0522 | 299(2) |
| *Argon* | ZnO | 2.1(4) | 3.2342 | 71.1(7) |
| | CoO:ZnO | 5(1) | 3.4034 | 153(2) |
| | $Co_3O_4$:ZnO | 4(2) | 3.2032 | 106(2) |
| | $m$Co:ZnO | 4(1) | 3.2997 | 101.8(5) |

**Table 2.** Grain growth kinetic parameters obtained in the range of temperatures above 900 °C (HT regime): grain growth kinetic exponent ($n$), grain growth apparent activation energy ($Q_{gHT}$), and log $K_0$ (for completeness).

| | Sample | log $K_0$ | $n$ | $Q_{gHT}$ (kJ/mol) |
|---|---|---|---|---|
| *Oxygen* | ZnO | 12.8(6) | 4.0665 | 316(2) |
| | CoO:ZnO | 23(1) | 5.2938 | 560(3) |
| | $Co_3O_4$:ZnO | 27(6) | 4.7108 | 662(5) |
| | $m$Co:ZnO | 30(5) | 6.3561 | 708(3) |
| *Argon* | ZnO | 7.5(3) | 2.4787 | 188(1) |
| | CoO:ZnO | 12.4(8) | 3.0351 | 325(2) |
| | $Co_3O_4$:ZnO | 13(1) | 2.1818 | 320(2) |
| | $m$Co:ZnO | 11(1) | 2.5421 | 276(1) |

Returning to the results concerning the obtained final grain sizes presented in Figure 8. Remember that for a given $T$ and $t$, the grain size ($G$) depends on both $n$ and $Q_g$ (Equation (9)), lager $n$ leads to larger $G$, and larger $Q_g$ leads to smaller $G$. Our results indicate a competition between these parameters. For samples sintered in $O_2$, adding Co to the system increases $Q_g$, however $n$ is also increased, with a final effect corresponding to a higher $G$ in comparison to the pure ZnO set of samples. For the samples sintered in Ar, $Q_g$ decreases in comparison to values for the $O_2$ atmosphere, however $n$ also decreases, in such a way that the final $G$ remains almost the same as for the samples sintered in $O_2$. Among the samples sintered in Ar, the addition of Co also increases $n$ and $Q_g$.

## 4 CONCLUSION

In summary, in this report we presented a systematic study concerning the processing of Co:ZnO ceramics. Different Co precursors ($Co_3O_4$, CoO and metallic Co) and sintering atmospheres ($O_2$ and Ar) was analyzed from the theoretical and experimental point of view. The theoretical calculation showed that the main point defects ($Zn_i$, $V_O$ and $V_{Zn}$) became energetically favorable in the presence of cobalt. Based on the theoretical finds a set of defect chemical reaction was proposed. The studies on the Co incorporation into the $w$-ZnO was conducted carefully in order to properly address the Co-doping of the *w*-ZnO. The analysis showed that the Ar sintering atmosphere promotes an incongruent decomposition of the ZnO, leading to a higher $Zn_i$ density and a higher Co solubility. Among the Co precursors the metallic Co proved to be the one with higher solubility into the *w*-ZnO (lower $Q_i$). The evaluation of the grain growth kinetics revealed a very rich dynamic. In order to achieve high grain growth and densities the $Co_3O_4$ demonstrated to be a good additive for the *w*-ZnO sintering. It was observed that due to the competition between different mass transport mechanism (*n*) and the different Co incorporation degrees ($Q_g$) the final grain size is almost independent of the sintering atmosphere. Evidences showed that in the Ar sintering atmosphere the metallic Co precursor, due to its higher solubility into the *w*-ZnO lattice, leads to a high Co homogeneous distribution over the volume of the *w*-ZnO grain. Besides, CoO precursor present the lowest solubility into *w*-ZnO lattice and final grain size, being the CoO the best precursor for top-down approach in processing of Co-doped *w*-ZnO nanopowders. Our findings give valuable contribution to the understanding of the preparation of Co-doped *w*-ZnO ceramics and the sintering growth kinetics, what would allow to improve the state of the art on the processing of the material at both bulk and nanometric scales.

## ACKNOWLEDGMENTS

Support from agencies CAPES and FAPEMIG is gratefully acknowledged. The authors also acknowledge Prof. Dr. A. C. Doriguetto coordinator of the Laboratório de Cristalografia of the Universidade Federal de Alfenas were the XRD measurements were performed.

## REFERENCES

[1] H.L. Tuller, S.R. Bishop, Point Defects in Oxides: Tailoring Materials Through Defect Engineering, Annual Review of Materials Research, Vol 41 41 (2011) 369-398.

[2] I. Zutic, J. Fabian, S. Das Sarma, Spintronics: Fundamentals and applications, Reviews of Modern Physics 76(2) (2004) 323-410.

[3] M. Knobel, J.C. Denardin, H.B. De Carvalho, M. Brasil, A.B. Pakhomov, F.P. Missell, Magnetic and magnetotransport properties of Co thin films on Si, Physica Status Solidi a-Applied Research 187(1) (2001) 177-188.

[4] H.B. de Carvalho, M. Brasil, J.C. Denardin, M. Knobel, Transport and magnetotransport transition of thin Co films grown on Si, Physica Status Solidi a-Applied Research 201(10) (2004) 2361-2365.

[5] M.Y. He, H.M. Sun, Q.L. He, Topological insulator: Spintronics and quantum computations, Frontiers of Physics 14(4) (2019).

[6] H.B. de Carvalho, M.P.F. de Godoy, R.W.D. Paes, M. Mir, A. Ortiz de Zevallos, F. Iikawa, M.J.S.P. Brasil, V.A. Chitta, W.B. Ferraz, M.A. Boselli, A.C.S. Sabioni, Absence of ferromagnetic order in high quality bulk Co-doped ZnO samples, J Appl. Phys. 108(3) (2010) 033914.

[7] V.M.A. Lage, R.T. da Silva, A. Mesquita, M.P.F. de Godoy, X. Gratens, V.A. Chitta, H.B. de Carvalho, Influence of reducing heat treatment on the structural and magnetic properties of MnO:ZnO ceramics, Journal of Alloys and Compounds 863 (2021).

[8] M.P.F. de Godoy, A. Mesquita, W. Avansi, P.P. Neves, V.A. Chitta, W.B. Ferraz, M.A. Boselli, A.C.S. Sabioni, H.B. de Carvalho, Evidence of defect-mediated magnetic coupling on hydrogenated Co-doped ZnO, Journal of Alloys and Compounds 555 (2013) 315-319.

[9] V.M. de Almeida, A. Mesquita, A.O. de Zevallos, N.C. Mamani, P.P. Neves, X. Gratens, V.A. Chitta, W.B. Ferraz, A.C. Doriguetto, A.C.S. Sabioni, H.B. de Carvalho, Room temperature ferromagnetism promoted by defects at zinc sites in Mn-doped ZnO, Journal of Alloys and Compounds 655 (2016) 406-414.

[10] N.C. Mamani, R.T. da Silva, A.O. de Zevallos, A.A.C. Cotta, W.A.D. Macedo, M.S. Li, M.I.B. Bernardi, A.C. Doriguetto, H.B. de Carvalho, On the nature of the room temperature ferromagnetism in nanoparticulate Co-doped ZnO thin films prepared by EB-PVD, Journal of Alloys and Compounds 695 (2017) 2682-2688.

[11] M.P.F. de Godoy, X. Gratens, V.A. Chitta, A. Mesquita, M.M. de Lima, A. Cantarero, G. Rahman, J.M. Morbec, H.B. de Carvalho, Defect induced room temperature ferromagnetism in high quality Co-doped ZnO bulk samples, Journal of Alloys and Compounds 859 (2021) 157772.

[12] Y. Belghazi, G. Schmerber, S. Colis, J.L. Rehspringer, A. Dinia, Extrinsic origin of ferromagnetism in ZnO and $Zn_{0.9}Co_{0.1}O$ magnetic semiconductor films prepared by sol-gel technique, Appl. Phys. Lett. 89(12) (2006) 122504.

[13] J.M.D. Coey, Dilute magnetic oxides, Curr. Opin. Solid State Mater. Sci. 10(2) (2006) 83-92.

[14] D.C. Kundaliya, S.B. Ogale, S.E. Lofland, S. Dhar, C.J. Metting, S.R. Shinde, Z. Ma, B. Varughese, K.V. Ramanujachary, L. Salamanca-Riba, T. Venkatesan, On the origin of high-temperature ferromagnetism in the low-temperature-processed Mn-Zn-O system, Nat. Mater. 3(10) (2004) 709-714.

[15] M. Tay, Y.H. Wu, G.C. Han, T.C. Chong, Y.K. Zheng, S.J. Wang, Y.B. Chen, X.Q. Pan, Ferromagnetism in inhomogeneous $Zn_{1-x}Co_xO$ thin films, J Appl. Phys. 100(6) (2006) 063910.

[16] J.M.D. Coey, M. Venkatesan, C.B. Fitzgerald, Donor impurity band exchange in dilute ferromagnetic oxides, Nature Materials 4(2) (2005) 173-179.

[17] K.R. Kittilstved, W.K. Liu, D.R. Gamelin, Electronic structure origins of polarity-dependent high-$T_C$ ferromagnetism in oxide-diluted magnetic semiconductors, Nature Materials 5(4) (2006) 291-297.

[18] M. Ivill, S.J. Pearton, S. Rawal, L. Leu, P. Sadik, R. Das, A.F. Hebard, M. Chisholm, J.D. Budai, D.P. Norton, Structure and magnetism of cobalt-doped ZnO thin films, New Journal of Physics 10 (2008).

[19] S. Ramachandran, J. Narayan, J.T. Prater, Effect of oxygen annealing on Mn doped ZnO diluted magnetic semiconductors, Applied Physics Letters 88(24) (2006).

[20] H.S. Hsu, J.C.A. Huang, Y.H. Huang, Y.F. Liao, M.Z. Lin, C.H. Lee, J.F. Lee, S.F. Chen, L.Y. Lai, C.P. Liu, Evidence of oxygen vacancy enhanced room-temperature ferromagnetism in Co-doped ZnO, Applied Physics Letters 88(24) (2006).

[21] T. Tietze, M. Gacic, G. Schutz, G. Jakob, S. Bruck, E. Goering, XMCD studies on Co and Li doped ZnO magnetic semiconductors, New Journal of Physics 10 (2008).

[22] P. Xu, Y. Sun, C. Shi, F. Xu, H. Pan, The electronic structure and spectral properties of ZnO and its defects, Nuclear Instruments & Methods in Physics Research Section B-Beam Interactions With Materials and Atoms 199 (2003) 286-290.

[23] A. Janotti, C.G. Van de Walle, Native point defects in ZnO, Physical Review B 76(16) (2007) 165202.

[24] D.C. Look, J.W. Hemsky, J.R. Sizelove, Residual native shallow donor in ZnO, Physical Review Letters 82(12) (1999) 2552-2555.

[25] H.Y. Zhang, W. Hao, Y.Q. Cao, X.F. Chang, M.X. Xu, X.L. Guo, K. Shen, D.H. Xiang, Q.Y. Xu, Room temperature ferromagnetic $Zn_{0.98}Co_{0.02}O$ powders with improved visible-light photocatalysis, Rsc Advances 6(8) (2016) 6761-6767.

[26] L.R. Shah, H. Zhu, W.G. Wang, B. Ali, T. Zhu, X. Fan, Y.Q. Song, Q.Y. Wen, H.W. Zhang, S.I. Shah, J.Q. Xiao, Effect of Zn interstitials on the magnetic and transport properties of bulk Co-doped ZnO, Journal of Physics D-Applied Physics 43(3) (2010).

[27] N. Khare, M.J. Kappers, M. Wei, M.G. Blamire, J.L. MacManus-Driscoll, Defect-induced ferromagnetism in co-doped ZnO, Advanced Materials 18(11) (2006) 1449.

[28] H.M. Xiong, ZnO Nanoparticles Applied to Bioimaging and Drug Delivery, Advanced Materials 25(37) (2013) 5329-5335.

[29] S.B. Rana, R.P.P. Singh, Investigation of structural, optical, magnetic properties and antibacterial activity of Ni-doped zinc oxide nanoparticles, Journal of Materials Science-Materials in Electronics 27(9) (2016) 9346-9355.

[30] D.J. Norris, A.L. Efros, S.C. Erwin, Doped nanocrystals, Science 319(5871) (2008) 1776-1779.

[31] J.F. Suyver, S.F. Wuister, J.J. Kelly, A. Meijerink, Luminescence of nanocrystalline ZnSe:$Mn^{2+}$, Physical Chemistry Chemical Physics 2(23) (2000) 5445-5448.

[32] R.T. da Silva, A. Mesquita, A.O. de Zevallos, T. Chiaramonte, X. Gratens, V.A. Chitta, J.M. Morbec, G. Rahman, V.M. Garcia-Suarez, A.C. Doriguetto, M.I.B. Bernardi, H.B. de Carvalho, Multifunctional nanostructured Co-doped ZnO: Co spatial distribution and correlated magnetic properties, Physical Chemistry Chemical Physics 20(30) (2018) 20257-20269.

[33] G. Gorrasi, A. Sorrentino, Mechanical milling as a technology to produce structural and functional bio-nanocomposites, Green Chemistry 17(5) (2015) 2610-2625.

[34] A.S. Bolokang, F.R. Cummings, B.P. Dhonge, H.M.I. Abdallah, T. Moyo, H.C. Swart, C.J. Arendse, T.F.G. Muller, D.E. Motaung, Characteristics of the mechanical milling on the room temperature ferromagnetism and sensing properties of $TiO_2$ nanoparticles, Applied Surface Science 331 (2015) 362-372.

[35] S. Kumar, S. Chatterjee, K.K. Chattopadhyay, A.K. Ghosh, Sol-Gel-Derived ZnO:Mn Nanocrystals: Study of Structural, Raman, and Optical Properties, Journal of Physical Chemistry C 116(31) (2012) 16700-16708.

[36] S.B. Rana, Influence of CTAB assisted capping on the structural and optical properties of ZnO nanoparticles, Journal of Materials Science-Materials in Electronics 28(18) (2017) 13787-13796.

[37] R.B.V.D. A.C. Larson, General Structure Analysis System (GSAS), Los Alamos National Laboratory Report LAUR, 1994.

[38] B. Toby, EXPGUI, a graphical user interface for GSAS, Journal of Applied Crystallography 34(2) (2001) 210-213.

[39] P. Hohenberg, W. Kohn, Inhomogeneous electron gas, Physical Review B 136(3B) (1964) B864- B871.

[40] J.M. Soler, E. Artacho, J.D. Gale, A. Garcia, J. Junquera, P. Ordejon, D. Sanchez-Portal, The SIESTA method for ab initio order-N materials simulation, Journal of Physics-Condensed Matter 14(11) (2002) 2745-2779.

[41] N. Troullier, J.L. Martins, Efficient pseudopotentials for plane-wave calculations, Physical Review B 43(3) (1991) 1993-2006.

[42] J. Han, P. Mantas, A. Senos, Defect chemistry and electrical characteristics of undoped and Mn-doped ZnO, Journal of the European Ceramic Society 22(1) (2002) 49-59.

[43] G.D. Mahan, Intrinsic defects in zno varistors, Journal of Applied Physics 54(7) (1983) 3825-3832.

[44] A.C.S. Sabioni, About the oxygen diffusion mechanism in ZnO, Solid State Ionics 170(1-2) (2004) 145-148.

[45] M. McCluskey, S. Jokela, Defects in ZnO, Journal of Applied Physics 106(7) (2009).

[46] A. Mesquita, F.P. Rhodes, R.T. da Silva, P.P. Neves, A.O. de Zevallos, M.R.B. Andreeta, M.M. de Lima, A. Cantarero, I.S. da Silva, M.A. Boselli, X. Gratens, V.A. Chitta, A.C. Doriguetto, W.B. Ferraz, A.C.S. Sabioni, H.B. de Carvalho, Dynamics of the incorporation of Co into the wurtzite ZnO matrix and its magnetic properties, Journal of Alloys and Compounds 637 (2015) 407-417.

[47] P. Bonasewicz, W. Hirschwald, G. Neumann, Influence of surface processes on electrical, photochemical, and thermodynamical properties of zinc-oxide films, Journal of the Electrochemical Society 133(11) (1986) 2270-2278.

[48] M.W. Barsoum, Fundamentals of ceramics, Taylor & Francis, New York; London, 2003.

[49] J. Jiang, L.C. Li, Synthesis of sphere-like $Co_3O_4$ nanocrystals via a simple polyol route, Materials Letters 61(27) (2007) 4894-4896.

[50] A.C.S. Sabioni, A. Daniel, R. Metz, A.M. Huntz, F. Jomard, First study of oxygen diffusion in a ZnO-based commercial varistor, Defect and Diffusion Forum 289-292 (2009) 339-345.

[51] R.D. Shannon, Revised effective ionic radii and systematic studies of interatomic distances in halides and chalcogenides, Acta Crystallographica Section A 32(5) (1976) 751-767.

[52] S. Kolesnik, B. Dabrowski, J. Mais, Structural and magnetic properties of transition metal substituted ZnO, Journal of Applied Physics 95(5) (2004) 2582-2586.

[53] J. M. Calleja, M. Cardona, Resonant Raman Scattering in ZnO, Physical Review B 16(8) (1977) 3753-3761.

[54] M. Schumm, M. Koerdel, S. Müller, H. Zutz, C. Ronning, J. Stehr, D.M. Hofmann, J. Geurts, Structural impact of Mn implantation on ZnO, New Journal of Physics 10(4) (2008) 043004.

[55] B. Sanches de Lima, P.R. Martínez-Alanis, F. Güell, W.A. dos Santos Silva, M.I.B. Bernardi, N.L. Marana, E. Longo, J.R. Sambrano, V.R. Mastelaro, Experimental and Theoretical Insights into the Structural Disorder and Gas Sensing Properties of ZnO, ACS Applied Electronic Materials 3(3) (2021) 1447-1457.

[56] L.R. Valerio, N.C. Mamani, A.O. de Zevallos, A. Mesquita, M.I.B. Bernardi, A.C. Doriguetto, H.B. de Carvalho, Preparation and structural-optical characterization of dip-coated nanostructured Co-doped ZnO dilute magnetic oxide thin films, Rsc Advances 7(33) (2017) 20611-20619.

[57] P. Koidl, Optical-absorption of $Co^{2+}$ in ZnO, Physical Review B 15(5) (1977) 2493-2499.

[58] K. Wang, Q.B. Yu, Q. Qin, W.J. Duan, Feasibility of a Co Oxygen Carrier for Chemical Looping Air Separation: Thermodynamics and Kinetics, Chemical Engineering & Technology 37(9) (2014) 1500-1506.

[59] M.I. Mendelson, Average grain size in polycrystalline ceramics, Journal of the American Ceramic Society 52(8) (1969) 443-446.

[60] T. Senda, R.C. Bradt, Grain Growth in Sintered ZnO and $ZnO-Bi_2O_3$ Ceramics, Journal of the American Ceramic Society 73(1) (1990) 106-114.

[61] R.M. German, Chapter Seven - Thermodynamic and Kinetic Treatments, in: R.M. German (Ed.), Sintering: from Empirical Observations to Scientific Principles, Butterworth-Heinemann, Boston, 2014, pp. 183-226.

[62] O.J. Whittemore, J.A. Varela, Initial sintering of ZnO, Journal of the American Ceramic Society 64(11) (1981) C154-C155.

[63] S.-D. Shin, C.-S. Sone, J.-H. Han, D.-Y. Kim, Effect of Sintering Atmosphere on the Densification and Abnormal Grain Growth of ZnO, Journal of the American Ceramic Society 79(2) (1996) 565-567.

[64] J. Han, P.Q. Mantas, A.M.R. Senos, Grain growth in Mn-doped ZnO, Journal of the European Ceramic Society 20(16) (2000) 2753-2758.

[65] G. Hardal, B.Y. Price, The Effect of $TiO_2$ and $B_2O_3$ Additions on the Grain Growth of ZnO, Metallurgical and Materials Transactions a-Physical Metallurgy and Materials Science 48A(4) (2017) 2090-2098.

[66] S. Roy, T.K. Roy, D. Das, Grain growth kinetics of $Er_2O_3$ doped $ZnO-V_2O_5$ based varistor ceramics, Ceramics International 45(18) (2019) 24835-24850.

# SUPPLEMENTARY DATA

**A comprehensive study on the processing Co:ZnO ceramics: defect engineering and grain growth kinetics**

R. T. da Silva[a], J. M. Morbec[b,‡], G. Rahman[c,†], and H. B. de Carvalho[d,*]

[a] *Universidade Federal de Ouro Preto – UFOP, 35400-000 Ouro Preto, MG, Brazil.*
[b] *School of Chemical and Physical Sciences, Keele University, Keele ST5 5BG, United Kingdom*
[c] *Department of Physics, Quaid-i-Azam University, 45320 Islamabad, Pakistan.*
[d] *Universidade Federal de Alfenas - UNIFAL, 37130-000 Alfenas, Brazil.*

*Corresponding Authors*:
‡ j.morbec@keele.ac.uk
† gulrahman@qau.edu.pk.
* hugo.carvalho@unifal-mg.edu.br.

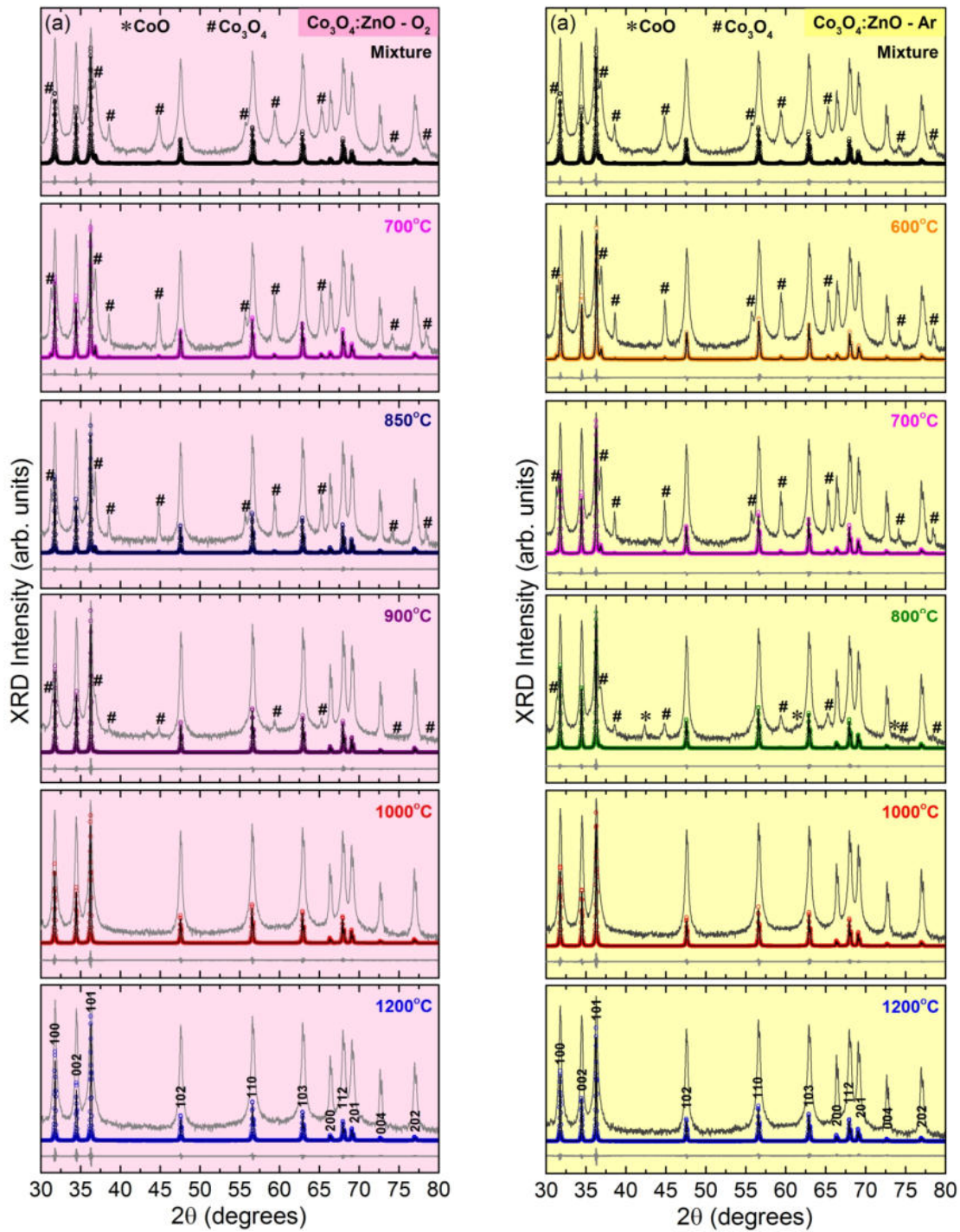

**Figure S1.** Refined XRD diffractograms of **$Co_3O_4$:ZnO** samples (8 at.%) synthesized at different temperatures in atmosphere of (a) oxygen and (b) argon. Each figure shows the observed pattern (symbols) and in a log scale (solid gray line), Rietveld calculated pattern (solid line), and the goodness of the fit or residual pattern (at the bottom). Diffraction peak associated with $Co_3O_4$ were indicated by the symbol (#).

**Table S1** - Elemental occupation factor for the Zn, Co (effective concentration, $x_E$) and O present in the **$Co_3O_4$:ZnO** wurtzite structure and the fractions of the ZnO, CoO e $Co_3O_4$ obtained through Rietveld refinement.

| | | Sample | Zn | Co | O | f. ZnO | f. CoO | f. $Co_3O_4$ |
|---|---|---|---|---|---|---|---|---|
| $Co_3O_4$:ZnO | Oxygen | Mixture | 1.000(1) | - | 0.980(5) | 0.9221 | - | 0.0778 |
| | | 600 °C | 0.998(1) | 0.002(1) | 0.981(4) | 0.9227 | - | 0.0773 |
| | | 700 °C | 0.990(1) | 0.010(1) | 0.978(5) | 0.9301 | - | 0.0699 |
| | | 800 °C | 0.982(1) | 0.018(1) | 0.986(4) | 0.9375 | - | 0.0625 |
| | | 850 °C | 0.962(1) | 0.039(1) | 0.986(4) | 0.9571 | - | 0.0428 |
| | | 900 °C | 0.925(1) | 0.075(1) | 0.985(4) | 0.9954 | - | 0.0046 |
| | | 1000 °C | 0.926(1) | 0.080(1) | 0.992(4) | 1.000 | - | - |
| | | 1200 °C | 0.922(1) | 0.080(1) | 0.991(5) | 1.000 | - | - |
| | Argon | Mixture | 1.000(1) | - | 0.980(5) | 0.9221 | - | 0.0778 |
| | | 600 °C | 0.992(1) | 0.008(1) | 0.987(4) | 0.9280 | - | 0.0720 |
| | | 700 °C | 0.975(1) | 0.026(1) | 0.983(4) | 0.9451 | - | 0.0549 |
| | | 800 °C | 0.935(1) | 0.065(1) | 0.981(4) | 0.9831 | - | 0.0169 |
| | | 1000 °C | 0.921(1) | 0.081(1) | 0.984(4) | 1.000 | - | - |
| | | 1200 °C | 0.919(1) | 0.079(1) | 0.972(5) | 1.000 | - | - |

**Table S2** - Structural data ($a$ and $c$) for the **$Co_3O_4$:ZnO** samples obtained through the Rietveld refinement. $V$ is the cell volume, $\chi^2$ is the square of the goodness-of-fit indicator, and $R_B$ is the refinement quality parameter.

| | | Sample | $a$ (Å) | $c$ (Å) | $V$ (Å$^3$) | $\chi^2$ | $R_{Bragg}$ |
|---|---|---|---|---|---|---|---|
| $Co_3O_4$:ZnO | Oxygen | Mixture | 3.25095(1) | 5.20885(3) | 47.675(1) | 6.9 | 4.6 |
| | | 600 °C | 3.25105(1) | 5.20868(3) | 47.677(1) | 3.7 | 2.9 |
| | | 700 °C | 3.25142(1) | 5.20855(3) | 47.686(1) | 4.0 | 4.7 |
| | | 800 °C | 3.25186(1) | 5.20736(2) | 47.688(1) | 3.2 | 2.8 |
| | | 850 °C | 3.25229(1) | 5.20737(2) | 47.701(1) | 2.5 | 3.1 |
| | | 900 °C | 3.25295(1) | 5.20579(2) | 47.706(1) | 4.3 | 3.2 |
| | | 1000 °C | 3.25359(1) | 5.20482(2) | 47.716(1) | 4.2 | 2.2 |
| | | 1200 °C | 3.25288(1) | 5.20709(3) | 47.716(1) | 6.6 | 3.6 |
| | Argon | Mixture | 3.25095(1) | 5.20885(3) | 47.675(1) | 6.9 | 4.6 |
| | | 600 °C | 3.25118(1) | 5.20877(3) | 47.681(1) | 4.1 | 5.4 |
| | | 700 °C | 3.25168(1) | 5.20792(2) | 47.688(1) | 3.0 | 2.6 |
| | | 800 °C | 3.25253(2) | 5.20689(3) | 47.706(1) | 3.7 | 2.6 |
| | | 1000 °C | 3.25353(1) | 5.20502(2) | 47.716(1) | 4.7 | 2.3 |
| | | 1200 °C | 3.25304(1) | 5.20663(3) | 47.716(1) | 5.2 | 5.0 |

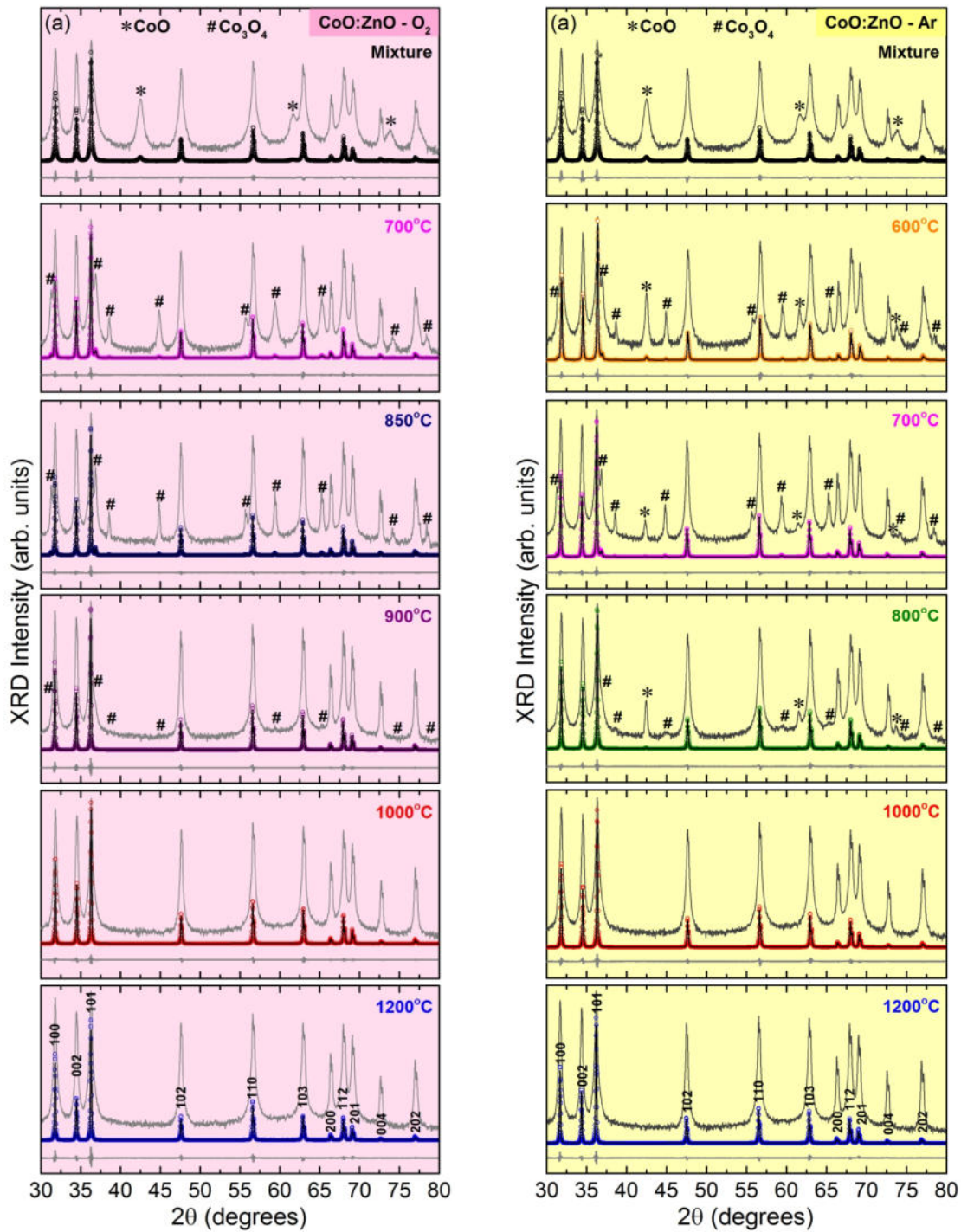

**Figure S2.** Refined XRD diffractograms of **CoO:ZnO** samples (8 at.%) synthesized at different temperatures in atmosphere of (a) oxygen and (b) argon. Each figure shows the observed pattern (symbols) and in a log scale (solid gray line), Rietveld calculated pattern (solid line), and the goodness of the fit or residual pattern (at the bottom). Diffraction peaks associated with CoO and $Co_3O_4$ were indicated by the symbol (∗) and (#), respectively.

**Table S3** - Elemental occupation factor for the Zn, Co (effective concentration, $x_E$) and O present in the **CoO:ZnO** wurtzite structure and the fractions of the ZnO, CoO e $Co_3O_4$ obtained through Rietveld refinement.

| | | Sample | Zn | Co | O | *f*. ZnO | *f*. CoO | *f*. $Co_3O_4$ |
|---|---|---|---|---|---|---|---|---|
| **CoO:ZnO** | *Oxygen* | Mixture | 1.000(1) | - | 0.985(7) | 0.9258 | 0.0742 | - |
| | | 600 °C | 0.998(1) | 0.002(1) | 0.981(4) | 0.9225 | - | 0.0775 |
| | | 700 °C | 0.992(1) | 0.008(1) | 0.987(5) | 0.9283 | - | 0.0717 |
| | | 800 °C | 0.981(1) | 0.019(1) | 0.984(4) | 0.9389 | - | 0.0611 |
| | | 850 °C | 0.960(1) | 0.040(1) | 0.981(4) | 0.9583 | - | 0.0417 |
| | | 900 °C | 0.922(1) | 0.078(1) | 0.978(4) | 0.9987 | - | 0.0013 |
| | | 1000 °C | 0.921(1) | 0.080(1) | 0.986(4) | 1.000 | - | - |
| | | 1200 °C | 0.919(1) | 0.079(1) | 0.995(4) | 1.000 | - | - |
| | *Argon* | Mixture | 1.000(1) | - | 0.985(7) | 0.9258 | 0.0742 | - |
| | | 600 °C | 0.996(1) | 0.007(1) | 0.972(5) | 0.9285 | 0.0327 | 0.0388 |
| | | 700 °C | 0.977(1) | 0.023(1) | 0.982(4) | 0.9425 | 0.0064 | 0.0510 |
| | | 800 °C | 0.935(1) | 0.067(1) | 0.985(4) | 0.9878 | 0.0122 | - |
| | | 1000 °C | 0.921(1) | 0.078(1) | 0.977(4) | 1.000 | - | - |
| | | 1200 °C | 0.920(1) | 0.077(1) | 0.978(4) | 1.000 | - | - |

**Table S4** - Structural data (*a* and *c*) for the **CoO:ZnO** samples obtained through the Rietveld refinement. *V* is the cell volume, $\chi^2$ is the square of the goodness-of-fit indicator, and $R_B$ is the refinement quality parameter.

| | | Sample | *a* (Å) | *c* (Å) | *V* (Å$^3$) | $\chi^2$ | $R_{Bragg}$ |
|---|---|---|---|---|---|---|---|
| **CoO:ZnO** | *Oxygen* | Mixture | 3.25094(2) | 5.20837(1) | 47.674(1) | 5.8 | 6.8 |
| | | 600 °C | 3.25098(1) | 5.20866(3) | 47.674(1) | 4.2 | 2.8 |
| | | 700 °C | 3.25129(1) | 5.20818(3) | 47.679(1) | 4.3 | 5.2 |
| | | 800 °C | 3.25185(1) | 5.20732(2) | 47.688(1) | 5.3 | 4.8 |
| | | 850 °C | 3.25205(1) | 5.20789(2) | 47.699(1) | 3.7 | 5.5 |
| | | 900 °C | 3.25314(1) | 5.20553(2) | 47.709(1) | 5.2 | 2.7 |
| | | 1000 °C | 3.25376(1) | 5.20429(2) | 47.716(1) | 3.2 | 2.0 |
| | | 1200 °C | 3.25293(1) | 5.20690(3) | 47.715(1) | 6.3 | 3.5 |
| | *Argon* | Mixture | 3.25094(2) | 5.20837(1) | 47.674(1) | 5.8 | 6.8 |
| | | 600 °C | 3.25108(1) | 5.20850(3) | 47.676(1) | 2.7 | 5.2 |
| | | 700 °C | 3.25157(1) | 5.20766(2) | 47.683(1) | 3.4 | 2.5 |
| | | 800 °C | 3.25269(1) | 5.20679(2) | 47.708(1) | 3.6 | 2.4 |
| | | 1000 °C | 3.25356(1) | 5.20487(3) | 47.715(1) | 4.8 | 2.4 |
| | | 1200 °C | 3.25323(1) | 5.20604(2) | 47.716(1) | 4.7 | 2.9 |

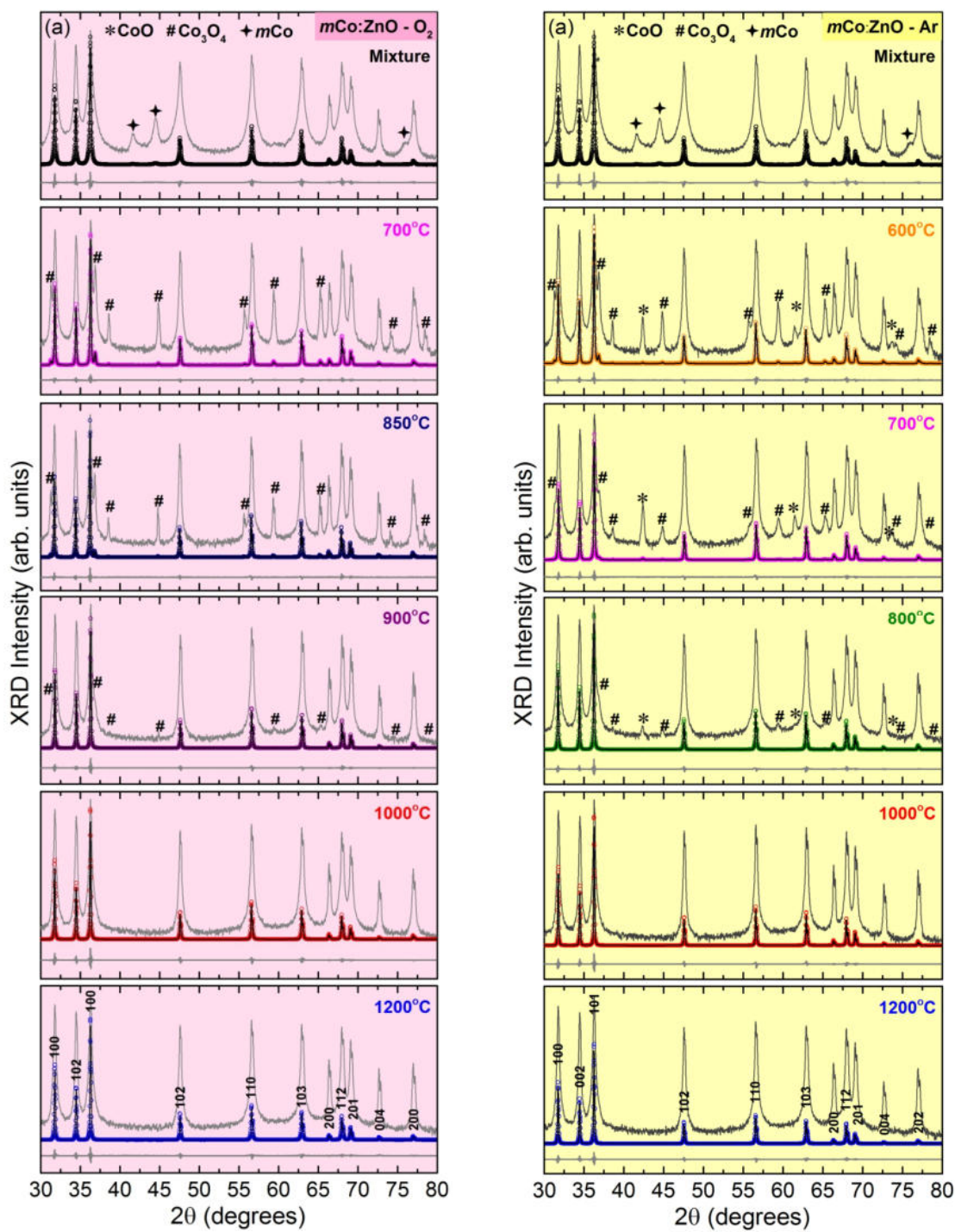

**Figure S3 -** Refined XRD diffractograms of *m*Co:ZnO samples (8 at.%) synthesized at different temperatures in atmosphere of (a) oxygen and (b) argon. Each figure shows the observed pattern (symbols) and in a log scale (solid gray line), Rietveld calculated pattern (solid line), and the goodness of the fit or residual pattern (at the bottom). Diffraction peaks associated with metallic Co, CoO and Co$_3$O$_4$ were indicated by the symbol (+), (∗) and (#), respectively.

**Table S1** - Elemental occupation factor for the Zn, Co (effective concentration, $x_E$) and O present in the *m*Co:ZnO wurtzite structure and the fractions of the ZnO, CoO, $Co_3O_4$ and metallic Co obtained through Rietveld refinement.

| | | Sample | Zn | Co | O | f. ZnO | f. CoO | f. $Co_3O_4$ | f. Co |
|---|---|---|---|---|---|---|---|---|---|
| *m*CoO:ZnO | Oxygen | Mixture | 0.997(1) | - | 0.977(6) | 0.9426 | - | - | 0.0573 |
| | | 600 °C | 0.997(1) | 0.005(1) | 0.978(5) | 0.9236 | - | 0.0764 | |
| | | 700 °C | 0.990(1) | 0.010(1) | 0.977(5) | 0.9305 | - | 0.0695 | |
| | | 800 °C | 0.984(1) | 0.016(1) | 0.986(4) | 0.9371 | - | 0.0629 | |
| | | **850 °C** | 0.960(1) | 0.040(1) | 0.987(4) | 0.9591 | - | 0.0409 | |
| | | 900 °C | 0.922(1) | 0.078(1) | 0.984(4) | 0.9986 | - | 0.0014 | |
| | | 1000 °C | 0.920(1) | 0.080(1) | 0.992(5) | 1.000 | - | - | |
| | | 1200 °C | 0.921(1) | 0.080(1) | 0.992(5) | 1.000 | - | - | |
| | Argon | Mixture | 0.997(1) | - | 0.977(6) | 0.9426 | - | - | 0.0573 |
| | | 500 °C | 0.996(1) | 0.004(1) | 0.977(4) | 0.9267 | 0.0113 | 0.0589 | 0.0029 |
| | | 600 °C | 0.991(1) | 0.011(1) | 0.973(4) | 0.9318 | 0.0096 | 0.0586 | - |
| | | 700 °C | 0.959(1) | 0.042(1) | 0.986(4) | 0.9593 | 0.0195 | 0.0211 | - |
| | | 800 °C | 0.929(1) | 0.071(1) | 0.984(4) | 0.9904 | 0.0064 | 0.0032 | - |
| | | 1000 °C | 0.920(1) | 0.079(1) | 0.974(4) | 1.000 | - | - | - |
| | | 1200 °C | 0.921(1) | 0.080(1) | 0.971(4) | 1.000 | - | - | - |

**Table S6** - Elemental occupation factor for the Zn, Co (effective concentration, $x_E$) and O present in the *m*Co:ZnO wurtzite structure and the fractions of the ZnO, CoO, $Co_3O_4$ and metallic Co obtained through Rietveld refinement.

| | | Sample | a (Å) | c (Å) | V (Å³) | $\chi^2$ | $R_{Bragg}$ |
|---|---|---|---|---|---|---|---|
| *m*CoO:ZnO | Oxygen | Mixture | 3.25105(2) | 5.20886(5) | 47.677(1) | 8.4 | 4.6 |
| | | 600 °C | 3.25108(1) | 5.20856(3) | 47.677(1) | 4.8 | 7.1 |
| | | 700 °C | 3.25137(1) | 5.20817(3) | 47.682(1) | 6.4 | 7.5 |
| | | 800 °C | 3.25202(1) | 5.20671(2) | 47.687(1) | 5.3 | 7.2 |
| | | 850 °C | 3.25248(1) | 5.20685(2) | 47.702(1) | 3.0 | 3.1 |
| | | 900 °C | 3.25319(1) | 5.20532(2) | 47.708(1) | 4.4 | 2.3 |
| | | 1000 °C | 3.25358(1) | 5.20490(3) | 47.716(1) | 5.6 | 2.3 |
| | | 1200 °C | 3.25328(1) | 5.20586(3) | 47.716(1) | 4.7 | 3.5 |
| | Argon | Mixture | 3.25105(2) | 5.20886(5) | 47.677(1) | 8.4 | 4.6 |
| | | 500 °C | 3.25107(1) | 5.20868(3) | 47.677(1) | 4.6 | 4.7 |
| | | 600 °C | 3.25130(1) | 5.20873(3) | 47.684(1) | 4.1 | 4.4 |
| | | 700 °C | 3.25190(1) | 5.20756(2) | 47.691(1) | 3.3 | 2.1 |
| | | 800 °C | 3.25269(1) | 5.20656(2) | 47.705(1) | 4.2 | 2.2 |
| | | 1000 °C | 3.25356(1) | 5.20491(2) | 47.716(1) | 4.7 | 2.7 |
| | | 1200 °C | 3.25355(1) | 5.20508(3) | 47.717(1) | 4.1 | 2.6 |

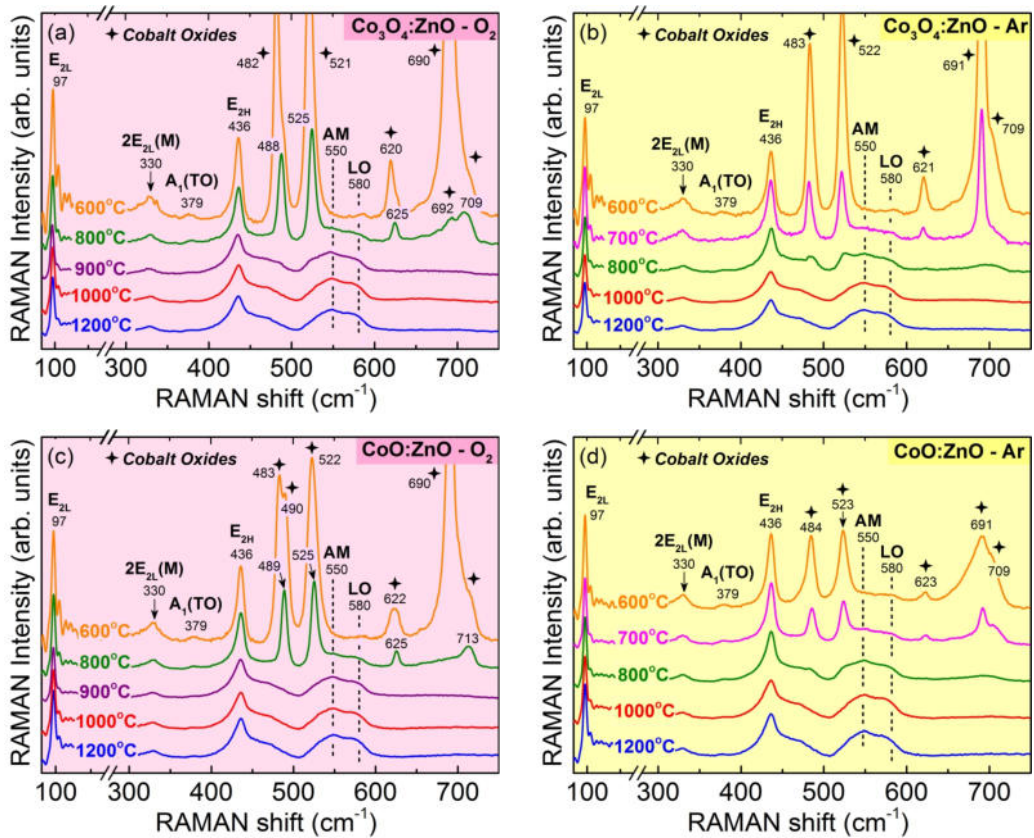

**Figure S4.** Raman scattering spectra obtained at room temperature for the samples (a) $Co_3O_4$:ZnO in $O_2$; (b) $Co_3O_4$:ZnO in Ar; (c) CoO:ZnO in $O_2$ and (d) CoO:ZnO in Ar. The spectra were acquired at room temperature and are normalized by integrated area of the $E_{2H}$ vibrational mode obtained by gaussian fit of the spectra.

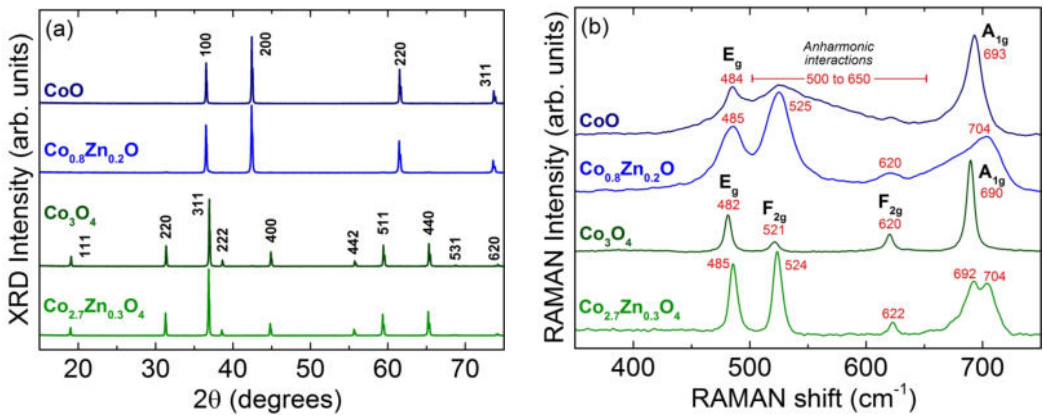

**Figure S5.** (a) X-ray diffraction patterns and (b) Raman scattering spectra obtained at room temperature for the CoO, $Co_{0.8}Zn_{0.2}O$ (20 at.% Zn-doped CoO), $Co_3O_4$, and $Co_{0.24}Zn_{0.6}O$ (10 at.% Zn-doped $Co_3O_4$).

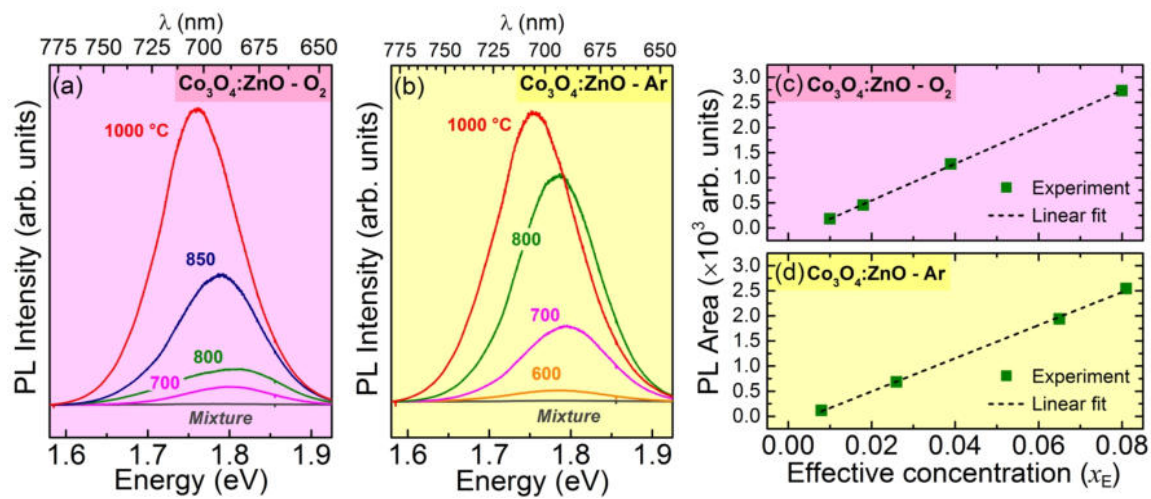

**Figure S6.** Room temperature PL spectra for part of the **$Co_3O_4$:ZnO** set of samples prepared in (a) oxygen ($O_2$) and (b) argon (Ar) atmospheres. It was used a 532 nm laser for excitation with optical power density of 50 kW/cm$^2$. Integrated area underneath the spectra plotted as a function of the calculated effective cobalt concentration ($x_E$) for the samples prepared (c) $O_2$ and in (d) Ar.

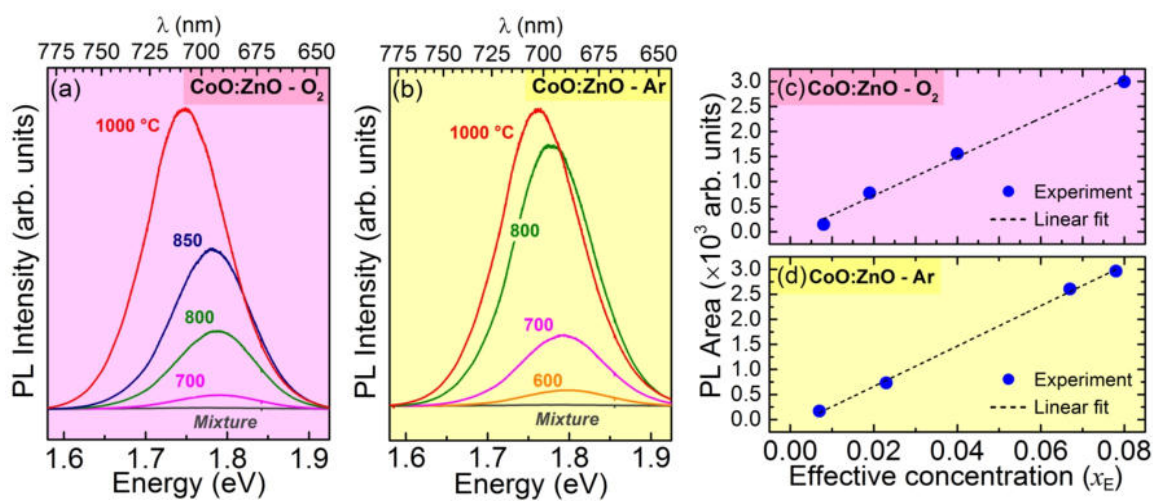

**Figure S7.** Room temperature PL spectra for part of the **CoO:ZnO** set of samples prepared in (a) oxygen ($O_2$) and (b) argon (Ar) atmospheres. It was used a 532 nm laser for excitation with optical power density of 50 kW/cm$^2$. Integrated area underneath the spectra plotted as a function of the calculated effective cobalt concentration ($x_E$) for the samples prepared (c) $O_2$ and in (d) Ar.

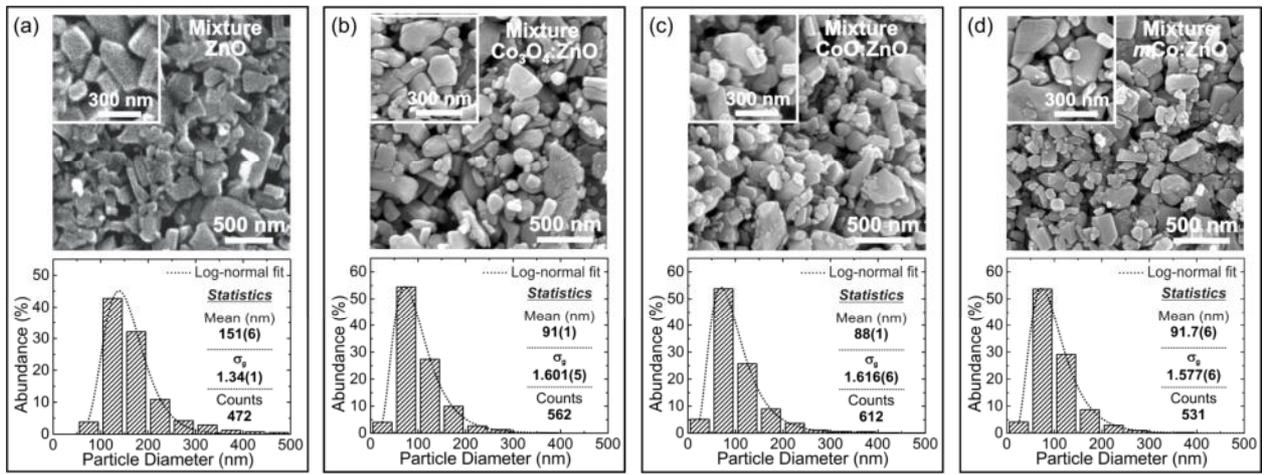

**Figure S8.** Representative SEM micrographs and obtained grain size statistical histograms for the powder mixtures (a) ZnO pure, (b) $Co_3O_4$:ZnO, (c) CoO:ZnO and (d) $m$Co:ZnO, before the heat treatments.

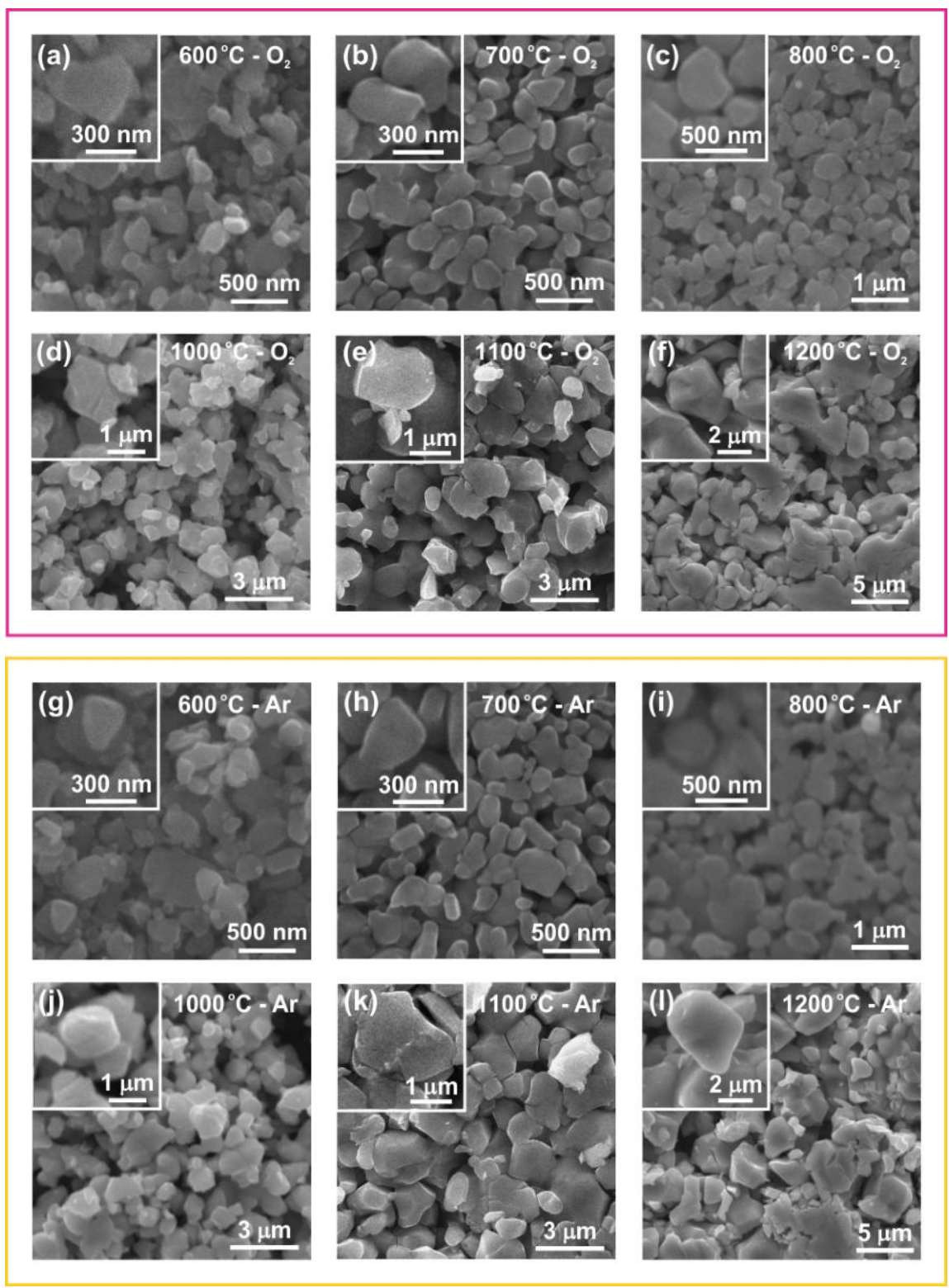

**Figure S9 -** Representative images by scanning electron microscopy of the surfaces of **ZnO** samples heat treated in $O_2$ at (a) 600 ºC, (b) 700 ºC, (c) 800 ºC, (d) 1000 ºC, (e) 1100 ºC, (f) 1200 ºC, and in Ar at (g) 600 ºC, (h) 700 ºC, (i) 800 ºC, (j) 1000 ºC, (k) 1100 ºC, and (l) 1200 ºC.

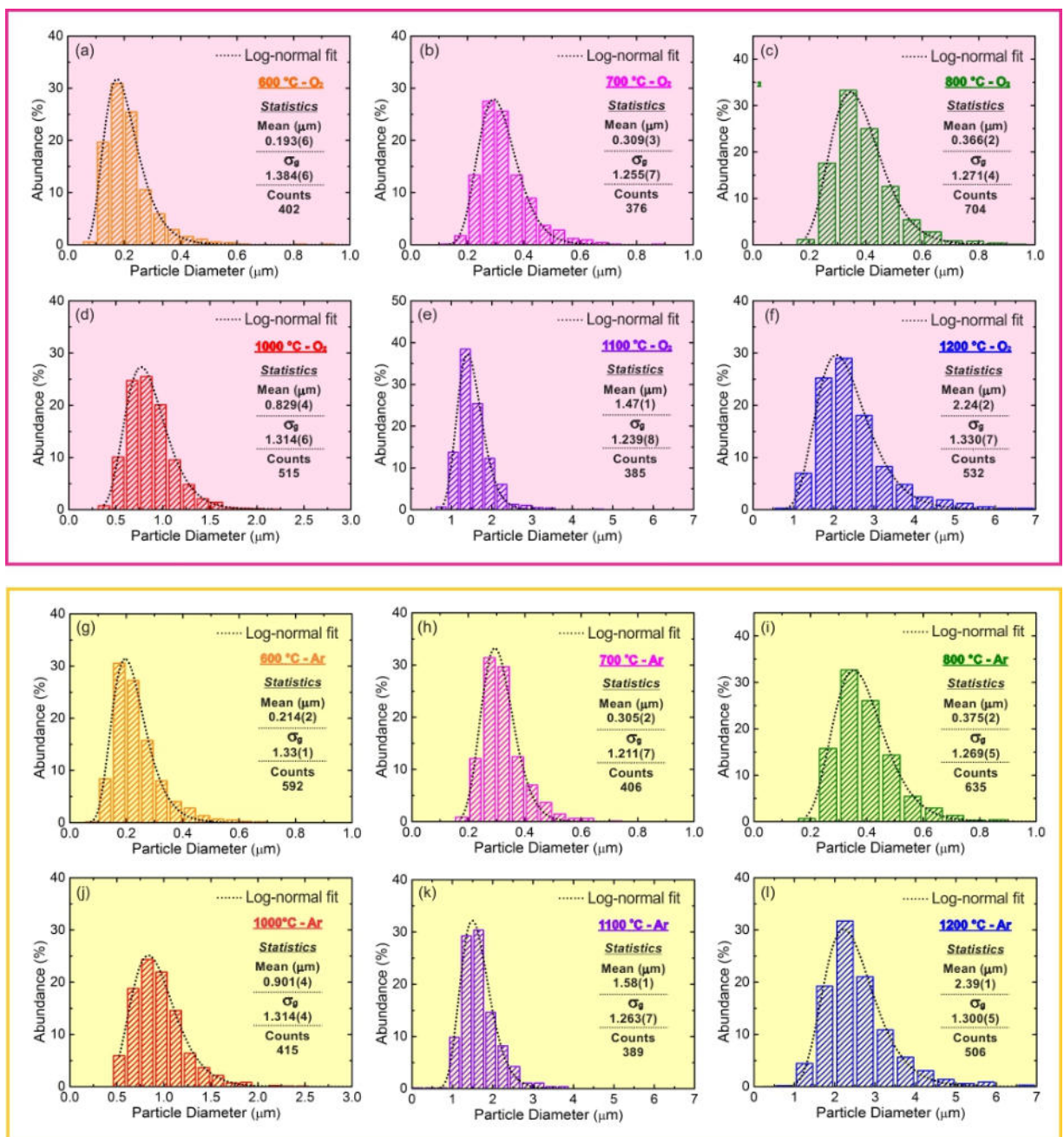

**Figure S10 -** Particle size distribution histograms of the **ZnO** samples heat treated in $O_2$ at (a) 600 ºC, (b) 700 ºC, (c) 800 ºC, (d) 1000 ºC, (e) 1100 ºC, (f) 1200 ºC, and in Ar at (g) 600 ºC, (h) 700 ºC, (i) 800 ºC, (j) 1000 ºC, (k) 1100 ºC, and (l) 1200 ºC. The distribution statistic obtained after log-normal fit is also presented in the panels.

**Table S7** - Particle size distribution analyses of the **ZnO** samples. $L$ is the mean value of the particle diameter and $\sigma_g$ is the geometric standard deviation obtained by the log-normal fit of particle size distribution histograms for each sample. $N$ is the total number of counted particles.

| | | Sample | $L$ (µm) | $\sigma_g$ | $N$ |
|---|---|---|---|---|---|
| **ZnO** | *Oxygen* | Mixture | 0.151(6) | 1.34(1) | 472 |
| | | 600 °C | 0.193(1) | 1.384(6) | 402 |
| | | 700 °C | 0.309(3) | 1.255(7) | 376 |
| | | 800 °C | 0.366(2) | 1.271(4) | 704 |
| | | 1000 °C | 0.829(4) | 1.314(6) | 515 |
| | | 1100 °C | 1.47(1) | 1.239(8) | 385 |
| | | 1200 °C | 2.24(2) | 1.330(7) | 532 |
| | *Argon* | Mixture | 0.151(6) | 1.34(1) | 472 |
| | | 600 °C | 0.214(2) | 1.33(1) | 592 |
| | | 700 °C | 0.305(2) | 1.211(7) | 406 |
| | | 800 °C | 0.375(2) | 1.269(5) | 635 |
| | | 1000 °C | 0.901(4) | 1.314(4) | 415 |
| | | 1100 °C | 1.58(1) | 1.263(7) | 389 |
| | | 1200 °C | 2.39(1) | 1.300(5) | 506 |

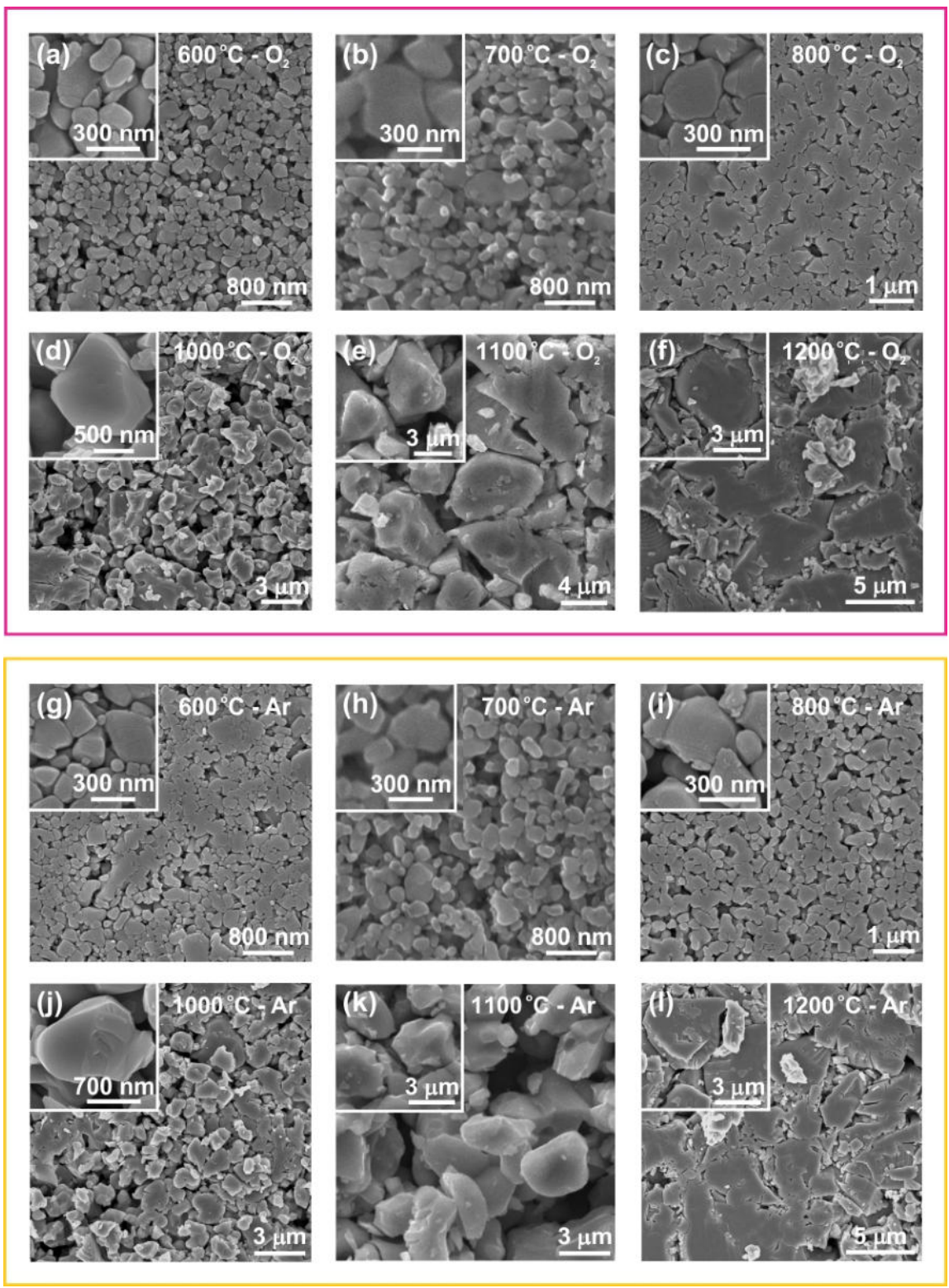

**Figure S11** - Representative images by scanning electron microscopy of the surfaces of **Co$_3$O$_4$:ZnO** samples heat treated in O$_2$ at (a) 600 ºC, (b) 700 ºC, (c) 800 ºC, (d) 1000 ºC, (e) 1100 ºC, (f) 1200 ºC, and in Ar at (g) 600 ºC, (h) 700 ºC, (i) 800 ºC, (j) 1000 ºC, (k) 1100 ºC, and (l) 1200 ºC.

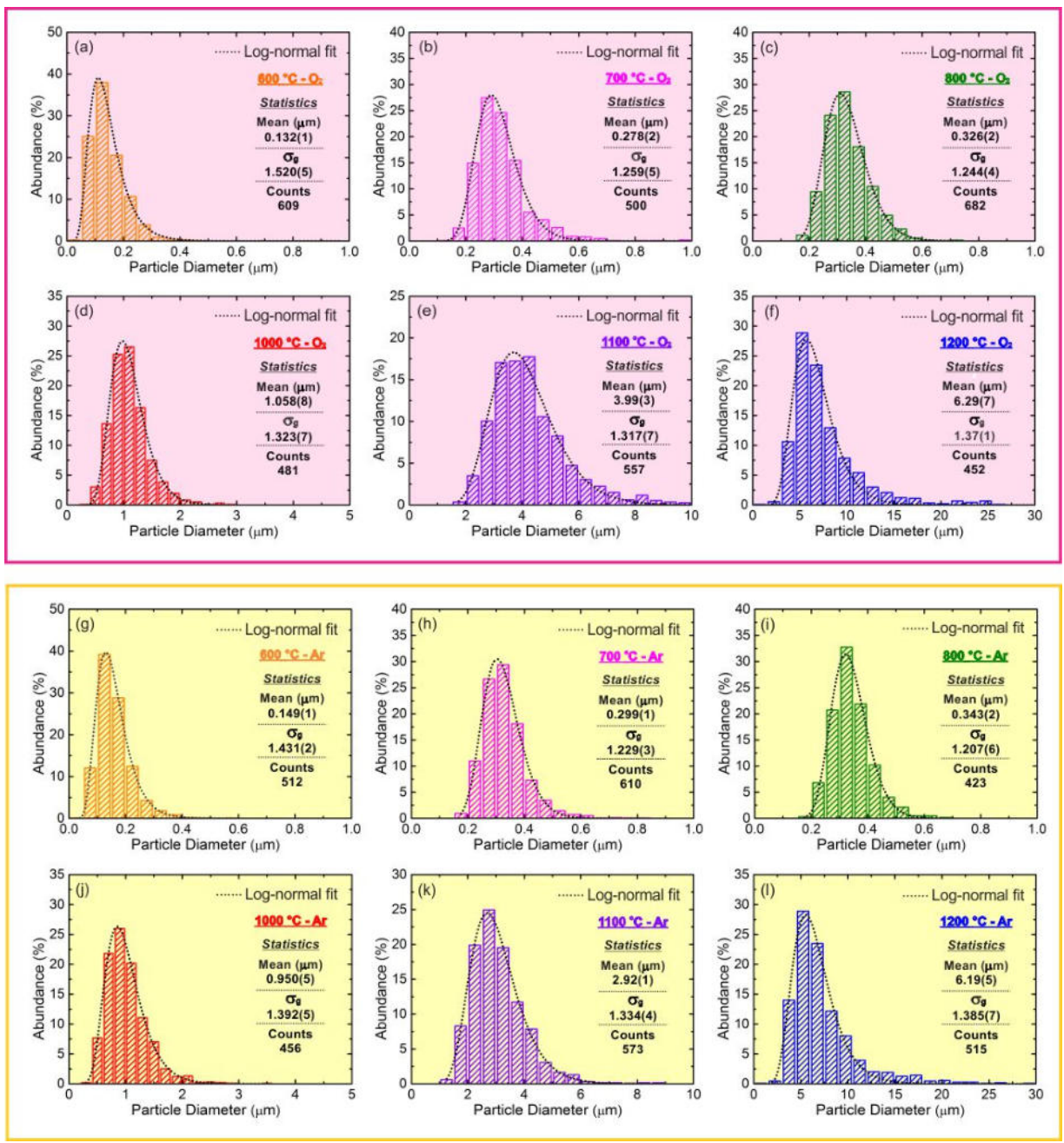

**Figure S12 -** Particle size distribution histograms of the **Co$_3$O$_4$:ZnO** samples heat treated in O$_2$ at (a) 600 ºC, (b) 700 ºC, (c) 800 ºC, (d) 1000 ºC, (e) 1100 ºC, (f) 1200 ºC, and in Ar at (g) 600 ºC, (h) 700 ºC, (i) 800 ºC, (j) 1000 ºC, (k) 1100 ºC, and (l) 1200 ºC. The distribution statistic obtained after log-normal fit is also presented in the panels.

**Table S8** - Particle size distribution analyses of the **Co$_3$O$_4$:ZnO** samples. $L$ is the mean value of the particle diameter and $\sigma_g$ is the geometric standard deviation obtained by the log-normal fit of particle size distribution histograms for each sample. $N$ is the total number of counted particles.

| | | Sample | $L(\mu m)$ | $\sigma_g$ | $N$ |
|---|---|---|---|---|---|
| Co$_3$O$_4$:ZnO | Oxygen | Mixture | 0.091(1) | 1.601(5) | 562 |
| | | 600 °C | 0.132(1) | 1.520(5) | 609 |
| | | 700 °C | 0.278(2) | 1.259(5) | 500 |
| | | 800 °C | 0.326(2) | 1.244(4) | 682 |
| | | 1000 °C | 1.058(8) | 1.323(7) | 481 |
| | | 1100 °C | 3.99(3) | 1.317(7) | 557 |
| | | 1200 °C | 6.29(7) | 1.37(1) | 452 |
| | Argon | Mixture | 0.091(1) | 1.601(5) | 562 |
| | | 600 °C | 0.149(1) | 1.431(2) | 512 |
| | | 700 °C | 0.299(1) | 1.229(3) | 610 |
| | | 800 °C | 0.343(2) | 1.207(6) | 423 |
| | | 1000 °C | 0.950(5) | 1.392(5) | 456 |
| | | 1100 °C | 2.92(1) | 1.334(4) | 573 |
| | | 1200 °C | 6.19(5) | 1.385(7) | 515 |

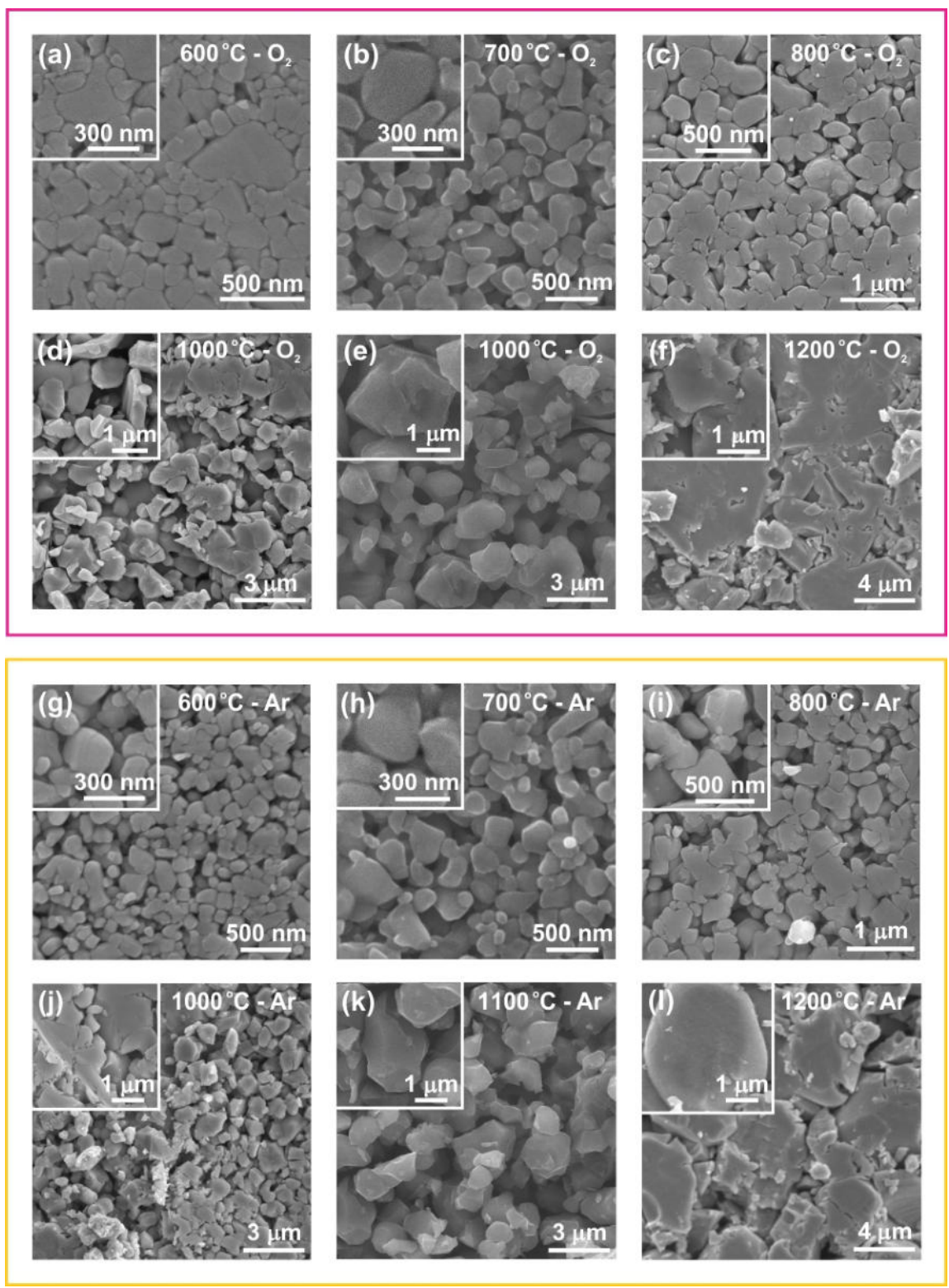

**Figure S13 -** Representative images by scanning electron microscopy of the surfaces of **CoO:ZnO** samples heat treated in $O_2$ at (a) 600 ºC, (b) 700 ºC, (c) 800 ºC, (d) 1000 ºC, (e) 1100 ºC, (f) 1200 ºC, and in Ar at (g) 600 ºC, (h) 700 ºC, (i) 800 ºC, (j) 1000 ºC, (k) 1100 ºC, and (l) 1200 ºC.

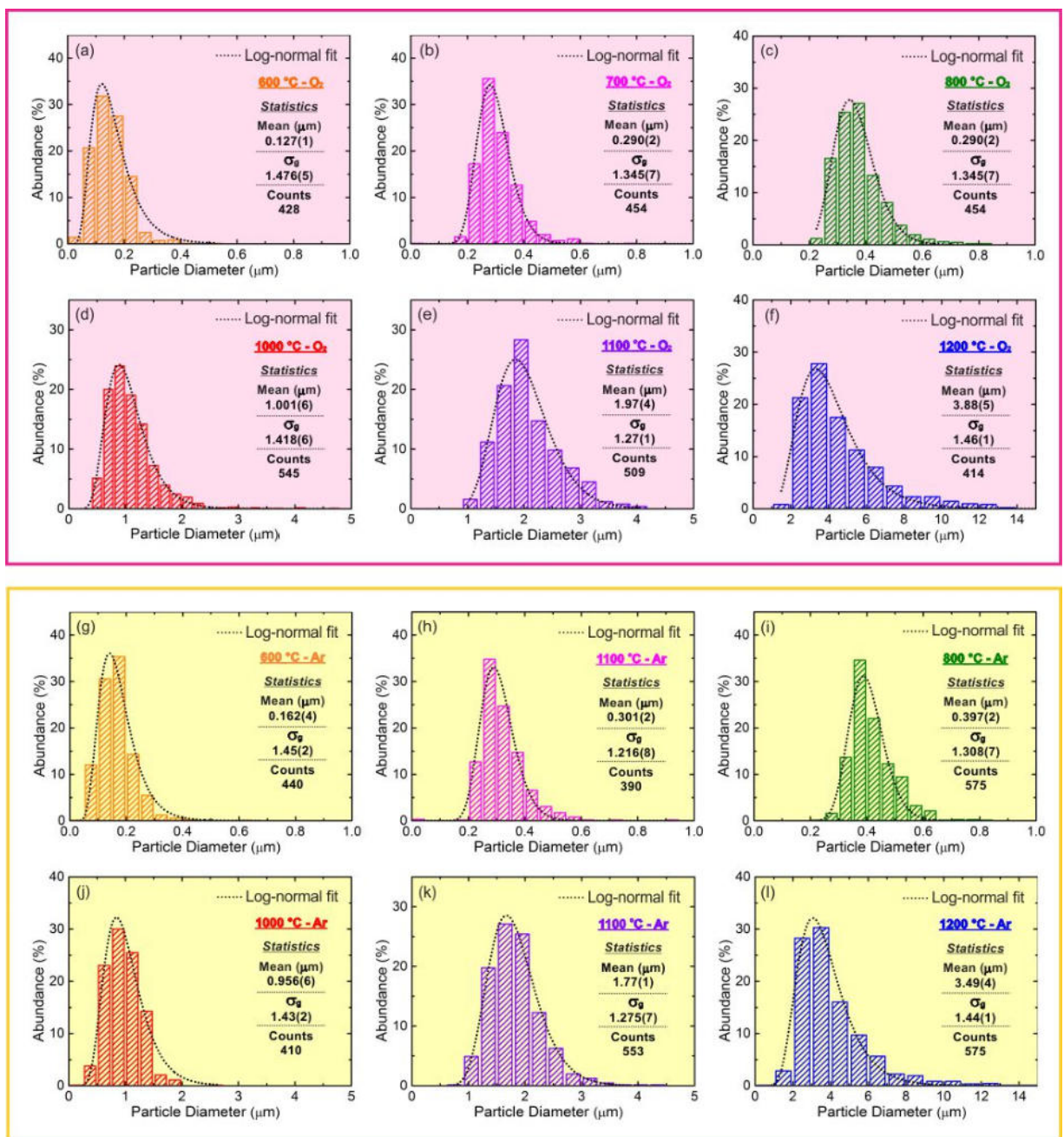

**Figure S14 -** Particle size distribution histograms of the **CoO:ZnO** samples heat treated in $O_2$ at (a) 600 ºC, (b) 700 ºC, (c) 800 ºC, (d) 1000 ºC, (e) 1100 ºC, (f) 1200 ºC, and in Ar at (g) 600 ºC, (h) 700 ºC, (i) 800 ºC, (j) 1000 ºC, (k) 1100 ºC, and (l) 1200 ºC. The distribution statistic obtained after log-normal fit is also presented in the panels.

**Table S9** - Particle size distribution analyses of the **CoO:ZnO** samples. $L$ is the mean value of the particle diameter and $\sigma_g$ is the geometric standard deviation obtained by the log-normal fit of particle size distribution histograms for each sample. $N$ is the total number of counted particles.

| | | Sample | $L$ (µm) | $\sigma_g$ | $N$ |
|---|---|---|---|---|---|
| **CoO:ZnO** | *Oxygen* | Mixture | 0.087(1) | 1.51(1) | 612 |
| | | 600 °C | 0.147(1) | 1.56(3) | 428 |
| | | 700 °C | 0.299(2) | 1.217(5) | 430 |
| | | 800 °C | 0.356(3) | 1.222(9) | 454 |
| | | 1000 °C | 1.001(6) | 1.418(6) | 545 |
| | | 1100 °C | 1.97(4) | 1.27(1) | 509 |
| | | 1200 °C | 3.88(5) | 1.46(1) | 414 |
| | *Argon* | Mixture | 0.088(1) | 1.616(5) | 612 |
| | | 600 °C | 0.162(4) | 1.45(2) | 440 |
| | | 700 °C | 0.301(2) | 1.216(8) | 390 |
| | | 800 °C | 0.397(2) | 1.308(7) | 402 |
| | | 1000 °C | 0.956(6) | 1.43(2) | 410 |
| | | 1100 °C | 1.77(1) | 1.275(7) | 553 |
| | | 1200 °C | 3.49(4) | 1.44(1) | 575 |

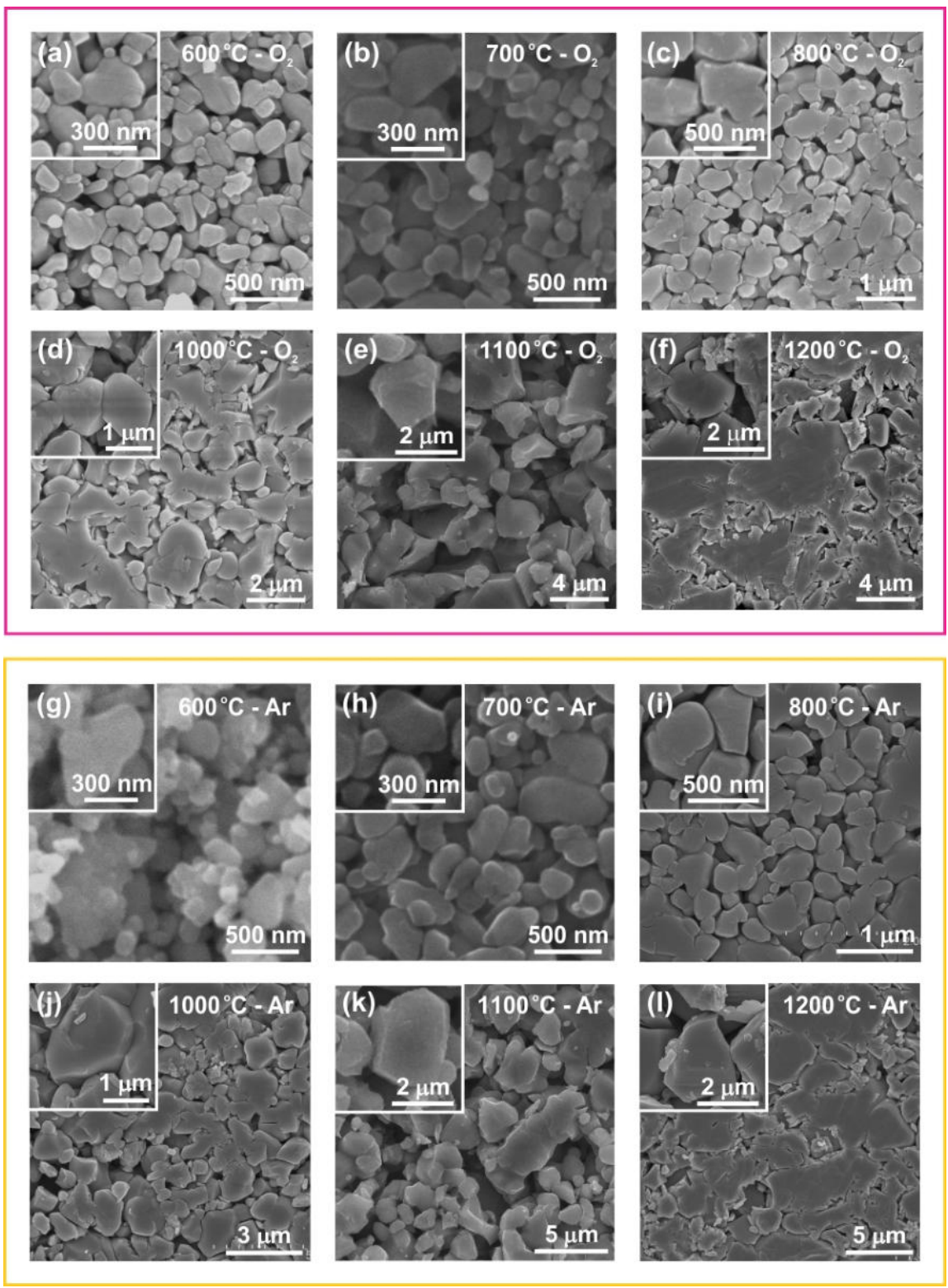

**Figure S15** - Representative images by scanning electron microscopy of the surfaces of **mCo:ZnO** samples heat treated in $O_2$ at (a) 600 °C, (b) 700 °C, (c) 800 °C, (d) 1000 °C, (e) 1100 °C, (f) 1200 °C, and in Ar at (g) 600 °C, (h) 700 °C, (i) 800 °C, (j) 1000 °C, (k) 1100 °C, and (l) 1200 °C.

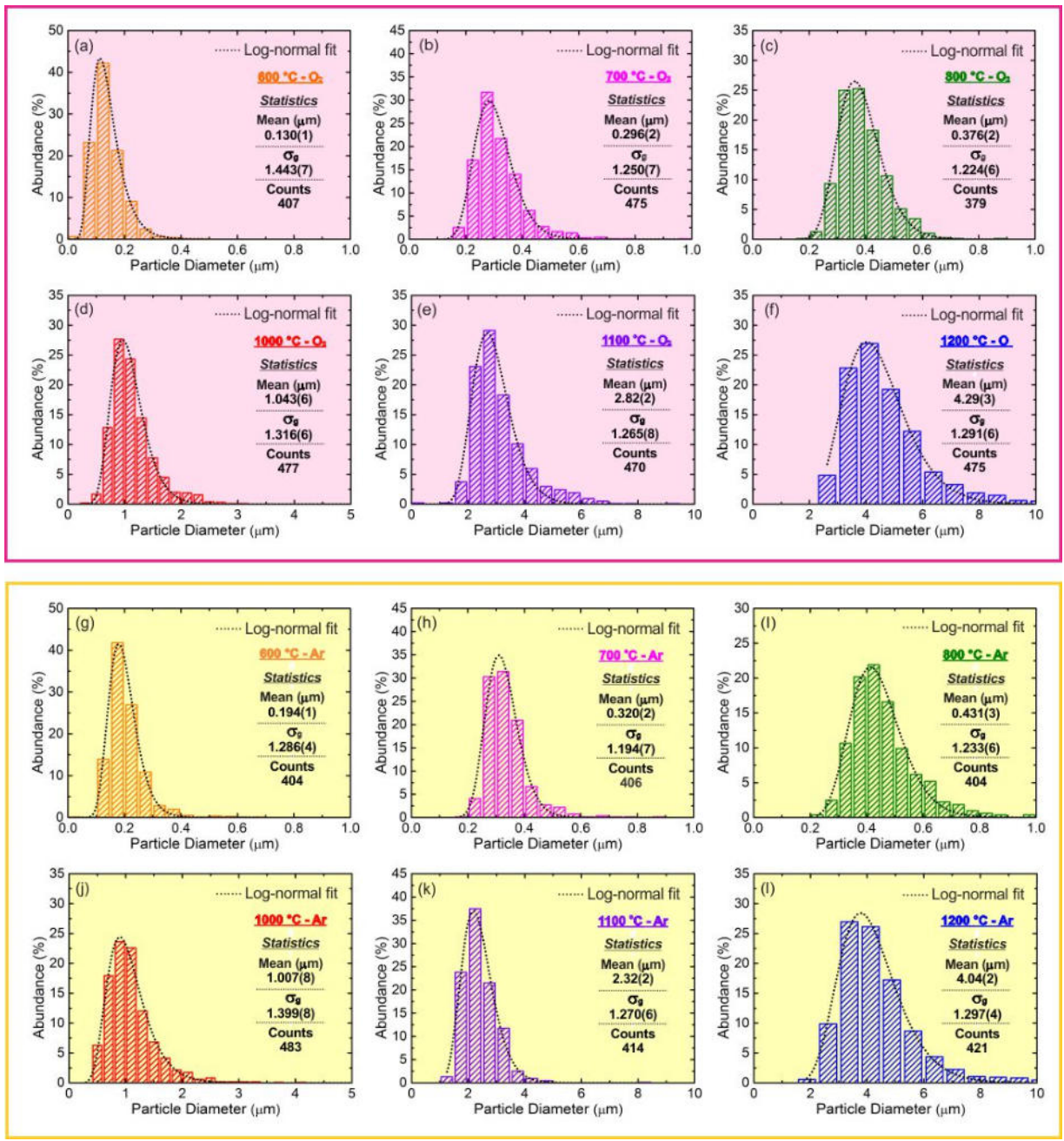

**Figure S16 -** Particle size distribution histograms of the *m*CoO:ZnO samples heat treated in $O_2$ at (a) 600 ºC, (b) 700 ºC, (c) 800 ºC, (d) 1000 ºC, (e) 1100 ºC, (f) 1200 ºC, and in Ar at (g) 600 ºC, (h) 700 ºC, (i) 800 ºC, (j) 1000 ºC, (k) 1100 ºC, and (l) 1200 ºC. The distribution statistic obtained after log-normal fit is also presented in the panels.

**Table S10** - Particle size distribution analyses of the *m*CoO:ZnO samples. $L$ is the mean value of the particle diameter and $\sigma_g$ is the geometric standard deviation obtained by the log-normal fit of particle size distribution histograms for each sample. $N$ is the total number of counted particles.

| | | Sample | $L$ (µm) | $\sigma_g$ | $N$ |
|---|---|---|---|---|---|
| *m*Co:ZnO | Oxygen | Mixture | 0.0917(6) | 1.477(6) | 531 |
| | | 600 ºC | 0.130(1) | 1.443(7) | 407 |
| | | 700 ºC | 0.296(2) | 1.250(7) | 475 |
| | | 800 ºC | 0.376(2) | 1.224(6) | 379 |
| | | 1000 ºC | 1.043(6) | 1.316(6) | 477 |
| | | 1100 ºC | 2.82(2) | 1.265(8) | 470 |
| | | 1200 ºC | 4.29(3) | 1.291(6) | 475 |
| | Argon | Mixture | 0.0917(6) | 1.477(6) | 531 |
| | | 600 ºC | 0.194(1) | 1.286(4) | 404 |
| | | 700 ºC | 0.320(2) | 1.194(7) | 406 |
| | | 800 ºC | 0.431(3) | 1.233(7) | 404 |
| | | 1000 ºC | 1.007(8) | 1.399(8) | 483 |
| | | 1100 ºC | 2.32(2) | 1.270(6) | 414 |
| | | 1200 ºC | 4.04(2) | 1.297(4) | 421 |

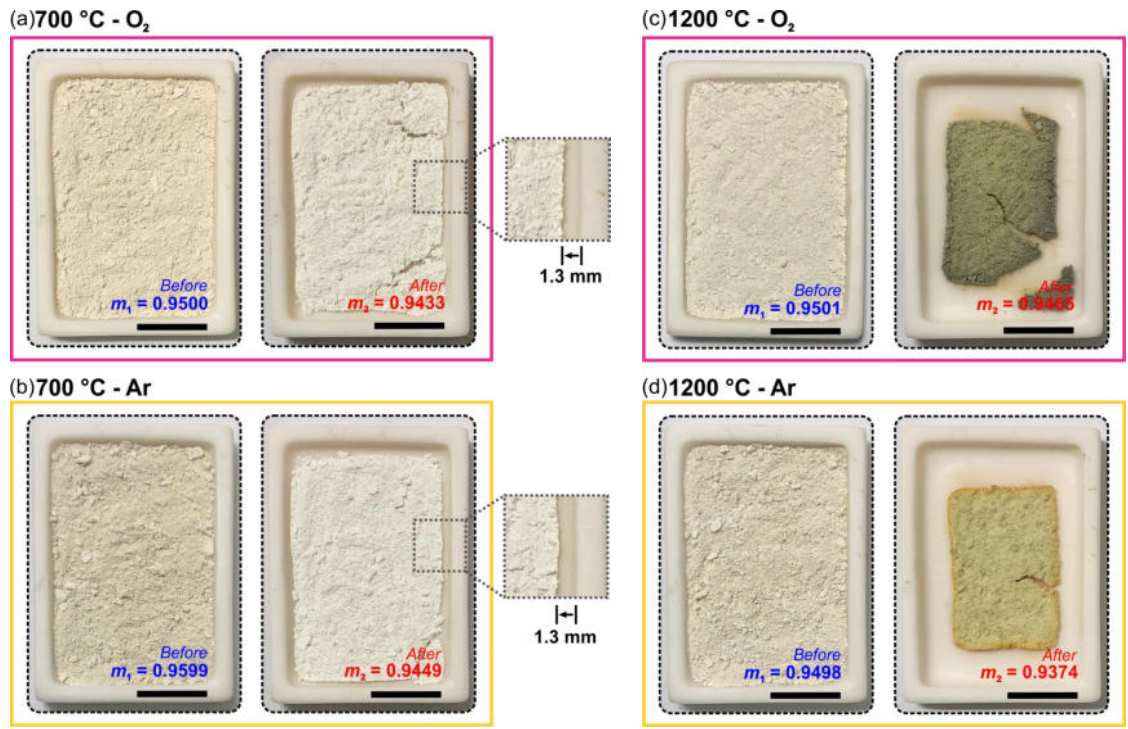

**Figure S17.** Representative micrographs for the pure ZnO samples before and after the heat treatments in O$_2$ and Ar at the temperatures of 700 °C and 1200 °C. A relatively small shrinkage is observed for the heat treatment at 700 °C, while an evident densification occurs at 1200 °C.

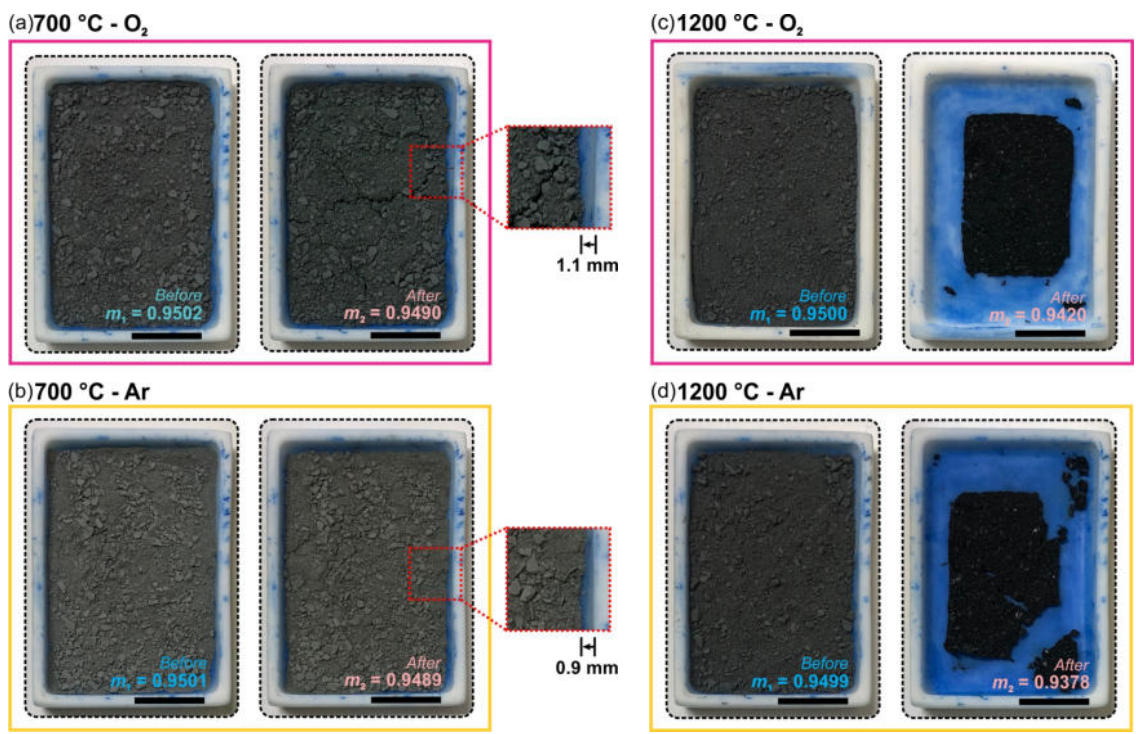

**Figure S18.** Representative micrographs for the *m*Co:ZnO samples before and after the heat treatments in O$_2$ and Ar at the temperatures of 700 °C and 1200 °C. A relatively small shrinkage is observed for the heat treatment at 700 °C, while an evident densification occurs at 1200 °C.

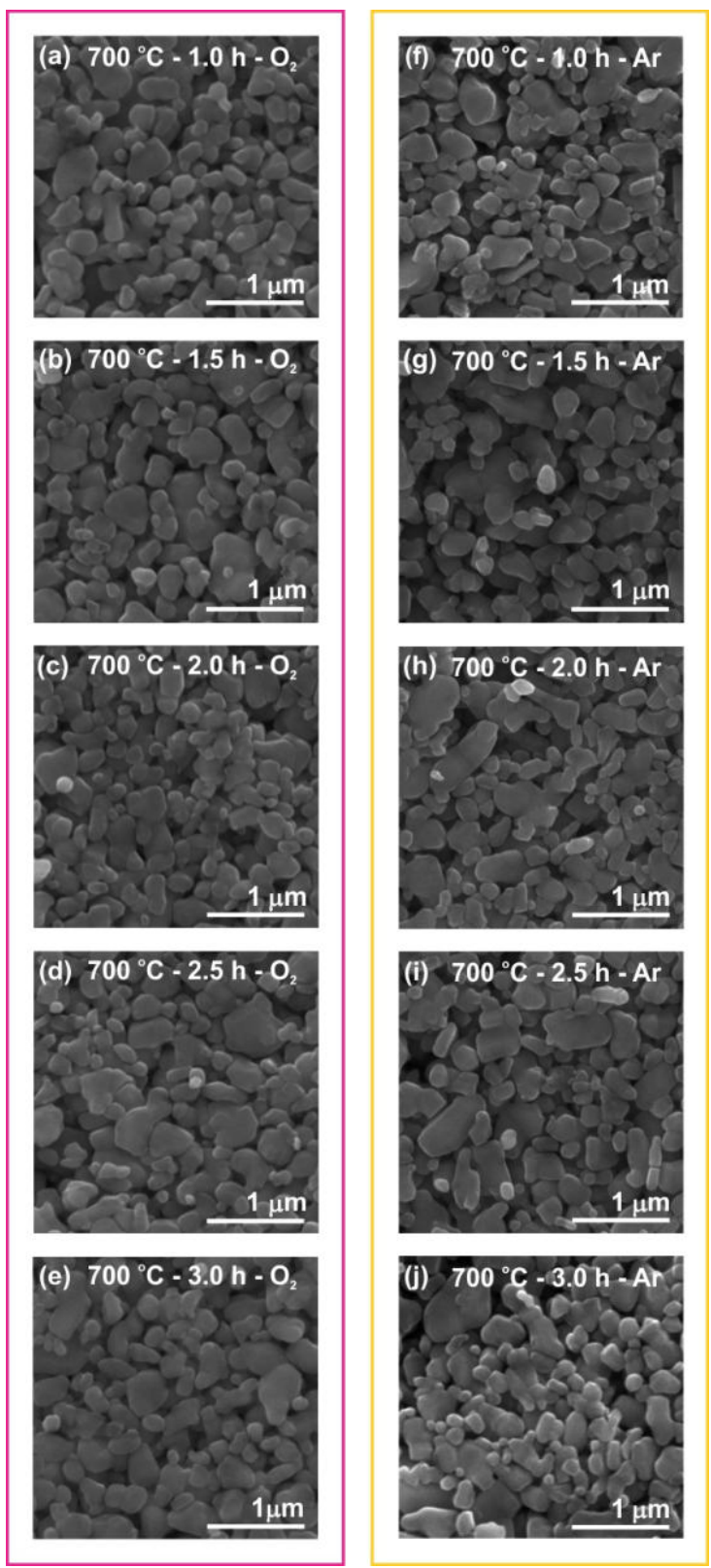

**Figure S19** - Representative images by scanning electron microscopy of the surfaces of **ZnO** samples heat treated at **700 ºC** in $O_2$ for (a) 1.0 h , (b) 1.5 h, (c) 2.0 h, (d) 2.5 h, (e) 3.0 h, and in Ar for (f) 1.0 h, (g) 1.5 h, (h) 2.0 h, (i) 2.5 h, and (j) 3.0 h.

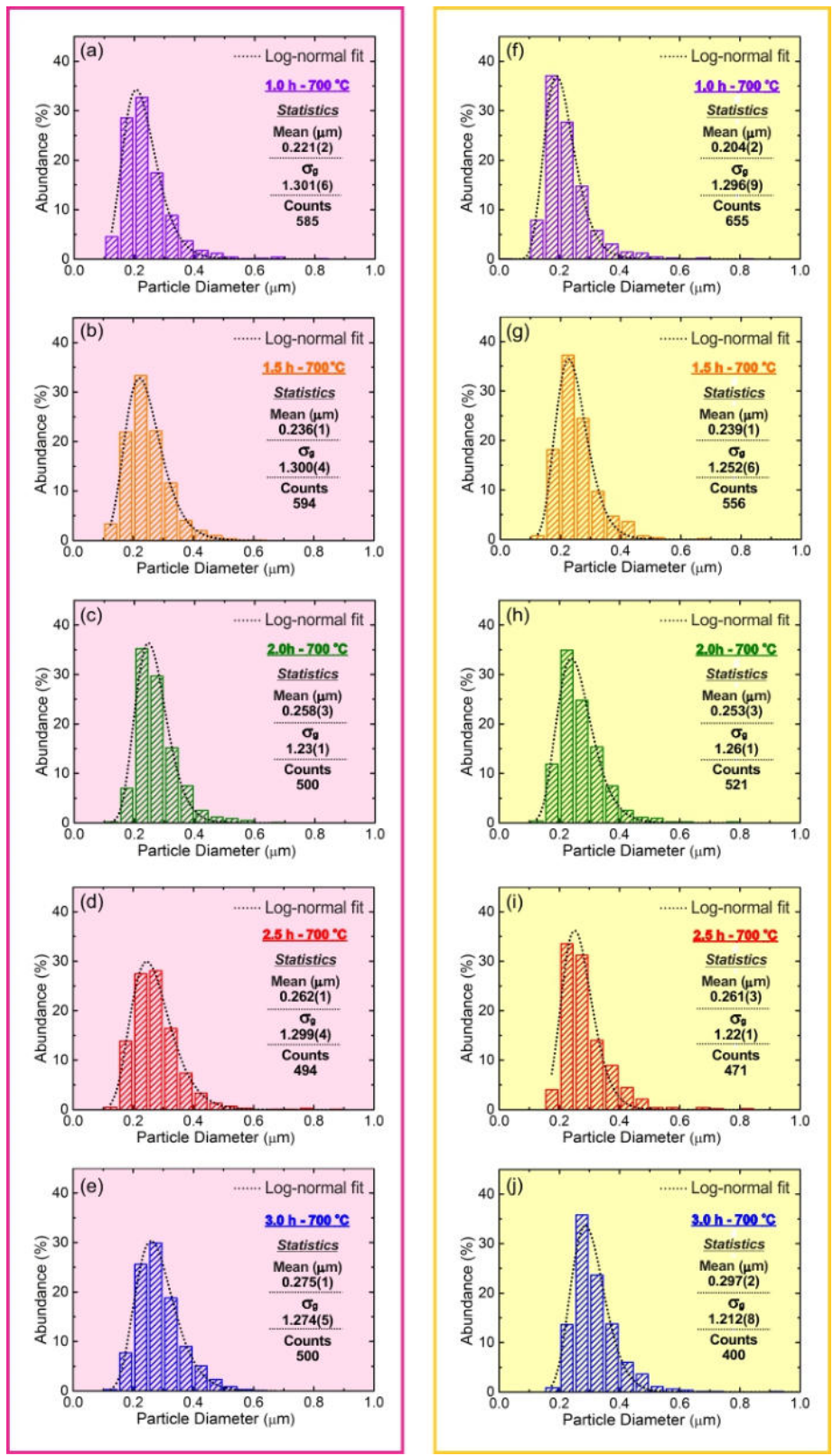

**Figure S20 -** Particle size distribution histograms of the **ZnO** samples heat treated at **700 °C** in O$_2$ for (a) 1.0 h, (b) 1.5 h, (c) 2.0 h, (d) 2.5 h, (e) 3.0 h, and in Ar for (f) 1.0 h, (g) 1.5 h, (h) 2.0 h, (i) 2.5 h, and (j) 3.0 h. The line in the panels correspond to the log-normal fit.

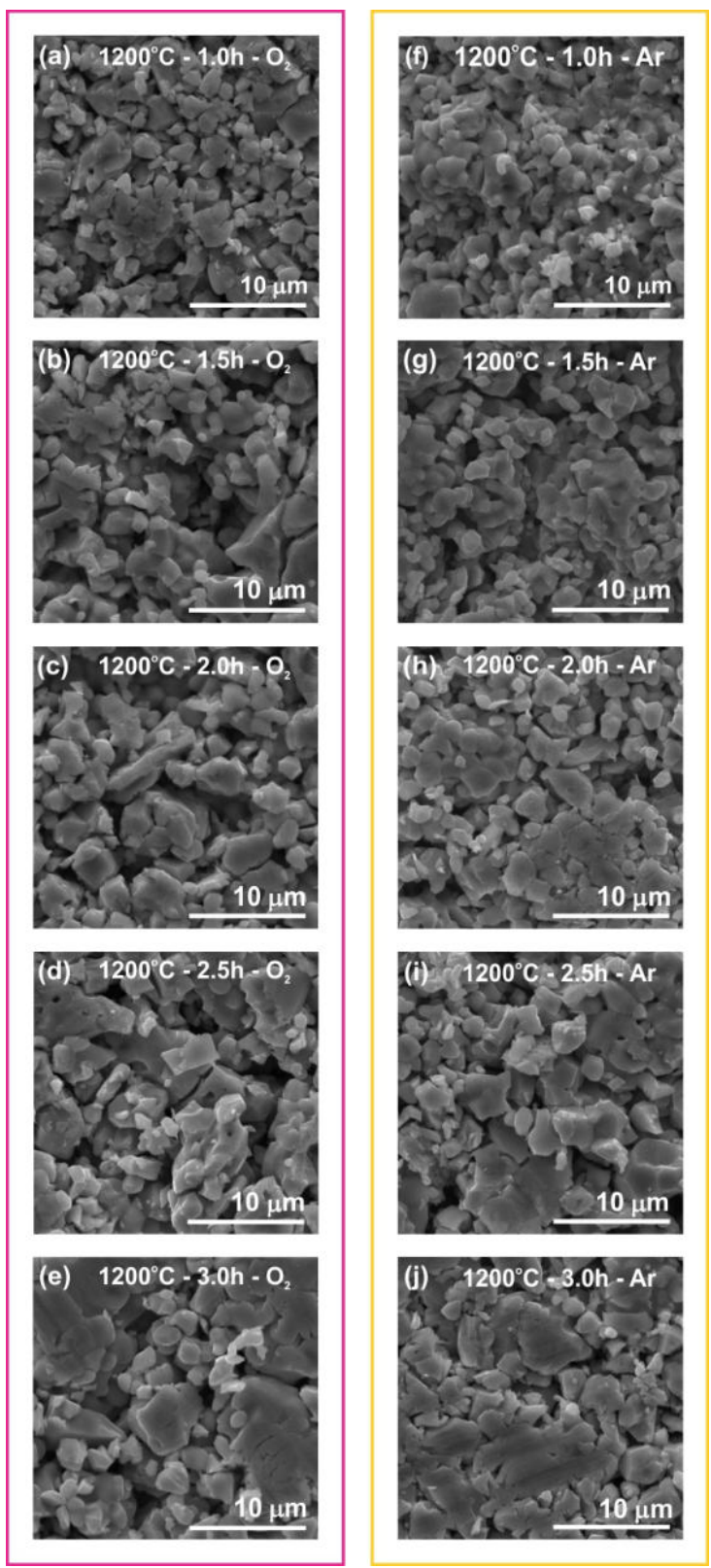

**Figure S21 -** Representative images by scanning electron microscopy of the surfaces of **ZnO** samples heat treated at **1200 ºC** in $O_2$ for (a) 1.0 h , (b) 1.5 h, (c) 2.0 h, (d) 2.5 h, (e) 3.0 h, and in Ar for (f) 1.0 h, (g) 1.5 h, (h) 2.0 h, (i) 2.5 h, and (j) 3.0 h.

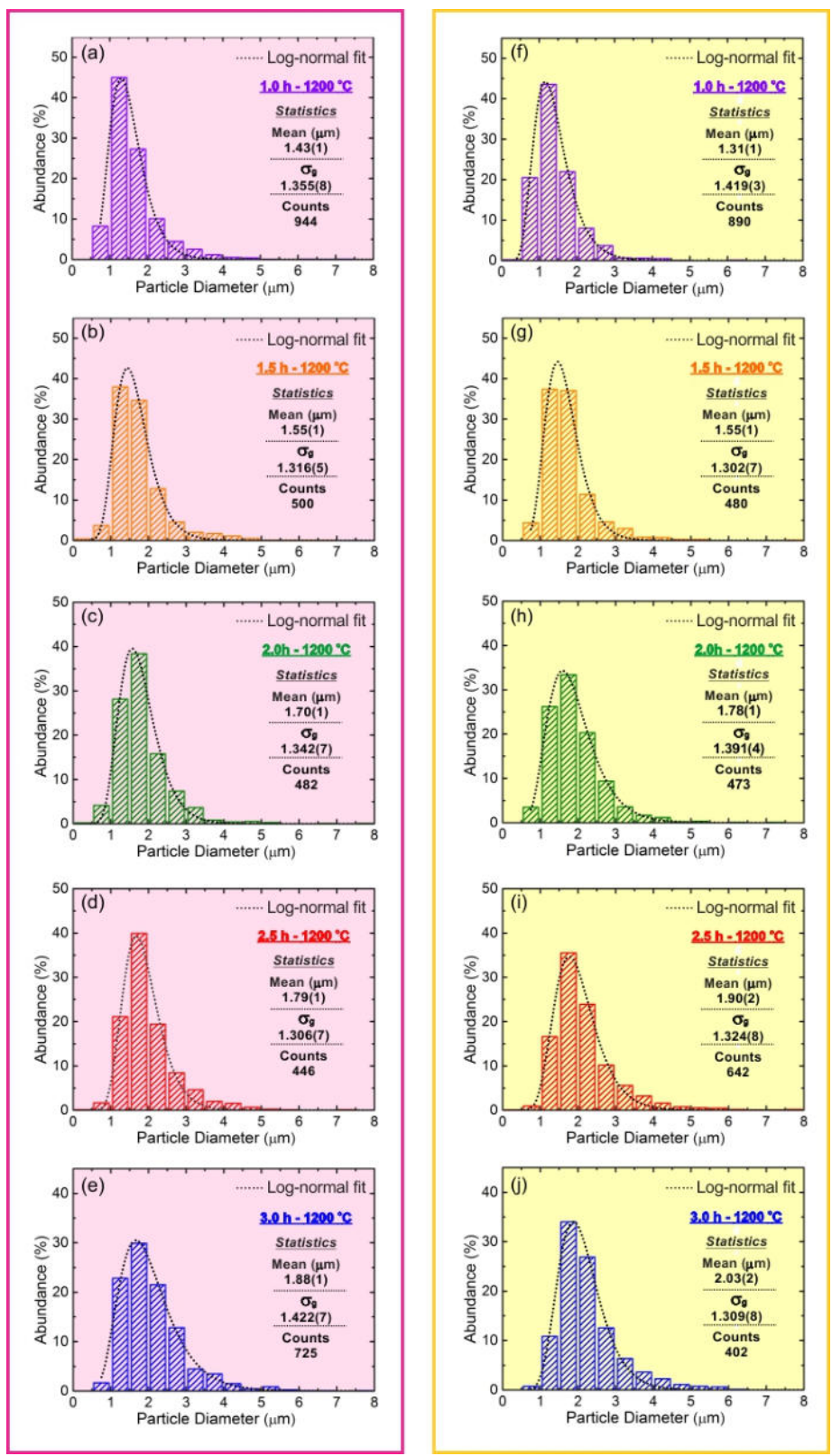

**Figure S22 -** Particle size distribution histograms of the **ZnO** samples heat treated at **1200 °C** in $O_2$ for (a) 1.0 h, (b) 1.5 h, (c) 2.0 h, (d) 2.5 h, (e) 3.0 h, and in Ar for (f) 1.0 h, (g) 1.5 h, (h) 2.0 h, (i) 2.5 h, and (j) 3.0 h. The line in the panels correspond to the log-normal fit.

**Table S11** - Particle size distribution analyses of the **ZnO** samples heat treated at **700 °C** at different times. *L* is the mean value of the particle diameter and σ$_g$ is the geometric standard deviation obtained by the log-normal fit of particle size distribution histograms for each sample. *N* is the total number of counted particles.

| | | Sample | *L* (µm) | σ$_g$ | *N* |
|---|---|---|---|---|---|
| **ZnO – 700 °C** | Oxygen | 1.0h | 0.221(2) | 1.301(6) | 585 |
| | | 1.5h | 0.236(1) | 1.300(4) | 594 |
| | | 2.0h | 0.258(3) | 1.23(1) | 500 |
| | | 2.5h | 0.262(1) | 1.299(4) | 494 |
| | | 3.0h | 0.275(1) | 1.274(5) | 500 |
| | Argon | 1.0h | 0.204(2) | 1.296(9) | 655 |
| | | 1.5h | 0.239(1) | 1.252(6) | 556 |
| | | 2.0h | 0.253(3) | 1.26(1) | 521 |
| | | 2.5h | 0.261(3) | 1.22(1) | 471 |
| | | 3.0h | 0.297(2) | 1.212(8) | 400 |

**Table S12** - Particle size distribution analyses of the **ZnO** samples heat treated at **1200 °C** at different times. *L* is the mean value of the particle diameter and σ$_g$ is the geometric standard deviation obtained by the log-normal fit of particle size distribution histograms for each sample. *N* is the total number of counted particles.

| | | Sample | *L* (µm) | σ$_g$ | *N* |
|---|---|---|---|---|---|
| **ZnO – 1200 °C** | Oxygen | 1.0h | 1.43(1) | 1.355(8) | 944 |
| | | 1.5h | 1.55(1) | 1.316(5) | 500 |
| | | 2.0h | 1.70(1) | 1.342(7) | 482 |
| | | 2.5h | 1.79(1) | 1.306(7) | 446 |
| | | 3.0h | 1.88(1) | 1.422(7) | 725 |
| | Argon | 1.0h | 1.31(1) | 1.419(3) | 890 |
| | | 1.5h | 1.55(1) | 1.302(7) | 480 |
| | | 2.0h | 1.78(1) | 1.391(4) | 473 |
| | | 2.5h | 1.90(2) | 1.324(8) | 642 |
| | | 3.0h | 2.03(2) | 1.309(8) | 402 |

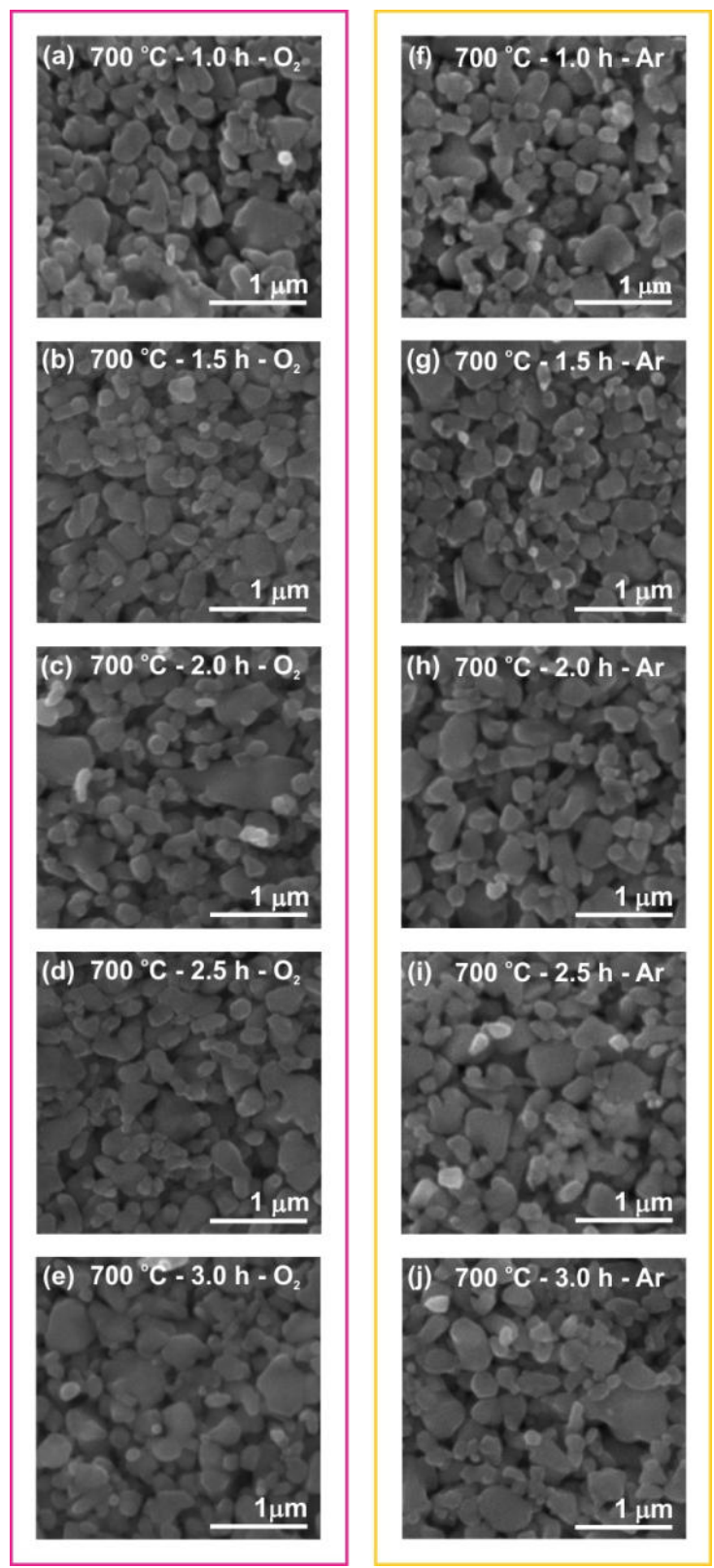

**Figure S23 -** Representative images by scanning electron microscopy of the surfaces of **Co$_3$O$_4$:ZnO** samples heat treated at **700 ºC** in O$_2$ for (a) 1.0 h , (b) 1.5 h, (c) 2.0 h, (d) 2.5 h, (e) 3.0 h, and in Ar for (f) 1.0 h, (g) 1.5 h, (h) 2.0 h, (i) 2.5 h, and (j) 3.0 h.

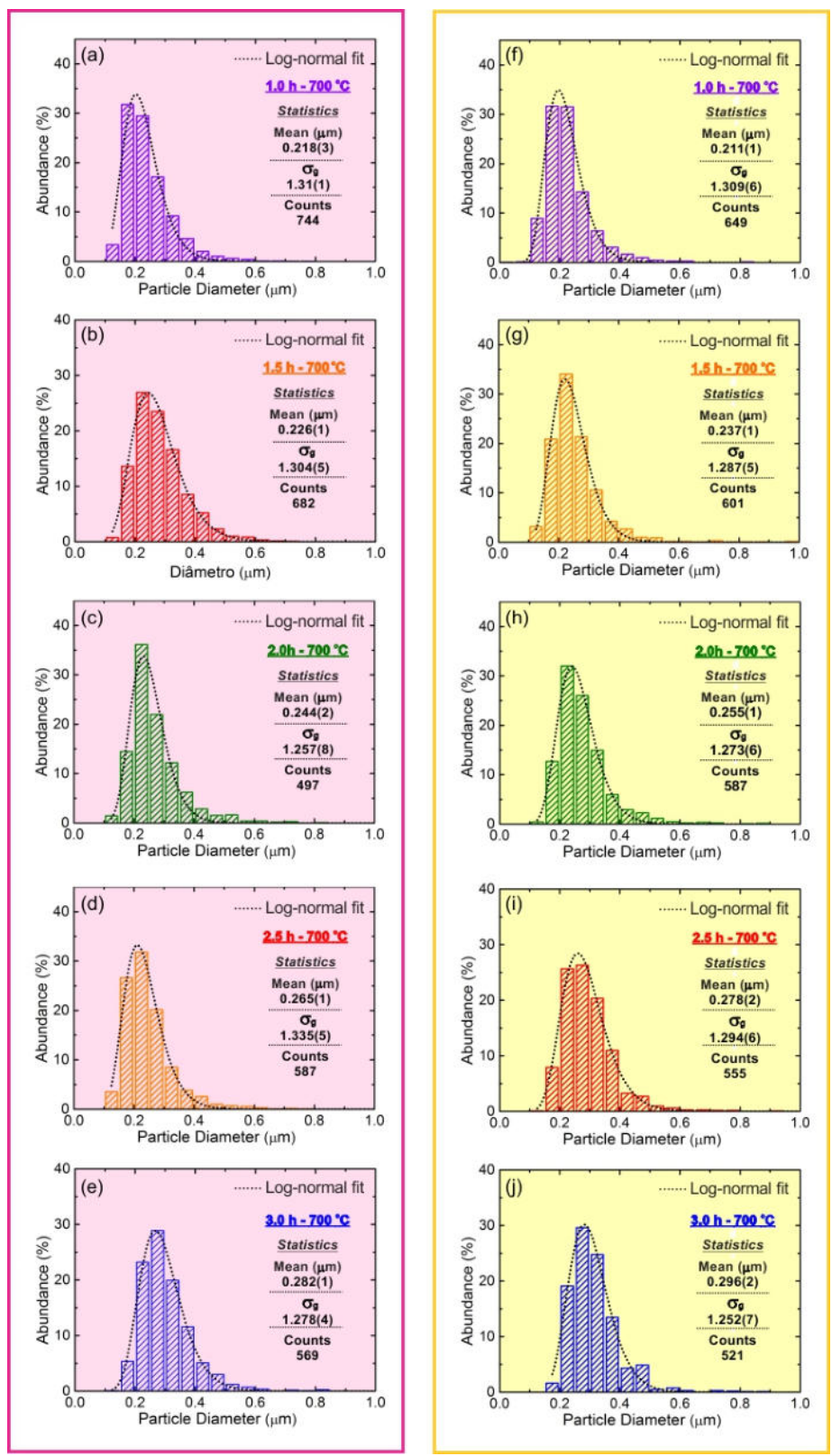

**Figure S24 -** Particle size distribution histograms of the **Co₃O₄:ZnO** samples heat treated at **700 °C** in $O_2$ for (a) 1.0 h, (b) 1.5 h, (c) 2.0 h, (d) 2.5 h, (e) 3.0 h, and in Ar for (f) 1.0 h, (g) 1.5 h, (h) 2.0 h, (i) 2.5 h, and (j) 3.0 h. The line in the panels correspond to the log-normal fit.

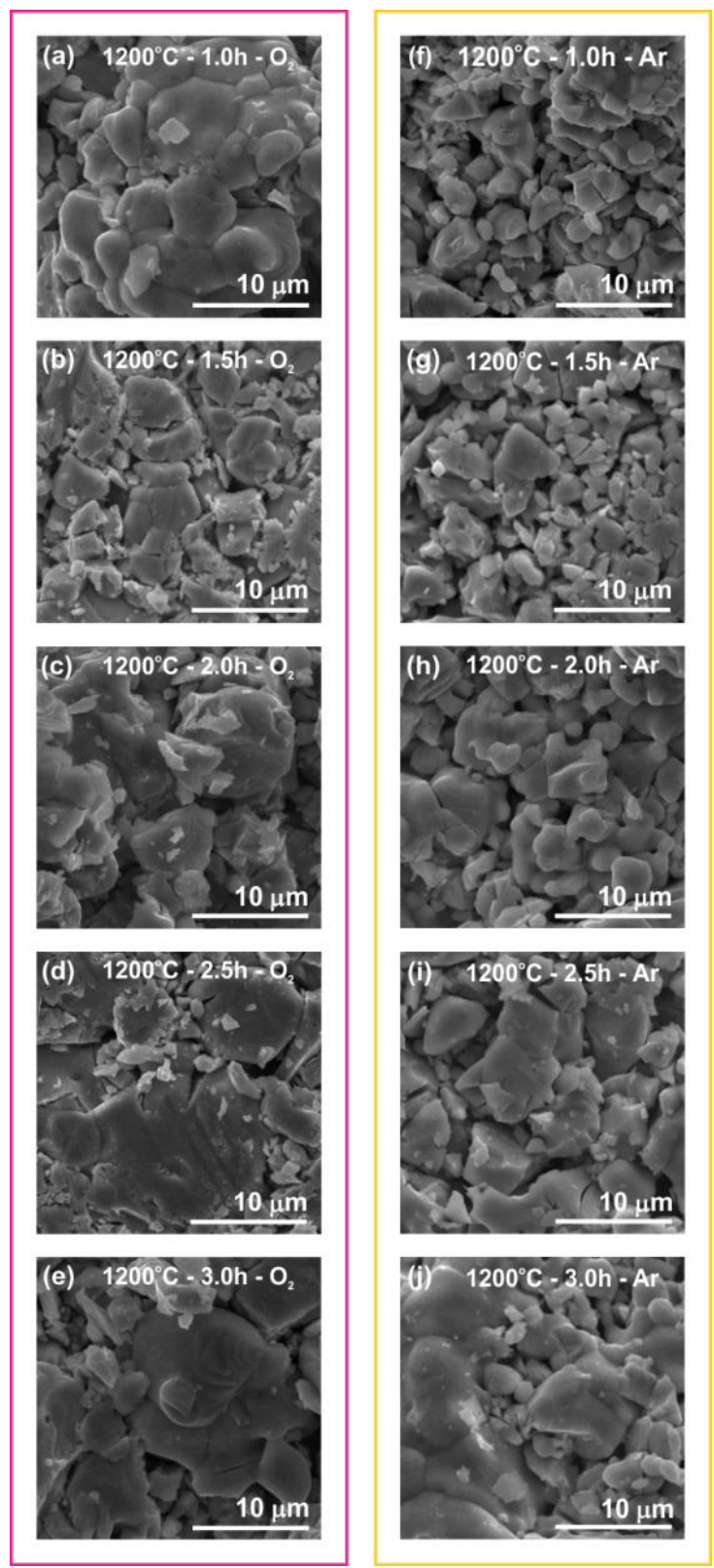

**Figure S25 -** Representative images by scanning electron microscopy of the surfaces of **Co$_3$O$_4$:ZnO** samples heat treated at **1200 ºC** in O$_2$ for (a) 1.0 h , (b) 1.5 h, (c) 2.0 h, (d) 2.5 h, (e) 3.0 h, and in Ar for (f) 1.0 h, (g) 1.5 h, (h) 2.0 h, (i) 2.5 h, and (j) 3.0 h.

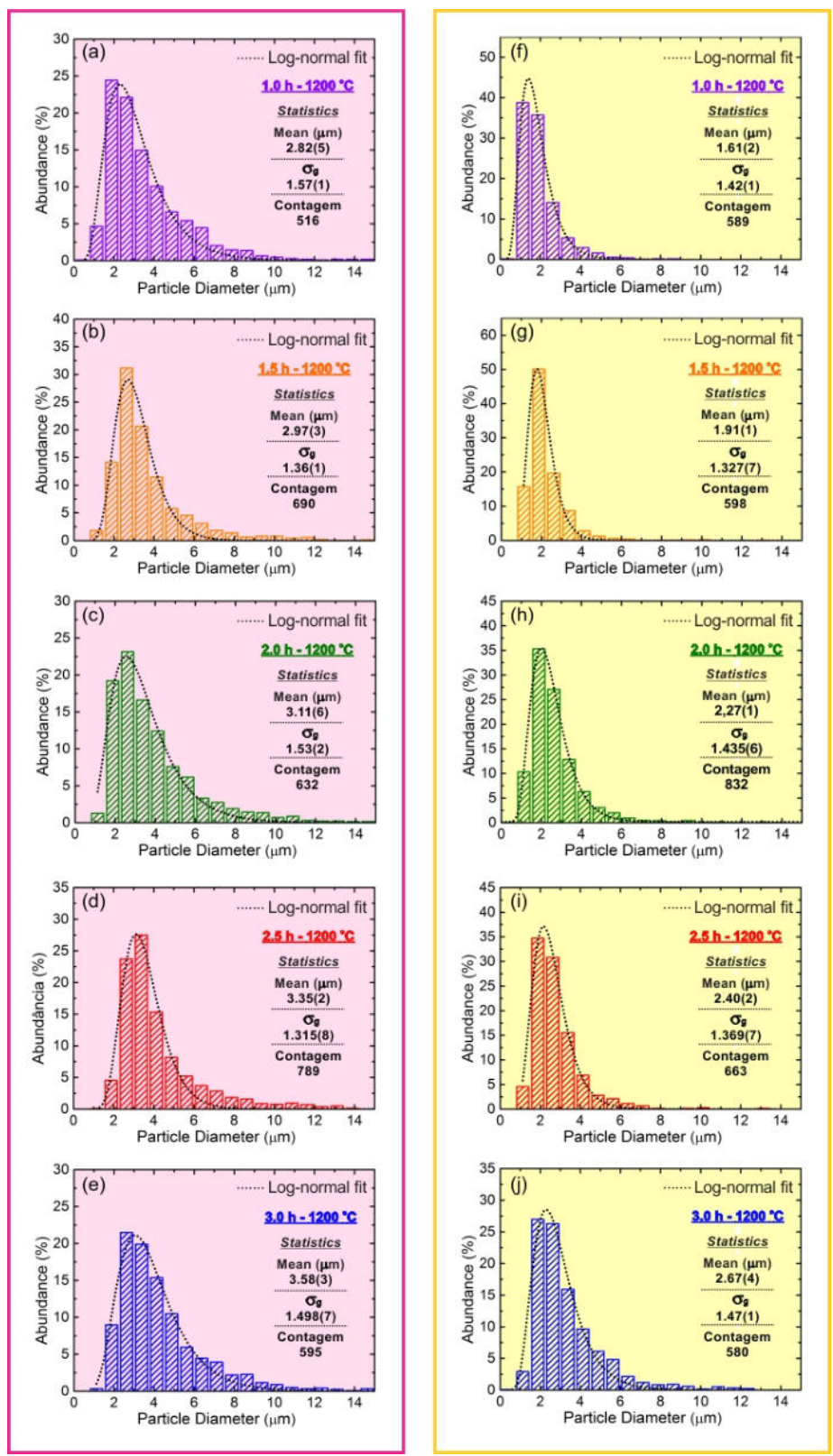

**Figure S26 -** Particle size distribution histograms of the **Co₃O₄:ZnO** samples heat treated at **1200 °C** in O$_2$ for (a) 1.0 h, (b) 1.5 h, (c) 2.0 h, (d) 2.5 h, (e) 3.0 h, and in Ar for (f) 1.0 h, (g) 1.5 h, (h) 2.0 h, (i) 2.5 h, and (j) 3.0 h. The line in the panels correspond to the log-normal fit.

**Table S13** - Particle size distribution analyses of the **Co₃O₄:ZnO** samples heat treated at **700 °C** at different times. $L$ is the mean value of the particle diameter and $\sigma_g$ is the geometric standard deviation obtained by the log-normal fit of particle size distribution histograms for each sample. $N$ is the total number of counted particles.

| | | Sample | $L$ (µm) | $\sigma_g$ | $N$ |
|---|---|---|---|---|---|
| Co₃O₄:ZnO – 700 °C | Oxygen | 1.0h | 0.218(3) | 1.31(1) | 744 |
| | | 1.5h | 0.226(1) | 1.304(5) | 682 |
| | | 2.0h | 0.244(2) | 1.257(8) | 497 |
| | | 2.5h | 0.265(1) | 1.335(5) | 587 |
| | | 3.0h | 0.282(1) | 1.278(4) | 569 |
| | Argon | 1.0h | 0.211(1) | 1.309(6) | 649 |
| | | 1.5h | 0.237(1) | 1.287(5) | 601 |
| | | 2.0h | 0.255(1) | 1.273(6) | 587 |
| | | 2.5h | 0.278(2) | 1.294(6) | 555 |
| | | 3.0h | 0.296(2) | 1.252(7) | 521 |

**Table S14** - Particle size distribution analyses of the **Co₃O₄:ZnO** samples heat treated at **1200 °C** at different times. $L$ is the mean value of the particle diameter and $\sigma_g$ is the geometric standard deviation obtained by the log-normal fit of particle size distribution histograms for each sample. $N$ is the total number of counted particles.

| | | Sample | $L$ (µm) | $\sigma_g$ | $N$ |
|---|---|---|---|---|---|
| Co₃O₄:ZnO – 1200 °C | Oxygen | 1.0h | 2.82(5) | 1.57(1) | 516 |
| | | 1.5h | 2.97(3) | 1.36(1) | 690 |
| | | 2.0h | 3.11(6) | 1.53(2) | 632 |
| | | 2.5h | 3.35(2) | 1.315(8) | 789 |
| | | 3.0h | 3.58(3) | 1.498(7) | 595 |
| | Argon | 1.0h | 1.61(2) | 1.42(1) | 589 |
| | | 1.5h | 1.91(1) | 1.327(7) | 598 |
| | | 2.0h | 2.27(1) | 1.435(6) | 832 |
| | | 2.5h | 2.40(2) | 1.369(7) | 663 |
| | | 3.0h | 2.67(4) | 1.47(1) | 580 |

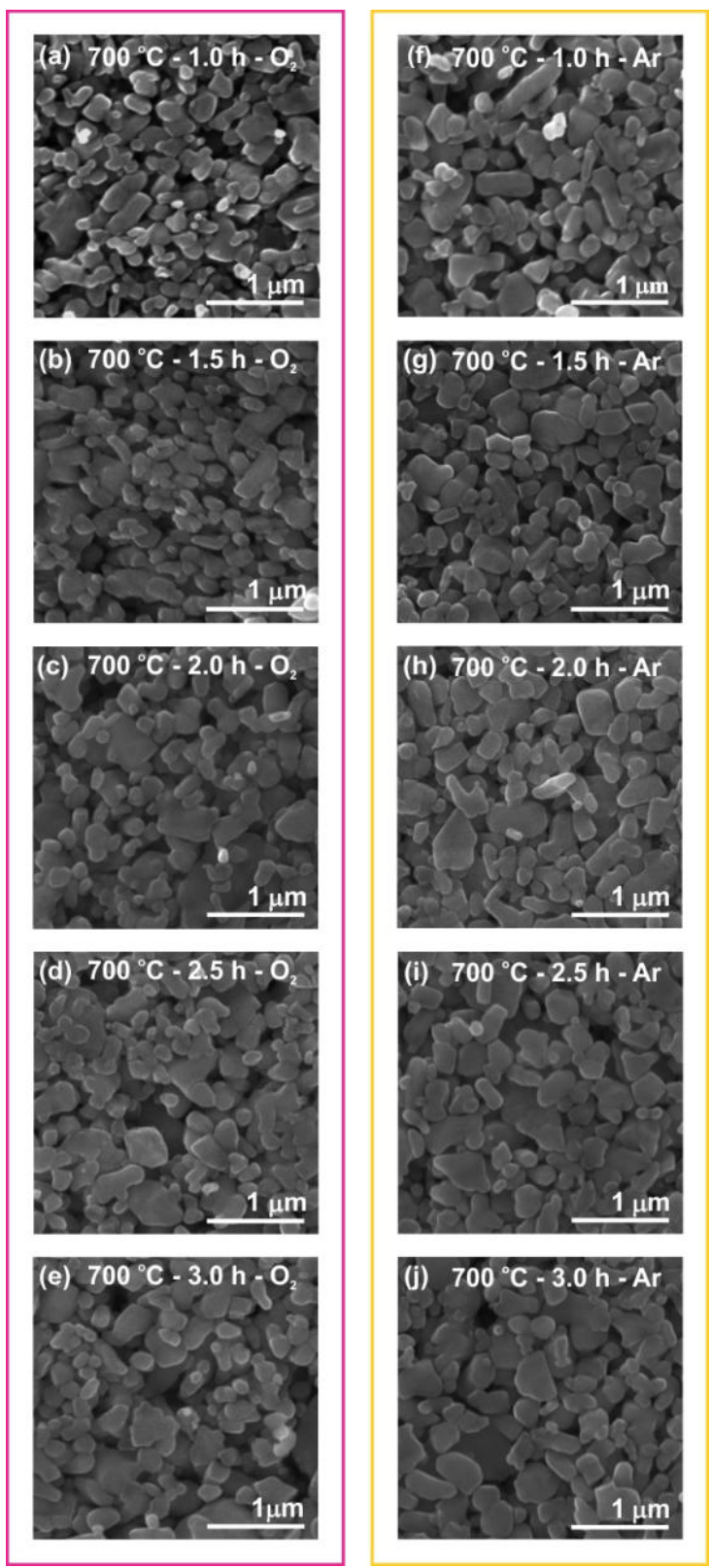

**Figure S27 -** Representative images by scanning electron microscopy of the surfaces of **CoO:ZnO** samples heat treated at **700 ºC** in $O_2$ for (a) 1.0 h , (b) 1.5 h, (c) 2.0 h, (d) 2.5 h, (e) 3.0 h, and in Ar for (f) 1.0 h, (g) 1.5 h, (h) 2.0 h, (i) 2.5 h, and (j) 3.0 h.

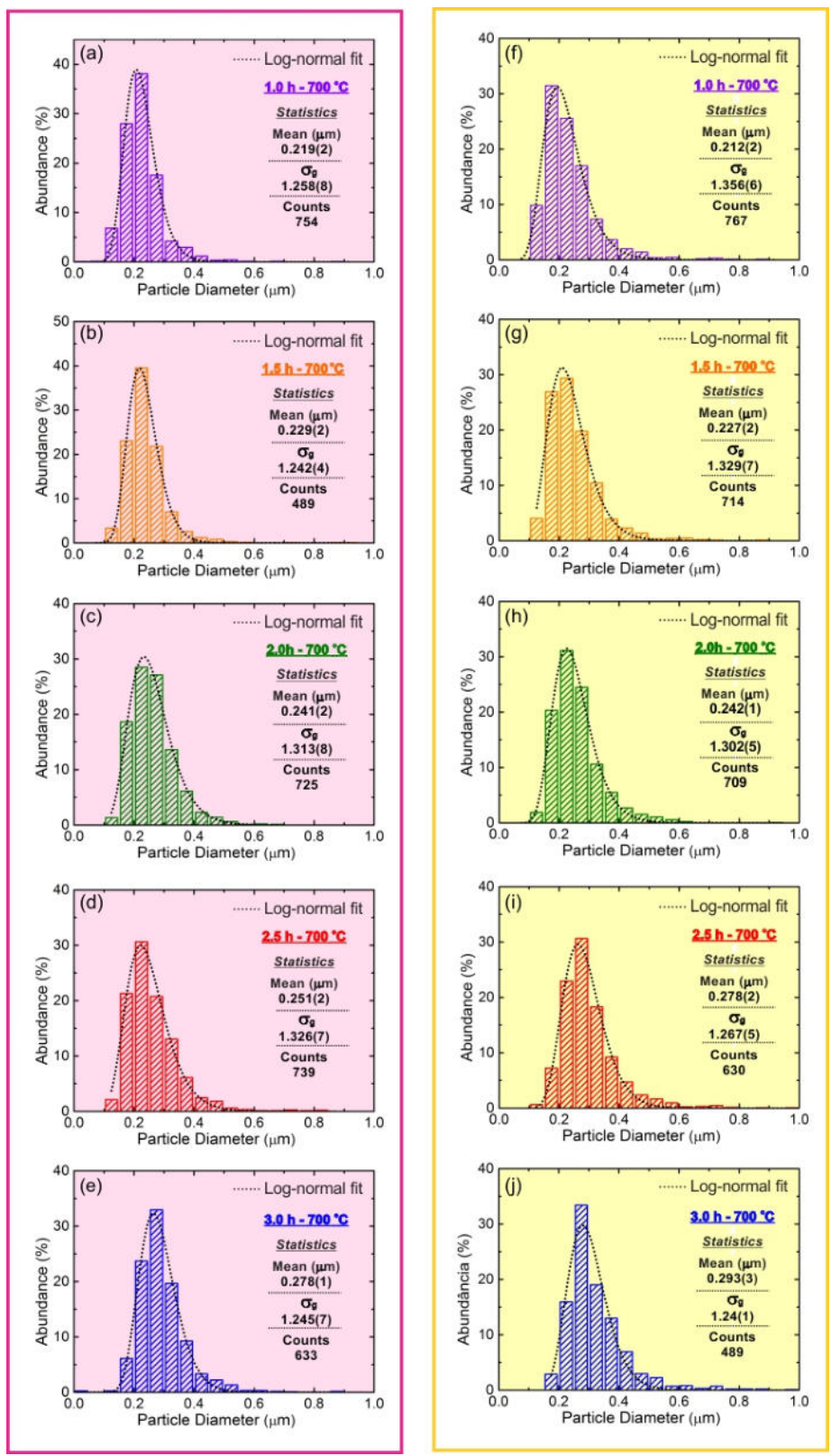

**Figure S28 -** Particle size distribution histograms of the **CoO:ZnO** samples heat treated at **700 ºC** in $O_2$ for (a) 1.0 h, (b) 1.5 h, (c) 2.0 h, (d) 2.5 h, (e) 3.0 h, and in Ar for (f) 1.0 h, (g) 1.5 h, (h) 2.0 h, (i) 2.5 h, and (j) 3.0 h. The line in the panels correspond to the log-normal fit.

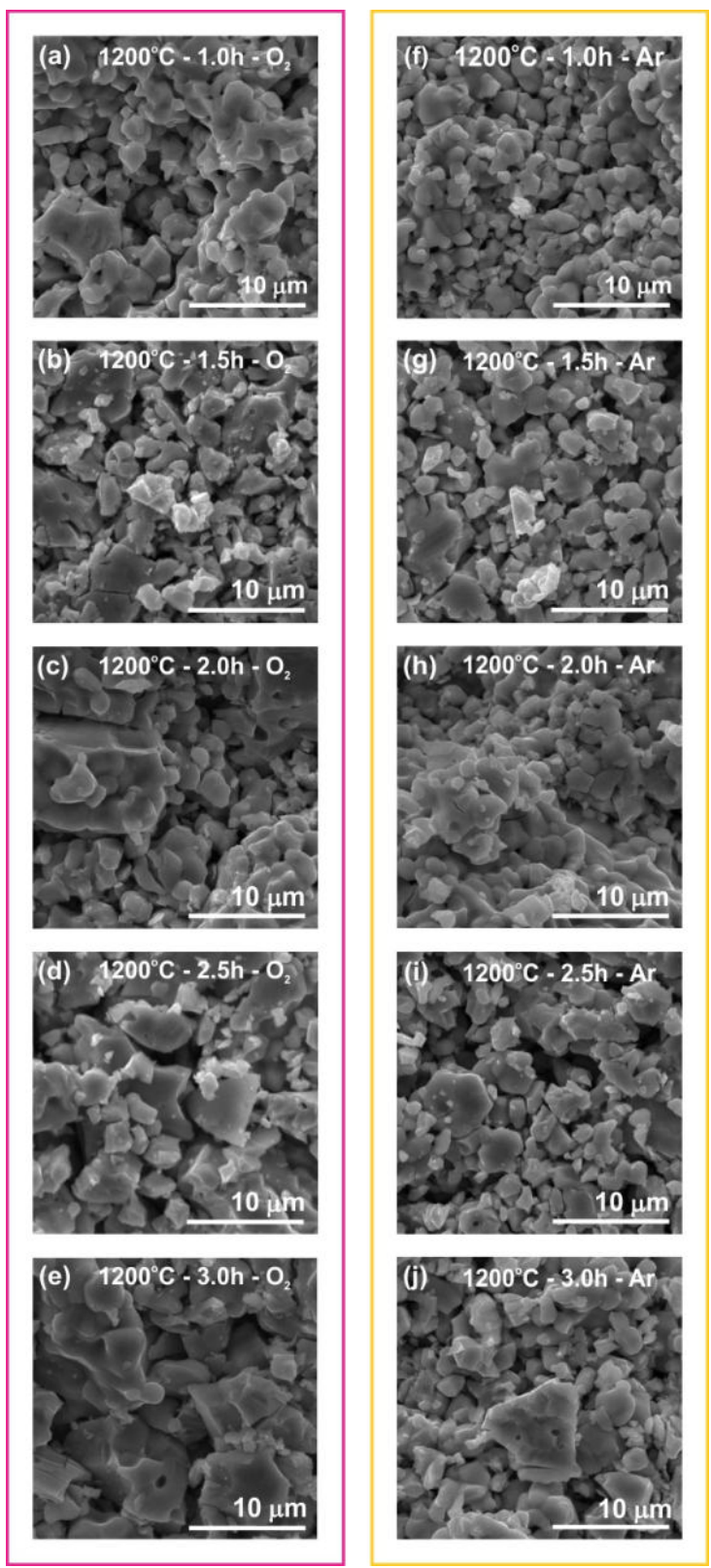

**Figure S29** - Representative images by scanning electron microscopy of the surfaces of **CoO:ZnO** samples heat treated at **1200 ºC** in $O_2$ for (a) 1.0 h , (b) 1.5 h, (c) 2.0 h, (d) 2.5 h, (e) 3.0 h, and in Ar for (f) 1.0 h, (g) 1.5 h, (h) 2.0 h, (i) 2.5 h, and (j) 3.0 h.

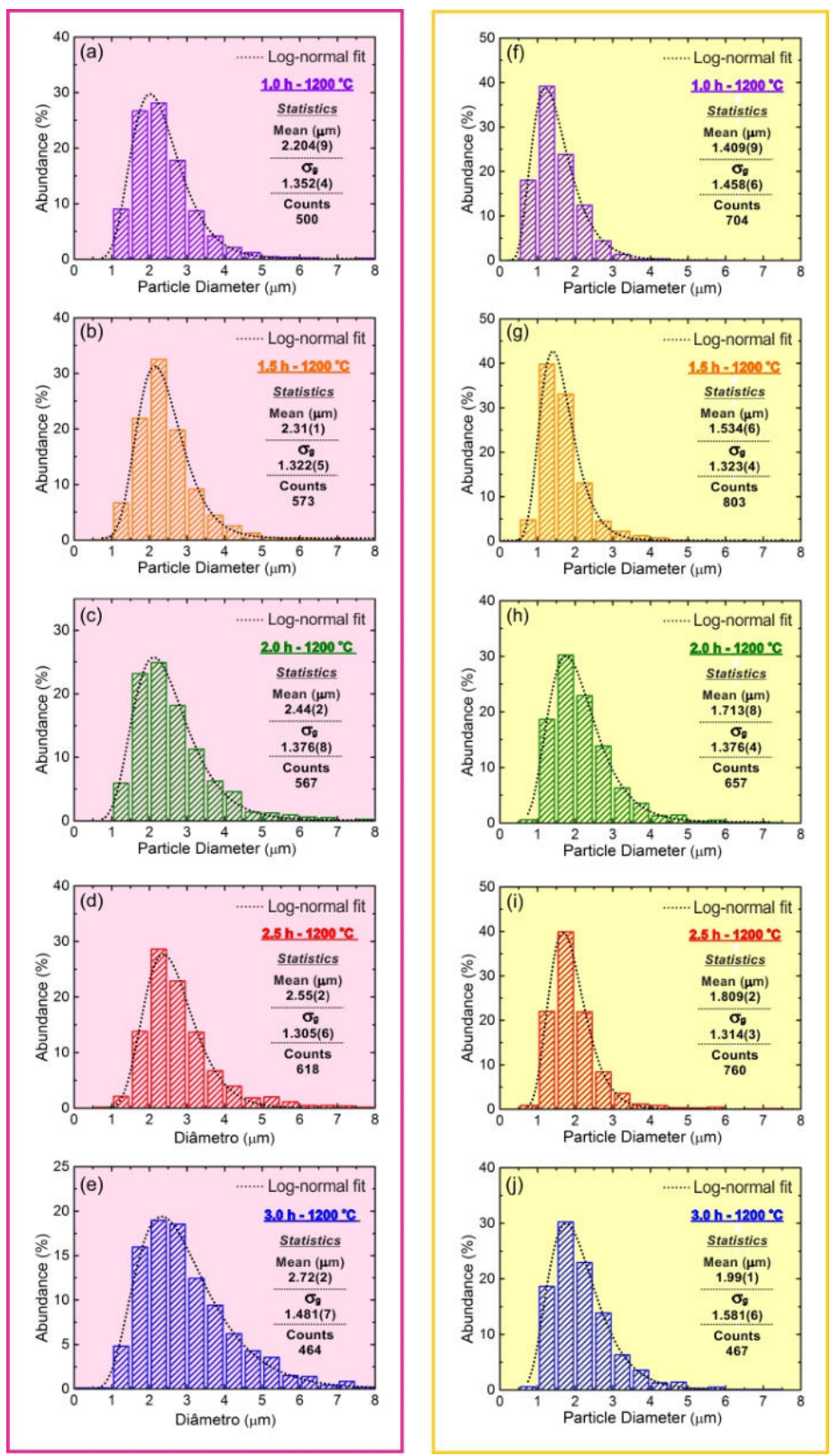

**Figure S30 -** Particle size distribution histograms of the **CoO:ZnO** samples heat treated at **1200 ºC** in $O_2$ for (a) 1.0 h, (b) 1.5 h, (c) 2.0 h, (d) 2.5 h, (e) 3.0 h, and in Ar for (f) 1.0 h, (g) 1.5 h, (h) 2.0 h, (i) 2.5 h, and (j) 3.0 h. The line in the panels correspond to the log-normal fit.

**Table S15** - Particle size distribution analyses of the **CoO:ZnO** samples heat treated at **700 °C** at different times. *L* is the mean value of the particle diameter and σ$_g$ is the geometric standard deviation obtained by the log-normal fit of particle size distribution histograms for each sample. *N* is the total number of counted particles.

| | | Sample | *L* (μm) | σ$_g$ | *N* |
|---|---|---|---|---|---|
| **CoO:ZnO – 700 °C** | Oxygen | 1.0h | 0.219(2) | 1.258(8) | 754 |
| | | 1.5h | 0.229(1) | 1.242(4) | 707 |
| | | 2.0h | 0.241(2) | 1.313(8) | 725 |
| | | 2.5h | 0.251(2) | 1.326(7) | 739 |
| | | 3.0h | 0.278(1) | 1.245(5) | 633 |
| | Argon | 1.0h | 0.212(2) | 1.356(6) | 767 |
| | | 1.5h | 0.227(2) | 1.329(7) | 714 |
| | | 2.0h | 0.242(1) | 1.302(5) | 709 |
| | | 2.5h | 0.278(2) | 1.267(5) | 630 |
| | | 3.0h | 0.293(3) | 1.24(1) | 489 |

**Table S16** - Particle size distribution analyses of the **CoO:ZnO** samples heat treated at **1200 °C** at different times. *L* is the mean value of the particle diameter and σ$_g$ is the geometric standard deviation obtained by the log-normal fit of particle size distribution histograms for each sample. *N* is the total number of counted particles.

| | | Sample | *L* (μm) | σ$_g$ | *N* |
|---|---|---|---|---|---|
| **CoO:ZnO – 1200 °C** | Oxygen | 1.0h | 2.204(9) | 1.352(4) | 500 |
| | | 1.5h | 2.31(1) | 1.322(5) | 573 |
| | | 2.0h | 2.44(2) | 1.376(8) | 580 |
| | | 2.5h | 2.55(2) | 1.305(6) | 618 |
| | | 3.0h | 2.72(2) | 1.481(7) | 467 |
| | Argon | 1.0h | 1.409(9) | 1.458(6) | 704 |
| | | 1.5h | 1.534(6) | 1.323(4) | 803 |
| | | 2.0h | 1.713(8) | 1.376(4) | 657 |
| | | 2.5h | 1.809(2) | 1.314(3) | 760 |
| | | 3.0h | 1.99(1) | 1.581(6) | 491 |

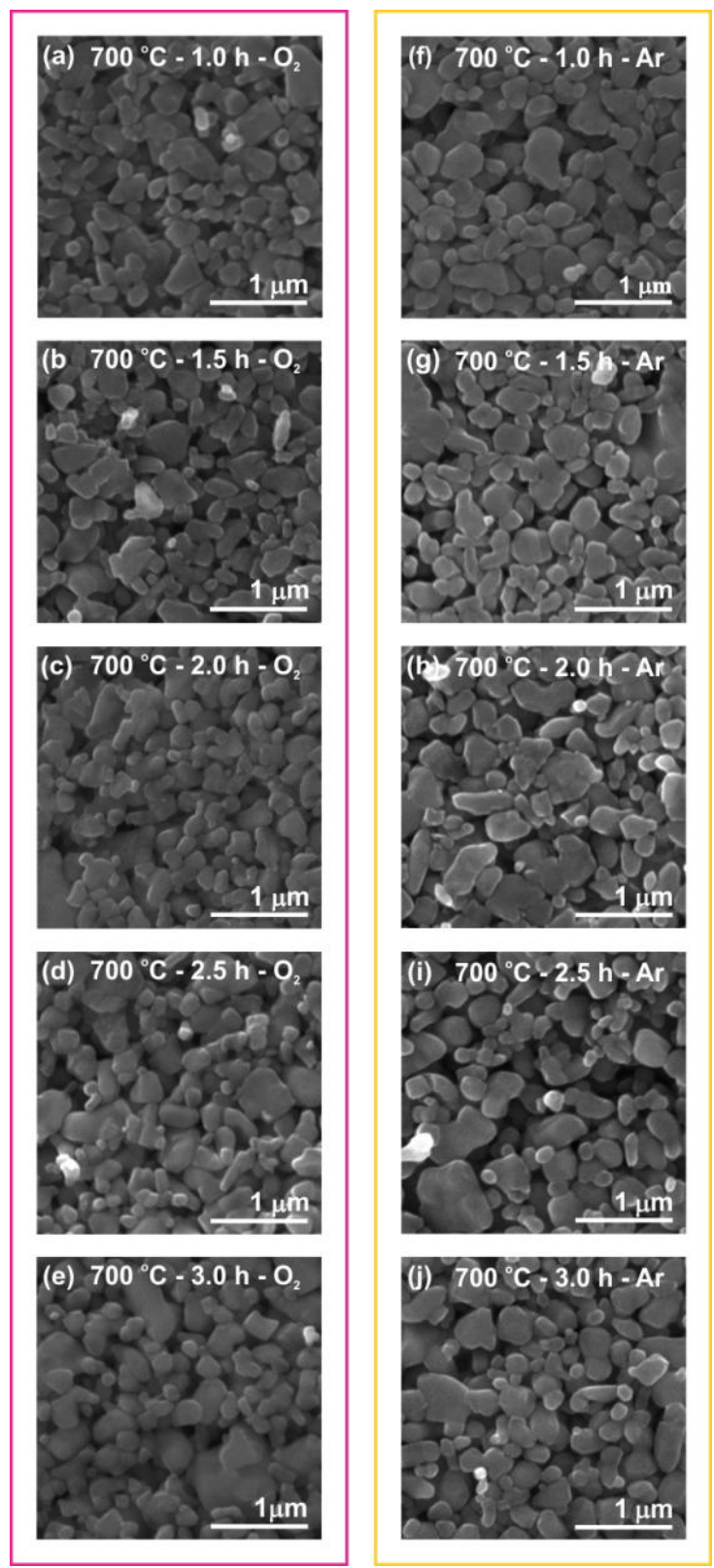

**Figure S31 -** Representative images by scanning electron microscopy of the surfaces of *m*Co:ZnO samples heat treated at **700 °C** in $O_2$ for (a) 1.0 h, (b) 1.5 h, (c) 2.0 h, (d) 2.5 h, (e) 3.0 h, and in Ar for (f) 1.0 h, (g) 1.5 h, (h) 2.0 h, (i) 2.5 h, and (j) 3.0 h.

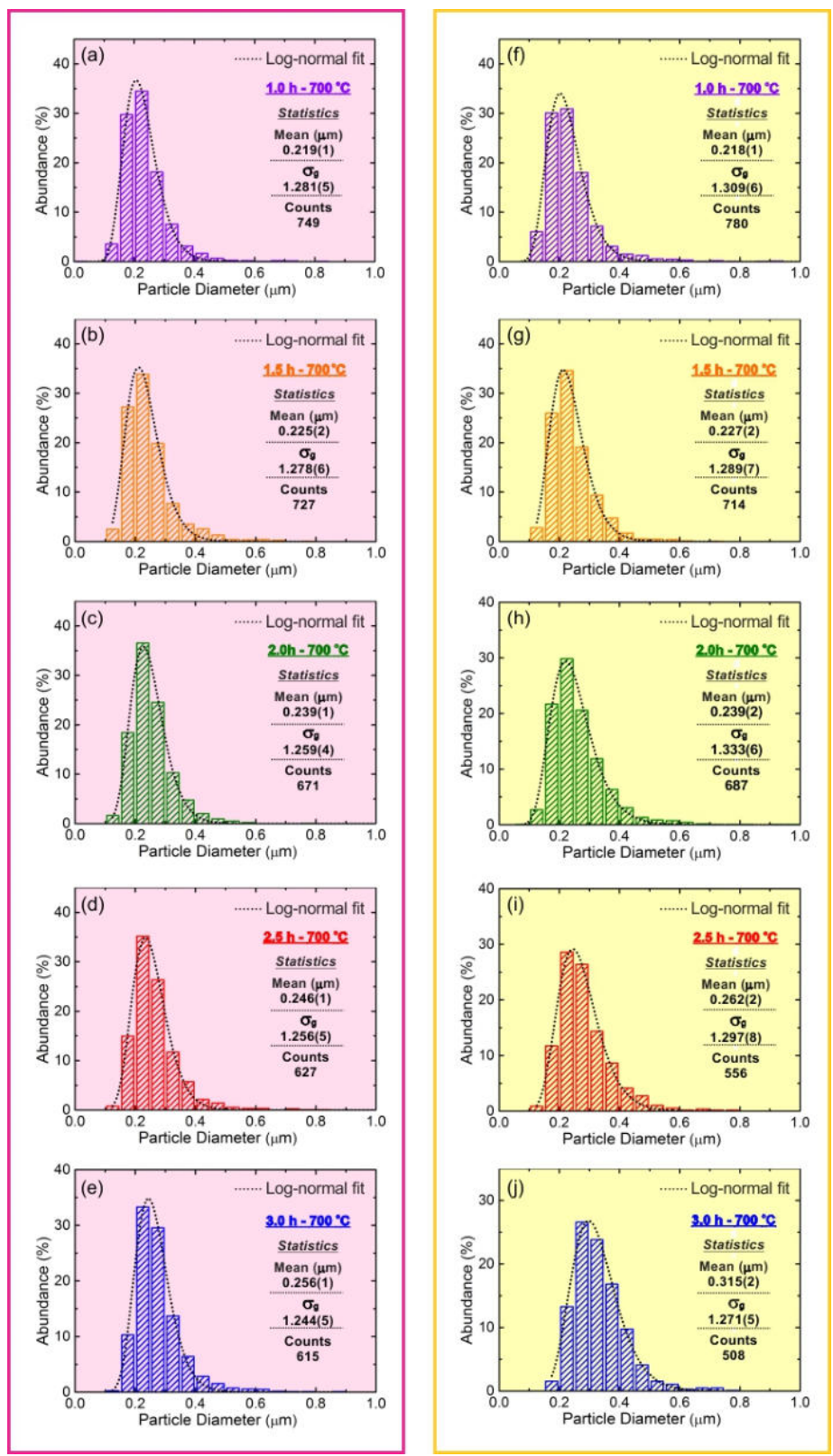

**Figure S32 -** Particle size distribution histograms of the *m*Co:ZnO samples heat treated at **700 ºC** in $O_2$ for (a) 1.0 h, (b) 1.5 h, (c) 2.0 h, (d) 2.5 h, (e) 3.0 h, and in Ar for (f) 1.0 h, (g) 1.5 h, (h) 2.0 h, (i) 2.5 h, and (j) 3.0 h. The line in the panels correspond to the log-normal fit.

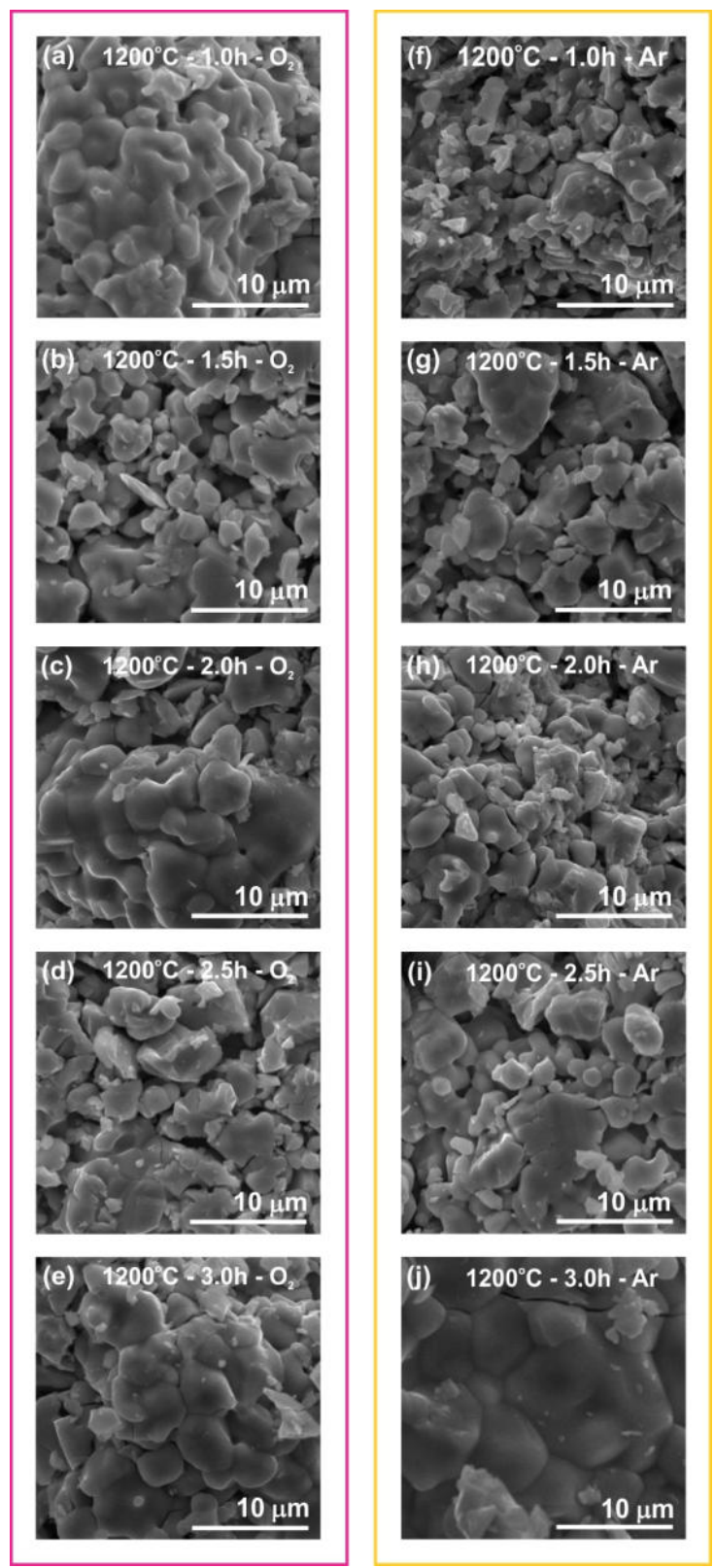

**Figure S33 -** Representative images by scanning electron microscopy of the surfaces of *m*Co:ZnO samples heat treated at **1200 ºC** in $O_2$ for (a) 1.0 h , (b) 1.5 h, (c) 2.0 h, (d) 2.5 h, (e) 3.0 h, and in Ar for (f) 1.0 h, (g) 1.5 h, (h) 2.0 h, (i) 2.5 h, and (j) 3.0 h.

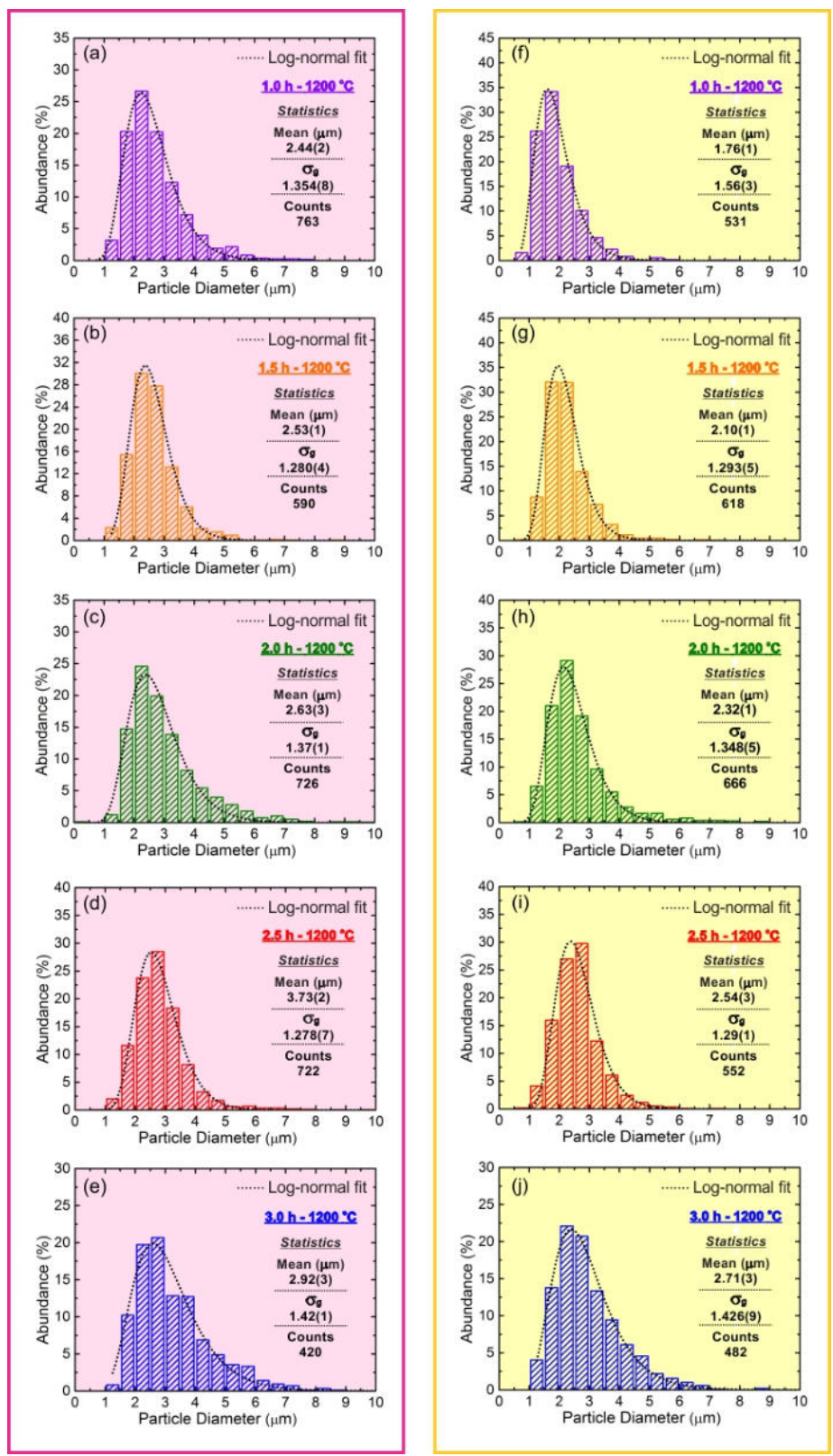

**Figure S34** - Particle size distribution histograms of the *m*Co:ZnO samples heat treated at **1200 ºC** in $O_2$ for (a) 1.0 h , (b) 1.5 h, (c) 2.0 h, (d) 2.5 h, (e) 3.0 h, and in Ar for (f) 1.0 h, (g) 1.5 h, (h) 2.0 h, (i) 2.5 h, and (j) 3.0 h. The line in the panels correspond to the log-normal fit.

**Table S17** - Particle size distribution analyses of the *m*Co:ZnO samples heat treated at **700 °C** at different times. $L$ is the mean value of the particle diameter and $\sigma_g$ is the geometric standard deviation obtained by the log-normal fit of particle size distribution histograms for each sample. $N$ is the total number of counted particles.

| | | Sample | $L$ (µm) | $\sigma_g$ | $N$ |
|---|---|---|---|---|---|
| *m*Co:ZnO – 700 °C | Oxygen | 1.0h | 0.219(1) | 1.281(5) | 749 |
| | | 1.5h | 0.225(2) | 1.278(6) | 727 |
| | | 2.0h | 0.239(1) | 1.259(4) | 671 |
| | | 2.5h | 0.246(1) | 1.256(5) | 627 |
| | | 3.0h | 0.256(1) | 1.244(5) | 615 |
| | Argon | 1.0h | 0.218(1) | 1.309(4) | 780 |
| | | 1.5h | 0.227(2) | 1.289(7) | 714 |
| | | 2.0h | 0.239(2) | 1.333(6) | 687 |
| | | 2.5h | 0.262(2) | 1.297(8) | 556 |
| | | 3.0h | 0.315(2) | 1.271(5) | 508 |

**Table S18** - Particle size distribution analyses of the *m*Co:ZnO samples heat treated at **1200 °C** at different times. $L$ is the mean value of the particle diameter and $\sigma_g$ is the geometric standard deviation obtained by the log-normal fit of particle size distribution histograms for each sample. $N$ is the total number of counted particles.

| | | Sample | $L$ (µm) | $\sigma_g$ | $N$ |
|---|---|---|---|---|---|
| *m*Co:ZnO – 1200 °C | Oxygen | 1.0h | 2.44(2) | 1.354(8) | 763 |
| | | 1.5h | 2.53(1) | 1.280(4) | 590 |
| | | 2.0h | 2.63(3) | 1.37(1) | 726 |
| | | 2.5h | 2.73(2) | 1.278(7) | 722 |
| | | 3.0h | 2.92(3) | 1.42(1) | 420 |
| | Argon | 1.0h | 1.76(1) | 1.56(3) | 531 |
| | | 1.5h | 2.10(1) | 1.293(6) | 618 |
| | | 2.0h | 2.32(1) | 1.348(5) | 666 |
| | | 2.5h | 2.54(3) | 1.29(1) | 552 |
| | | 3.0h | 2.71(3) | 1.426(9) | 482 |